\documentclass[preprintnumbers,floatfix,aps]{revtex4}
\usepackage{amsmath}
\usepackage{amsfonts}
\usepackage{amssymb}
\usepackage{physymb}
\usepackage{graphicx}
\usepackage{epsfig}
\usepackage[hang,nooneline]{subfigure}

\newcounter{fig}

\begin{document}

\title{Vortex--soliton complexes in coupled nonlinear Schr\"{o}dinger
equations with unequal dispersion coefficients}
\author{E. G. Charalampidis\thanks{%
Email: charalamp@math.umass.edu}}
\affiliation{Department of Mathematics and Statistics, University of Massachusetts
Amherst, Amherst, Massachusetts 01003-4515, USA}
\author{P. G. Kevrekidis\thanks{%
Email: kevrekid@math.umass.edu}}
\affiliation{Department of Mathematics and Statistics, University of Massachusetts
Amherst, Amherst, Massachusetts 01003-4515, USA}
\author{D. J. Frantzeskakis\thanks{%
Email: dfrantz@phys.uoa.gr}}
\affiliation{Department of Physics, University of Athens, Panepistimiopolis, Zografos,
Athens 15784, Greece}
\author{B. A. Malomed\thanks{%
Email: malomed@post.tau.ac.il}}
\affiliation{Department of Physical Electronics, School of Electrical Engineering,
Faculty of Engineering, Tel Aviv University, Tel Aviv 69978, Israel }
\affiliation{Laboratory of Nonlinear-Optical Informatics, ITMO University,
St. Petersburg 197101, Russia}
\date{\today}

\begin{abstract}
We consider a two-component, two-dimensional nonlinear Schr{\"{o}}dinger
system with unequal dispersion coefficients and self-defocusing
nonlinearities, chiefly with equal strengths of the self- and
cross-interactions. In this setting, a natural waveform with a nonvanishing
background in one component is a vortex, which induces an effective
potential well in the second component, via the nonlinear coupling of the
two components. We show that the potential well may support not only the
fundamental bound state, 
but also multi-ring excited radial state complexes for
suitable ranges of values of the dispersion coefficient of the second
component. We systematically explore the existence, stability, and nonlinear
dynamics of these states. The complexes involving the excited radial states
are weakly unstable, with a growth rate depending on the dispersion of the
second component. Their evolution leads to transformation of the multi-ring
complexes into stable VB solitons ones with the fundamental state in the
second component. The excited states may be stabilized by a
harmonic-oscillator trapping potential, as well as by unequal strengths of
the self- and cross-repulsive nonlinearities.
\end{abstract}

\maketitle

\section{Introduction}

Multi-component nonlinear Schr{\"{o}}dinger (NLS) systems emerge in a
variety of contexts of optical~\cite{kivshar} and atomic physics~\cite%
{book2a,book2}. In the former setting, they model, in particular, the
interaction of waves with different carrier wavelengths~\cite%
{manakov,intman1}, while in atomic Bose-Einstein condensates (BECs) they
apply [in the form of coupled Gross-Pitaevskii (GP) equations] to spinor (or
pseudo-spinor) systems, which 
represent mixed condensates formed by different hyperfine states of the same
atomic species \cite{kawueda,stampueda,siambook}, as well as to
heteronuclear mixtures composed by different species \cite{hetero}.

In such multi-component settings, when the nonlinearity is self-defocusing
(self-repulsive), a prototypical example of a one-dimensional self-trapped
structure is given by dark-bright (DB) solitons. These are ubiquitous in
two-component systems with the self- and cross-repulsion (alias self- and
cross-phase-modulation, SPM and XPM, respectively) represented by cubic
terms. Since long ago, the DB solitons have drawn much interest in nonlinear
optics~\cite{christo,vdbysk1,vddyuri,ralak,dbysk2,shepkiv,parkshin},
including their realization in pioneering experiments reported in Refs.~\cite%
{seg1,seg2}. More recently, the remarkable control available in pristine
experimental settings of atomic BECs in ultracold gases, such as $^{87}$Rb
and $^{23}$Na, with a multitude of available co-trapped hyperfine states, as
well as in heteronuclear mixtures, such as $^{87}$Rb-$^{41}$K \cite{hetero},
has opened a new gateway to the realization of DB solitons. 
Indeed, these structures were created in a multitude of state-of-the-art
experiments either controllably, or spontaneously, and their pairwise
interactions, as well as interactions with external potentials, were studied~%
\cite{hamburg,pe1,pe2,pe3,azu}. Related $SO(2)$ rotated DB soliton states,
in the form of dark-dark solitons, were also experimentally produced~\cite%
{pe4,pe5}.

The formation of the DB solitons is based on the fact that dark solitons are
fundamental modes in single-component one-dimensional (1D) self-defocusing
media. These modes, when created in one component of a two-component system,
induce an effective potential well in the other component. This potential
well gives rise to its fundamental bound state, i.e., the ground state (GS),
which represents the bright component of the DB soliton complex. The
knowledge of the explicit form of the dark soliton enables one to explore
the induced potential well, which is generically of the P{\"{o}}schl-Teller
type \cite{LDL}. It is possible to demonstrate, as done recently \cite%
{dbs_coupled_Manakov}, that, if the dispersion coefficient in the second
component is different from its counterpart in the first component, not only
the GS, but also excited states can be trapped by the potential well in the
second component. When the DB soliton states emerge at their bifurcation
point, they have an infinitesimal amplitude of the bright component in the
effective potential induced by the action of the SPM term. However, they can
be readily continued numerically to finite values of the amplitude.
Heteronuclear BEC mixtures with different atomic masses of the components
provide a straightforward realization of the coupled GP equations with
different dispersion coefficients (inverse atomic masses). In addition,
spin-orbit coupled BECs \cite{SOCBEC} offer the same possibility, for states
that coexist in the upper- and lower-energy bands of the linear spectrum
\cite{vaszb}. In terms of optics, a similar realization is provided by the
copropagation of two beams carried by widely different wavelengths in a
self-defocusing medium. Another recently developed ramification of the topic
of the DB solitons in systems with unequal effective dispersion coefficients
is the consideration of systems of equations with quintic SPM and cubic XPM
repulsive interactions. They model the immiscibility regimes in
heteronuclear binary Tonks-Girardeau (TG) gases \cite{Mario}, as well as
BEC-TG mixtures \cite{Giofil}.

It is natural and quite interesting to extend the concept of DB soliton
states to higher dimensions. In particular, the fact that the component
carrying patterns supported by nonzero background induces an effective
trapping potential in the other component, remains valid in this case. In
the two-dimensional (2D) setting, such patterns are well-known stable
vortices \cite{vortex} (vortices were studied in multi-component systems too
\cite{Japan}). A vortex in one component generates an effective axisymmetric
potential well in the 
other, which may trap a bright 2D solitary wave, producing a complex that
was given different names -- in particular, a vortex-bright (VB) soliton~%
\cite{kodyprl}, a half-quantum vortex~\cite{tsubota}, 
a filled-core~vortex \cite{anderson}, as well as a baby Skyrmion~\cite%
{cooper}. Similar stable two-component modes are \textquotedblleft
semi-vortices\textquotedblright\ in the free 2D \cite{Fukuoka} and 3D \cite%
{Han Pu} space with the attractive SPM and XPM terms, which are made stable
by the spin-orbit coupling; they are composed of a {bright vortex soliton}
in one component, and a bright fundamental one in the other.

It is important to note that the 
VB soliton complexes in the
self-repulsive setting of a mixture of internal states of $^{87}$Rb atoms
were created experimentally in the early work of~\cite{anderson}.
Subsequently, their stability~\cite{skryabin,kodyprl} and dynamics~\cite%
{kodyprl,tsubota} have been examined theoretically. It was shown that these
states feature intriguing interactions that decay with the distance $r$
between them as $1/r^{3}$~\cite{tsubota}. Pairs of VB soliton complexes can
form bound states in atomic BECs, as shown in detail in Ref.~\cite{pola}.

Our objective in the present work is to consider VB soliton complexes in the
system featuring repulsive SPM and XPM interactions, and different
dispersion coefficients of the two components (i.e., different atomic masses
in the respective coupled GP equations, or different propagation constants
in the coupled NLS equations for optical beams). We aim to generate a broad
set of novel families of excited complexes, with the vortex in the first
component potentially trapping not only the fundamental bright solitons, but
also excited radial states in the second component, represented by confined
multi-ring shaped waveforms. We demonstrate that such complexes are possible
in the two-component NLS/GP system. The fundamental state among them, the VB
soliton complex, is generically found to be stable. On the other hand, the
complexes whose bright component is represented by the excited ring-shaped
modes are found to be unstable. However, varying the dispersion coefficient
of the second component, we can identify scenarios where it is possible to
render the corresponding instability very weak, and the associated
structures very long-lived. Furthermore, if an additional
harmonic-oscillator trapping potential is added to the system, which is, as
a matter of fact, a mandatory ingredient of the experimental realization of
the setting in BEC, we show that it is possible to render such structures
completely stable in suitable parametric intervals. Lastly, we showcase
basic scenarios of the instability development, inferring that the unstable
(in the free space) multi-ring states are typically transformed into the
stable VB fundamental ones.

The presentation in the paper is structured as follows. The model is
introduced in Section II.
In Sec.~III, we discuss the computational analysis of the model, presenting
both the numerical methods and results. Finally, in Sec.~IV we summarize
our findings and mention possible directions for future studies.


\section{The model and analytical considerations}

Motivated by the above-mentioned realizations in BECs and nonlinear optics,
we consider the coupled defocusing GP/NLS system in $(2+1)$ dimensions (two
spatial and one temporal). In the scaled form, the system is
%
\begin{eqnarray}
&&i\partial _{t}{\Phi _{-}}=-\frac{D_{-}}{2}\nabla ^{2}\Phi _{-}+\gamma
\left( g_{1}|\Phi _{-}|^{2}+\sigma _{12}|\Phi _{+}|^{2}\right) \Phi
_{-}+V(x,y)\Phi _{-},  \label{start_gps_2Da} \\
&&i\partial _{t}{\Phi _{+}}=-\frac{D_{+}}{2}\nabla ^{2}\Phi _{+}+\gamma
\left( \sigma _{12}|\Phi _{-}|^{2}+g_{2}|\Phi _{+}|^{2}\right) \Phi
_{+}+V(x,y)\Phi _{+},  \label{start_gps_2Db}
\end{eqnarray}%
where $\nabla ^{2}=\partial _{x}^{2}+\partial _{y}^{2}$ is the Laplacian in
2D, $D_{\pm }$ are the dispersion coefficients, $\gamma $ is the overall
nonlinearity strength, with relative SPM and XPM interaction coefficients $%
g_{j}$ ($j=1,2$) and $\sigma _{12}$, respectively. Equations~(\ref%
{start_gps_2Da}) and (\ref{start_gps_2Db}) include the usual parabolic trapping
potential, 
\begin{equation}
V(x,y)=\frac{1}{2}\Omega ^{2}(x^{2}+y^{2}), 
\end{equation}
with normalized trap strength $\Omega $.
Fields $\Phi _{-}$ and $\Phi _{+}$ carry the vortex and bright-soliton
components, respectively. From now on, we focus on the basic case of equal
interaction coefficients,
\begin{equation}
g_{1,2}=\sigma _{12}=1,  \label{1}
\end{equation}
and use rescaling to fix $D_{-}=\gamma =1$, while $D_{+}\equiv D\,\geq 0$ is
the relative dispersion coefficient in the second component.

In the case of the binary heteronuclear BECs, coefficient $D$ is determined
by the two atomic masses, $D=m_{-}/m_{+}$, while in the case of the
spin-orbit coupled BECs, is given by the ratio of the group-velocity
dispersion coefficients, as found by the corresponding dispersion relation
of the two-component branches \cite{vaszb}. On the other hand, in the optics
model, time $t$ is replaced by the propagation coordinate, $z$, in the
corresponding bulk waveguide \cite{kivshar}, and $D$ is determined by the
carrier wavelengths of the two beams, $D=\Lambda _{+}/\Lambda _{-}$. In
particular, referring to $^{87}$Ru-$^{7}$Li BEC mixtures, which are
available to current experiments (see Ref.~\cite{RuLi} and references
therein), the relative dispersion coefficient may reach values as large as $%
\simeq 12$, and as small as $\simeq 0.08$. In optics, the use of materials
with broadband transparency may give rise to a roughly similar range of $D$.
However, in the case of very large or very small $D$, 
Eq.~(\ref{1}) is not relevant, and the analysis will need to be 
adjusted to other values of the SPM and XPM coefficients.

Stationary solutions to Eqs.~(\ref{start_gps_2Da})-(\ref{start_gps_2Db})
with chemical potentials $\mu _{\pm }$ (or propagation constants $-\mu _{\pm
}$, in terms of the optical beams) are looked for as $\Phi _{\pm
}(x,y,t)=\phi _{\pm }(x,y)\exp (-i\mu _{\pm }t)$, reducing Eqs.~(\ref%
{start_gps_2Da})-(\ref{start_gps_2Db}) to the coupled system of stationary
equations:
\begin{eqnarray}
\mu _{-}{\phi _{-}} &=&-\frac{1}{2}\nabla ^{2}\phi _{-}+\left( |\phi
_{-}|^{2}+|\phi _{+}|^{2}\right) \phi _{-}+V(x,y)\phi _{-},
\label{stat_gps_2Da} \\
\mu _{+}{\phi _{+}} &=&-\frac{D}{2}\nabla ^{2}\phi _{+}+\left( |\phi
_{-}|^{2}+|\phi _{+}|^{2}\right) \phi _{+}+V(x,y)\phi _{+}.
\label{stat_gps_2Db}
\end{eqnarray}%
Further, we introduce the following \textit{Ans\"{a}tze} for the stationary
fields, with real radial functions $f_{\pm }(r)$:
\begin{eqnarray}
&&\phi _{-}(r,\theta )=f_{-}(r)e^{iS\theta },  \label{-} \\
&&\phi _{+}(r,\theta )=f_{+}(r)e^{in\theta },  \label{+}
\end{eqnarray}%
%
%
%
where $r=\sqrt{x^{2}+y^{2}}$ is the radial distance, $\theta =\tan ^{-1}(y/x)
$ is the polar angle, and the integer topological charges of the vortex and
bright solitons are $S$ and $n$, respectively. Thus, Eqs.~%
\eqref{stat_gps_2Da}-\eqref{stat_gps_2Db} reduce to the radial equations:
\begin{gather}
\frac{d^{2}f_{-}}{dr^{2}}+\frac{1}{r}\frac{df_{-}}{dr}-\frac{S^{2}f_{-}}{%
r^{2}}-2\left( f_{-}^{~2}+f_{+}^{~2}-\mu _{-}+V(r)\right) f_{-}=0,
\label{stat_gps_1Da} \\
D\left( \frac{d^{2}f_{+}}{dr^{2}}+\frac{1}{r}\frac{df_{+}}{dr}-\frac{%
n^{2}f_{+}}{r^{2}}\right) -2\left( f_{-}^{~2}+f_{+}^{~2}-\mu
_{+}+V(r)\right) f_{+}=0.  \label{stat_gps_1Db}
\end{gather}%
It should be noted in passing that in the majority of
cases studied below we consider Eqs.~(\ref{start_gps_2Da})-(\ref%
{start_gps_2Db}), (\ref{stat_gps_2Da})-(\ref{stat_gps_2Db}), as well as
Eqs.~(\ref{stat_gps_1Da})-(\ref{stat_gps_1Db}) in the absence of the
trapping potential. Therefore, we set $V(r)=0$, unless it is said otherwise.

As indicated above, our fundamental premise, similar to that adopted in the
study of the 1D setting in Ref. \cite{dbs_coupled_Manakov}, is that the dark
mode (dark soliton in 1D, and vortex in this present case) of the defocusing
NLS equation induces an effective potential (via the XPM interaction) in the
other component, which in turn gives rise to trapping of the bright-soliton
state in it. Thus, Eq.~(\ref{stat_gps_2Da}) and, in particular, its radial
version simplifies to the single-component equation,
\begin{equation}
\nabla _{r}^{2}f_{-}-\frac{S^{2}f_{-}}{r^{2}}-2\left( f_{-}^{~2}-\mu
_{-}\right) f_{-}=0,  \label{bvp_vortex}
\end{equation}%
in the absence of the bright component, i.e., for $f_{+}=0$;  
here, $\nabla _{r}^{2}=d^{2}/dr^{2}+r^{-1}d/dr$ is the radial part of the
Laplace operator. Equation~(\ref{bvp_vortex}) was solved numerically via fixed-point 
(Newton-Raphson) iterations (see Sec.~\ref{Compu_Anaysis} below for details on the
computational methods employed in this work). Suitable approximate solutions
for the vortical waveform are known too (see, e.g. Ref. \cite{berloff}), and
they may be useful as initial guesses for the iterative process described
numerically below. From here on, we assume that this iterative process
converges to a radial solution for the vortex. This is different from the 1D
case, where the dark soliton is available in the commonly known analytical
form, and the P{\"{o}}schl-Teller potential~\cite{LDL} that it induces in
the other component is analytically tractable~\cite{dbs_coupled_Manakov}. In
the 2D system presented in this work, the analysis has to be completed
numerically.

Thus, the resulting vortex profile $f_{-}$ (or $\phi _{-}$, for given $S$)
of Eq.~(\ref{bvp_vortex}) plays the role of the background for the weak
component $f_{+}$ (or $\phi _{+}$, for given $n$). As follows from Eq.~(\ref%
{bvp_vortex}), the amplitude of the background for the vortex is%
\begin{equation}
f_{-}(r\rightarrow\infty )=\sqrt{\mu _{-}}  \label{asympt}
\end{equation}
which (upon rescaling) is set to be $\mu _{-}=1$, in our numerical
computations below. Then, when the solution for the component $f_{+}$
bifurcates from its linear limit 
corresponding to $f_{+}\rightarrow 0$, the linearized form of Eq.~(\ref%
{stat_gps_1Db}) amounts to an eigenvalue problem
%
\begin{equation}
\mathcal{L}f_{+}=\mu _{+}f_{+},  \label{bright_linear_polar_eig_prob}
\end{equation}%
for known $f_{-}$, where $\mathcal{L}=-\left( D/2\right) \,\left( \nabla
_{r}^{2}-n^{2}/r^{2}\right) +f_{-}^{~2}$ is a linear operator and $(\mu
_{+},f_{+})$ is the eigenvalue-eigenvector pair. Armed with the set of
profiles for $f_{\pm }$ obtained from Eqs.~(\ref{bvp_vortex}) and (\ref%
{bright_linear_polar_eig_prob}) as initial guesses, we utilize an iterative
scheme towards the solution of the full nonlinear system of Eqs.~(\ref%
{stat_gps_1Da})-(\ref{stat_gps_1Db}) (see Sec.~\ref{Compu_Anaysis} below for
details).

It is natural to expect that nonlinear solutions to Eqs.~(\ref{stat_gps_1Da}%
)-(\ref{stat_gps_1Db}) and~(\ref{stat_gps_2Da})-(\ref{stat_gps_2Db}),
corresponding to the ground and excited states in the linear limit for
component $f_{+}$, emerge (bifurcate) at some critical values of $D$ with
the corresponding eigenvalues $\mu _{+}$ of the linear problem based on Eq.~(%
\ref{bright_linear_polar_eig_prob}). These values are found below by
performing numerical continuations over the aforementioned parameters.

\section{Numerical Analysis}

\label{Compu_Anaysis}

\subsection{Computational methods}

In this section, numerical results are presented for the coupled GP/NLS
system (\ref{start_gps_2Da})-(\ref{start_gps_2Db}). Our analysis addresses
the \textit{existence}, \textit{stability}, and \textit{dynamical evolution}
of the nonlinear modes under consideration. As concerns the existence
and stability, a parametric continuation is performed in chemical potential $%
\mu _{+}$ of the bright component for given values of relative dispersion
coefficient $D$. The corresponding states are thus identified along with
their stability spectra. When the solutions are predicted to be stable, this
is verified by direct simulations. For unstable states, the simulations aim
to reveal the eventual states into which they are spontaneously transformed.

In our numerical computations, a 1D uniform spatial grid is employed along
the radial direction, consisting of $N$ points $r_{j}=j\delta {r}$, with $%
j=1,\dots ,N$ and lattice spacing $\delta {r}=0.05$. The origin is located
at $j=0$, whereas the domain cut-off ($r_{\text{max}}$) is set at $j=N+1$
(from now on, we fix $r_{\text{max}}=50$). In this way, both fields $f_{\pm
}(r)$ are replaced by their discrete counterparts on the spatial grid, $%
f_{j,\pm }=f_{\pm }(r_{j})$. Then, the radial Laplacian $\nabla _{r}^{2}$ in
Eqs.~(\ref{stat_gps_1Da})-(\ref{stat_gps_1Db}), (\ref{bvp_vortex}) and (\ref%
{bright_linear_polar_eig_prob}) 
is replaced by second-order central-finite-difference formulas for the first
and second derivatives. To secure a well-posed problem, we employ the
boundary conditions (BCs) $f_{-}(r=0)=0$ and $df_{-}/dr(r \rightarrow \infty )=0$ for
the vortex soliton component, and $df_{+}/dr(r=0)=0$ and $f_{+}({r \rightarrow \infty }%
)=0$ for the bright-soliton one. In particular, the zero-derivative
(Neumann) BCs are incorporated into the internal discretization scheme using
the first-order backward and forward difference formulas, respectively.
Essentially, the zero-derivative (no-flux) BCs are enforced by requiring $%
f_{N+1,-}=f_{N,-}$ and $f_{0,+}=f_{1,+}$, whereas $f_{0,-}=f_{N+1,+}=0$, as
per the corresponding homogeneous Dirichlet 
BCs.

The starting point is Eq.~(\ref{bvp_vortex}) for radial profile $f_{-}$ of
the vortex component. From now on, we fix the vorticities of the vortex- and
bright-soliton components to be $S=1$ and $n=0$, respectively, given that
our emphasis is on VB soliton complexes. We solve Eq.~(\ref{bvp_vortex}) by
means of the standard Newton-Raphson method, which converges to a vortex
profile as long as a sufficiently good initial guess is used. An example of
an input, which ensures both the convergence and compliance with
error-tolerance criteria, is $f_{-}(r)=\tanh r$. The resulting converged
waveform for different vorticities $S\geq 1$ features the correct asymptotic
form at $r\rightarrow 0$, $f_{-}(r\rightarrow 0)\sim r^{S}$, and its density
asymptotes to $\mu _{-}$ for large $r$.
%
%
Subsequently, with background field $f_{-}$ at hand,
we solve the eigenvalue problem~(\ref{bright_linear_polar_eig_prob})
numerically, to obtain the corresponding bright component, $f_{+}(r)$, with
an infinitesimal amplitude, along with the associated chemical potential, $%
\mu _{+}$. Our study is organized according to the order of the bound states
(the ground state, first excited state, and so on), and the value of $D$.
Specifically, we determine chemical potentials $\mu_{+}$ corresponding to
one of the lower eigenstates (the ground state corresponds to lowest $\mu
_{+}$, the first excited state pertains to the second lowest eigenvalue, and
so on) and the corresponding bright eigenfunction $f_{+}$ is obtained
afterwards. This way, the fully nonlinear self-trapped states of 
system (\ref{stat_gps_1Da})-(\ref{stat_gps_1Db}) can be obtained with the
help of the Newton-Raphson scheme, the seed for the respective iterations
consisting of the vortex radial profile $f_{-}(r)$ together with the
eigenvalue-eigenvector pair $(\mu_{+},f_{+})$. \ Essentially, the seed fed
to our nonlinear solver originates from the underlying linear limit
discussed in the previous Section. In addition, we trace the stationary
solutions, for a given value of dispersion coefficient $D$, by performing a
numerical continuation with respect to chemical potential $\mu _{+}$, by
dint of the \textit{sequential continuation} method, i.e., using the
solution for given $\mu _{+}$, found by the solver, as the seed for the next
continuation step. We are thus able to numerically determine not only the
range of dispersion coefficient $D$, but also the range of chemical
potential $\mu _{+}$ for each case of interest. The validity of the
stationary solutions produced by the Newton-Raphson code has been
corroborated upon using a collocation method \cite{COLSYS} for solving
boundary value problems. 


Having identified the stationary states, we turn to the study of their
stability. Motivated by the decomposition described in Ref. \cite%
{Kollar_Pego}, we start with the generalized perturbation ansatz around
stationary solutions, writing in the polar coordinates:
\begin{eqnarray}
\widetilde{\Phi }_{-}(r,\theta ,t) &=&e^{-i\mu _{-}t}e^{iS\theta }%
\Big\lbrace f_{-}+\varepsilon \sum_{|m|=0}^{\infty }\left[
a_{m}(r)e^{\lambda t}e^{im\theta }+b_{m}^{\ast }(r)e^{\lambda ^{\ast
}t}e^{-im\theta }\right] \Big\rbrace,  \label{lin_ansatz_polara} \\
\widetilde{\Phi }_{+}(r,\theta ,t) &=&e^{-i\mu _{+}t}e^{in\theta }%
\Big\lbrace f_{+}+\varepsilon \sum_{|m|=0}^{\infty }\left[
c_{m}(r)e^{\lambda t}e^{im\theta }+d_{m}^{\ast }(r)e^{\lambda ^{\ast
}t}e^{-im\theta }\right] \Big\rbrace,  \label{lin_ansatz_polarb}
\end{eqnarray}%
where $\lambda $ is a (complex) eigenvalue, $\varepsilon $ is an
infinitesimal amplitude of the perturbation, and the asterisk stands for
complex conjugate. We insert Eqs.~(\ref{lin_ansatz_polara})-(\ref%
{lin_ansatz_polarb}) into the radial version of Eqs.~(\ref{start_gps_2Da})-(%
\ref{start_gps_2Db}) and thus obtain, at order $\varepsilon $, an eigenvalue
problem in the following matrix form:
\begin{equation}
\tilde{\lambda}%
\begin{pmatrix}
a_{m} \\
b_{m} \\
c_{m} \\
d_{m}%
\end{pmatrix}%
=%
\begin{pmatrix}
A_{11} & A_{12} & A_{13} & A_{14} \\
-A_{12}^{\ast } & A_{22} & -A_{14}^{\ast } & -A_{13}^{\ast } \\
A_{13}^{\ast } & A_{14} & A_{33} & A_{34} \\
-A_{14}^{\ast } & -A_{13} & -A_{34}^{\ast } & A_{44}%
\end{pmatrix}%
\begin{pmatrix}
a_{m} \\
b_{m} \\
c_{m} \\
d_{m}%
\end{pmatrix}%
,  \label{eig_prob_polar}
\end{equation}%
with eigenvalues 
$\tilde{\lambda}=i\lambda $, eigenvectors $\mathcal{V}%
=(a_{m},b_{m},c_{m},d_{m})^{T}$, and matrix elements
\begin{eqnarray}
A_{11} &=&-\frac{D_{-}}{2}\left[ \nabla _{r}^{2}-\frac{\left( S+m\right) ^{2}%
}{r^{2}}\right] +\gamma \left[ 2g_{1}|f_{-}|^{2}+\sigma _{12}|f_{+}|^{2}%
\right] +V-\mu _{-},  \label{A11_polar} \\
A_{12} &=&\gamma \,g_{1}\,\left( f_{-}\right) ^{2}, \\
A_{13} &=&\gamma \,\sigma _{12}\,f_{-}\left( f_{+}\right) ^{\ast }, \\
A_{14} &=&\gamma \,\sigma _{12}\,f_{-}\,f_{+}, \\
A_{22} &=&\frac{D_{-}}{2}\left[ \nabla _{r}^{2}-\frac{\left( S-m\right) ^{2}%
}{r^{2}}\right] -\gamma \left[ 2g_{1}|f_{-}|^{2}+\sigma _{12}|f_{+}|^{2}%
\right] -\left( V-\mu _{-}\right) ,  \label{A22_polar} \\
A_{33} &=&-\frac{D_{+}}{2}\left[ \nabla _{r}^{2}-\frac{\left( n+m\right) ^{2}%
}{r^{2}}\right] +\gamma \left[ \sigma _{12}|f_{-}|^{2}+2g_{2}|f_{+}|^{2}%
\right] +V-\mu _{+},  \label{A33_polar} \\
A_{34} &=&\gamma \,g_{2}\,\left( f_{+}\right) ^{2}, \\
A_{44} &=&\frac{D_{+}}{2}\left[ \nabla _{r}^{2}-\frac{\left( n-m\right) ^{2}%
}{r^{2}}\right] -\gamma \left[ \sigma _{12}|f_{-}|^{2}+2g_{2}|f_{+}|^{2}%
\right] -\left( V-\mu _{+}\right) .  \label{A44_polar}
\end{eqnarray}%
As a result, the full spectrum of nonlinear solutions $f_{\pm }$ is obtained
by putting together the spectra for different integer values of $m$. This
was done for $m\geq 0$ only ($m=0,1,2,3,4$ and $5$ in this work), since the
sets of the eigenvalues for $\pm m$ (with $m>0$) are complex conjugates. The
corresponding steady state is classified as a stable one if none of the
eigenvalues $\lambda =\lambda _{r}+i\,\lambda _{i}$ has $\lambda _{r}\neq 0$%
, given the Hamiltonian nature of our system. Note that two types of
instabilities can be thus identified: (i) \textit{exponential}, characterized
by a pair of real eigenvalues with $\lambda _{i}=0$, and (ii) \textit{%
oscillatory instabilities}, characterized by complex eigenvalue quartets.

Finally, the results for the spectral stability, obtained from the solution
of the eigenvalue problem [see, Eq.~(\ref{eig_prob_polar})] were
corroborated by means of direct simulations of the coupled GP/NLS system~(%
\ref{start_gps_2Da})-(\ref{start_gps_2Db}). To do so, a parallel version
(using OpenMP) of the standard fourth-order Runge-Kutta method (RK4), with a
fixed time-step of $\delta {t}=10^{-4}$, was employed. The simulations were
initialized at $t=0$ using the available stationary solutions. To obtain the
latter, we employed the Newton-Raphson method in a two-dimensional
rectangular domain for system~(\ref{stat_gps_2Da})-(\ref{stat_gps_2Db}),
using the NITSOL package \cite{NITSOL}. The 2D uniform spatial grid was
built of $N_{x}\times N_{y}$ grid points with $N_{x}\equiv N_{y}=251$ and
resolution $\delta {x}\equiv \delta {y}=0.08$. Both fields $\phi _{\pm
}(x,y) $ were replaced by their discrete counterparts on the 2D spatial
grid, i.e., $\phi _{i,j,\pm }=\phi _{\pm }(x_{i},y_{j})$ with $i=1,\dots
,N_{x}$ and $j=1,\dots ,N_{y}$. Then, the Laplacian on the rectangular grid
is replaced by the second-order central-finite-difference formula. As
mentioned above, the Neumann BCs for both fields at edges of the grid were
replaced by the first-order forward and backward difference formulas.
Furthermore, the steady states on the 2D Cartesian grid are obtained from
the radial ones by utilizing the following standard numerical procedure.
Having the 2D grid and (discrete) radial nonlinear states $f_{j,\pm }$ at
hand, we build interpolants $f_{\pm }(r)$ using a cubic spline
interpolation. Then, fields $\phi _{\pm }(r,\theta )$ given by Eqs.~(\ref{-}%
)-(\ref{+}) are constructed (for given vorticities $S$ and $n$) and
transformed from polar to Cartesian coordinates, $(r,\theta )\rightarrow
(x,y)$, by utilizing relations $r=\sqrt{x^{2}+y^{2}}$ and $\theta =\tan
^{-1}\left( y/x\right) $ once again. The resulting approximate solutions are
fed as initial guesses to our 2D Newton-Raphson method on the Cartesian grid
$\phi _{\pm }(x,y)$, and the resulting iteration process converges within a
few steps.

Two possible initializations of the direct simulations can be distinguished.
On the one hand, we initialized the dynamics in the presence of small
(uniformly distributed) random perturbations with amplitude $\varepsilon
=10^{-3}$, added to presumably stable stationary states. An alternative
approach was to initialize the evolution using the linearization ansatz~(\ref%
{lin_ansatz_polara})-(\ref{lin_ansatz_polarb}) for unstable solutions, with $%
\varepsilon =10^{-3}$, eigenvector $\mathcal{V}$ and given $m$ corresponding
to a (complex) eigenvalue being responsible for the instability. The latter
approach helps to stimulate the onset of the expected instability and
observe the ensuing dynamics. The underlying eigenvector $V$ is transformed
from polar to Cartesian coordinates using the same interpolation technique
mentioned previously.

\subsection{Numerical results}

\label{numer_res}

%

\begin{figure}[tp]
\begin{center}
\vspace{-0.1cm}
\mbox{\hspace{-0.1cm}
\subfigure[][]{\hspace{-0.3cm}
\includegraphics[height=.16\textheight, angle =0]{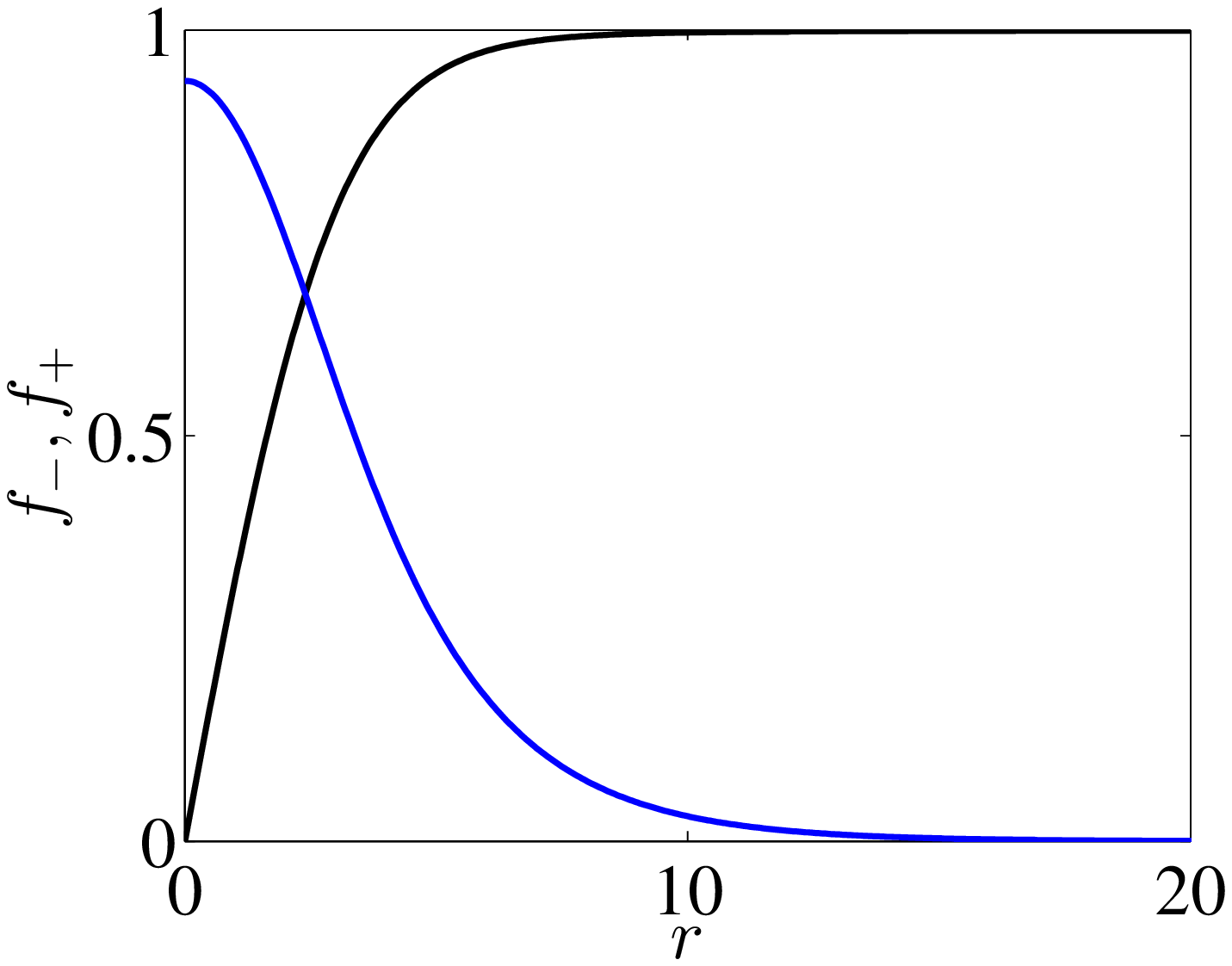}
\label{fig2a}
}
\subfigure[][]{\hspace{-0.3cm}
\includegraphics[height=.16\textheight, angle =0]{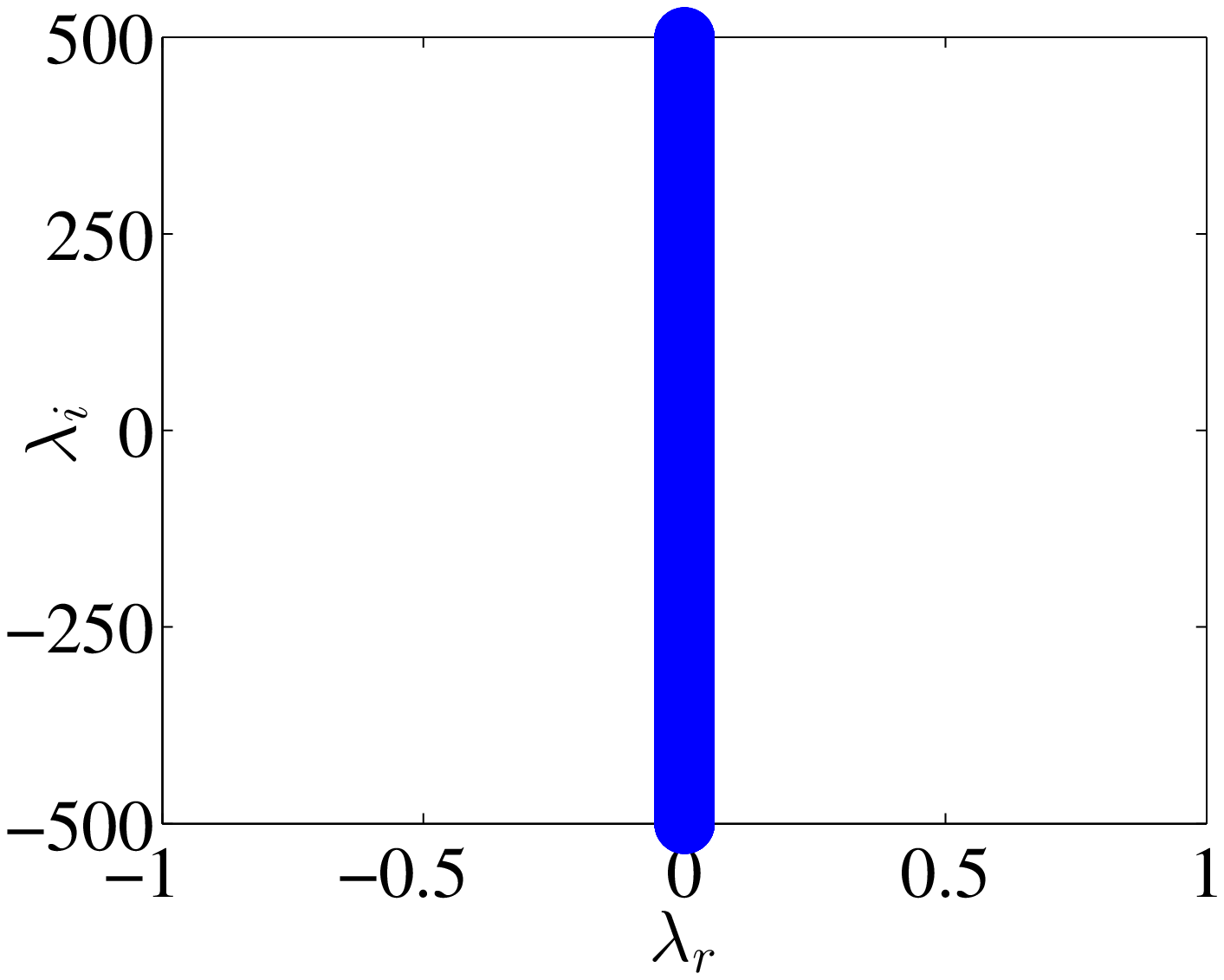}
\label{fig2b}
}
}
\end{center}
\par
\vspace{-0.55cm}
\caption{
(Color online) Steady-state profiles (a) of the vortex and bright soliton 
components (black and blue lines, respectively), and the eigenvalue spectrum
(b) corresponding to a ground state with $D=0.6$ and $\protect\mu _{+}=0.95$.
}
\label{fig2}
\end{figure}

We start by considering the most fundamental state, namely the VB
soliton one in Fig.~\ref{fig2}. In this case, as well as in all the
other cases that we have considered herein, we find the VB structure
(shown, e.g., in the left panel of Fig.~\ref{fig2}) to be stable, as
evidenced by the absence of eigenvalues with nonzero real part in the
right panel of Fig.~\ref{fig2}.

\begin{figure}[tbp]
\begin{center}
\vspace{-0.1cm}
\mbox{\hspace{-0.1cm}
\subfigure[][]{\hspace{-0.3cm}
\includegraphics[height=.16\textheight, angle =0]{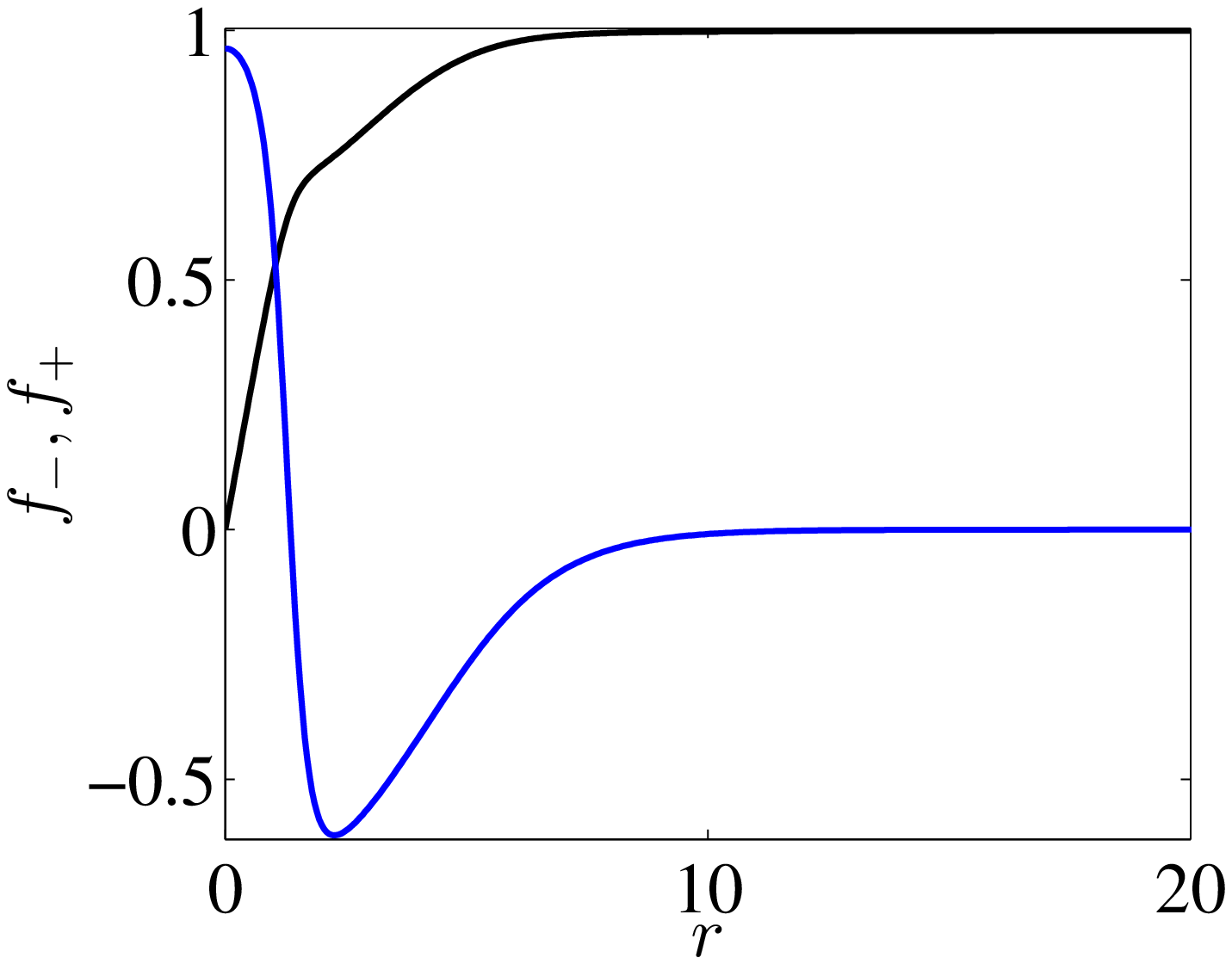}
\label{fig3a}
}
\subfigure[][]{\hspace{-0.3cm}
\includegraphics[height=.16\textheight, angle =0]{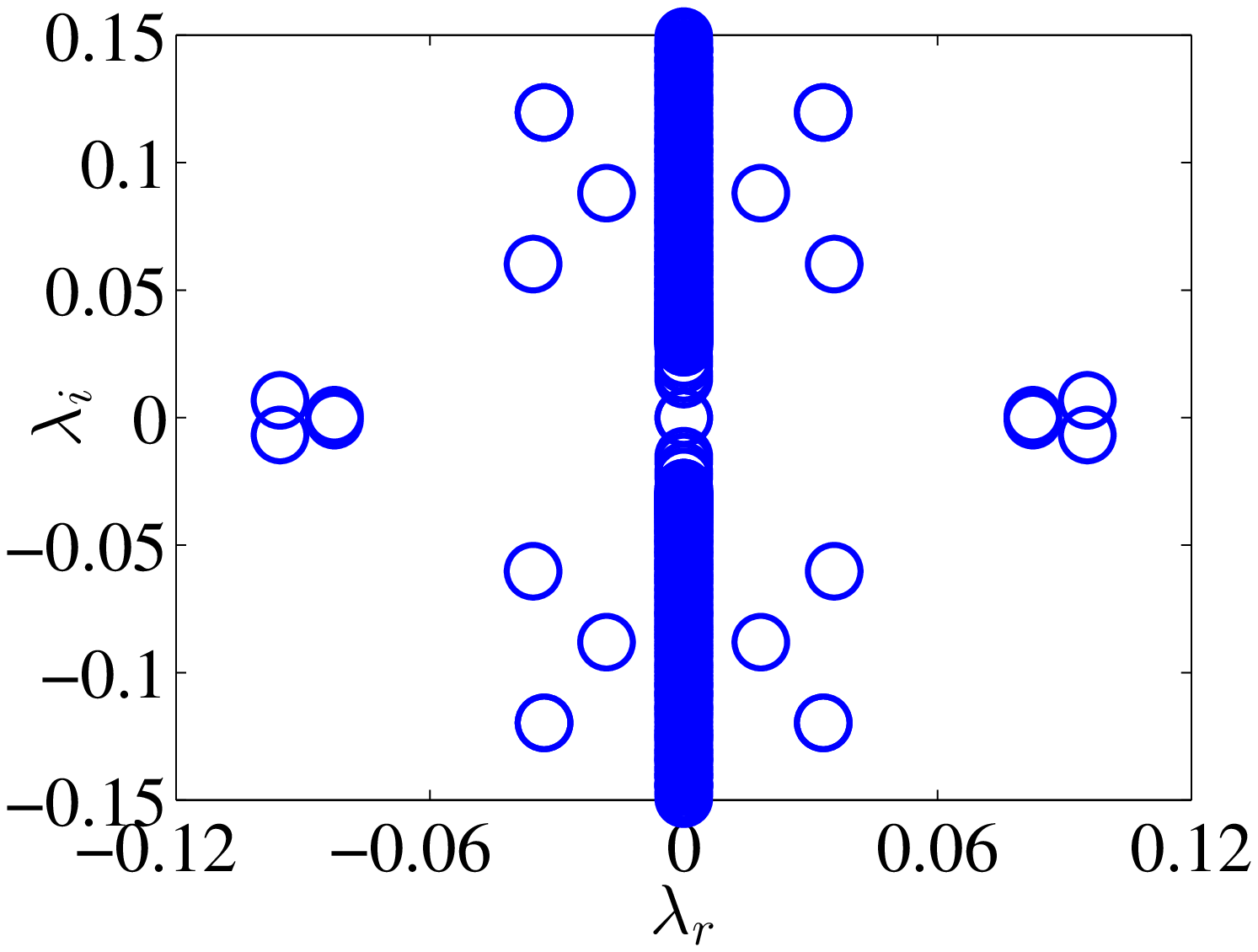}
\label{fig3b}
}
}
\mbox{\hspace{-0.1cm}
\subfigure[][]{\hspace{-0.3cm}
\includegraphics[height=.16\textheight, angle =0]{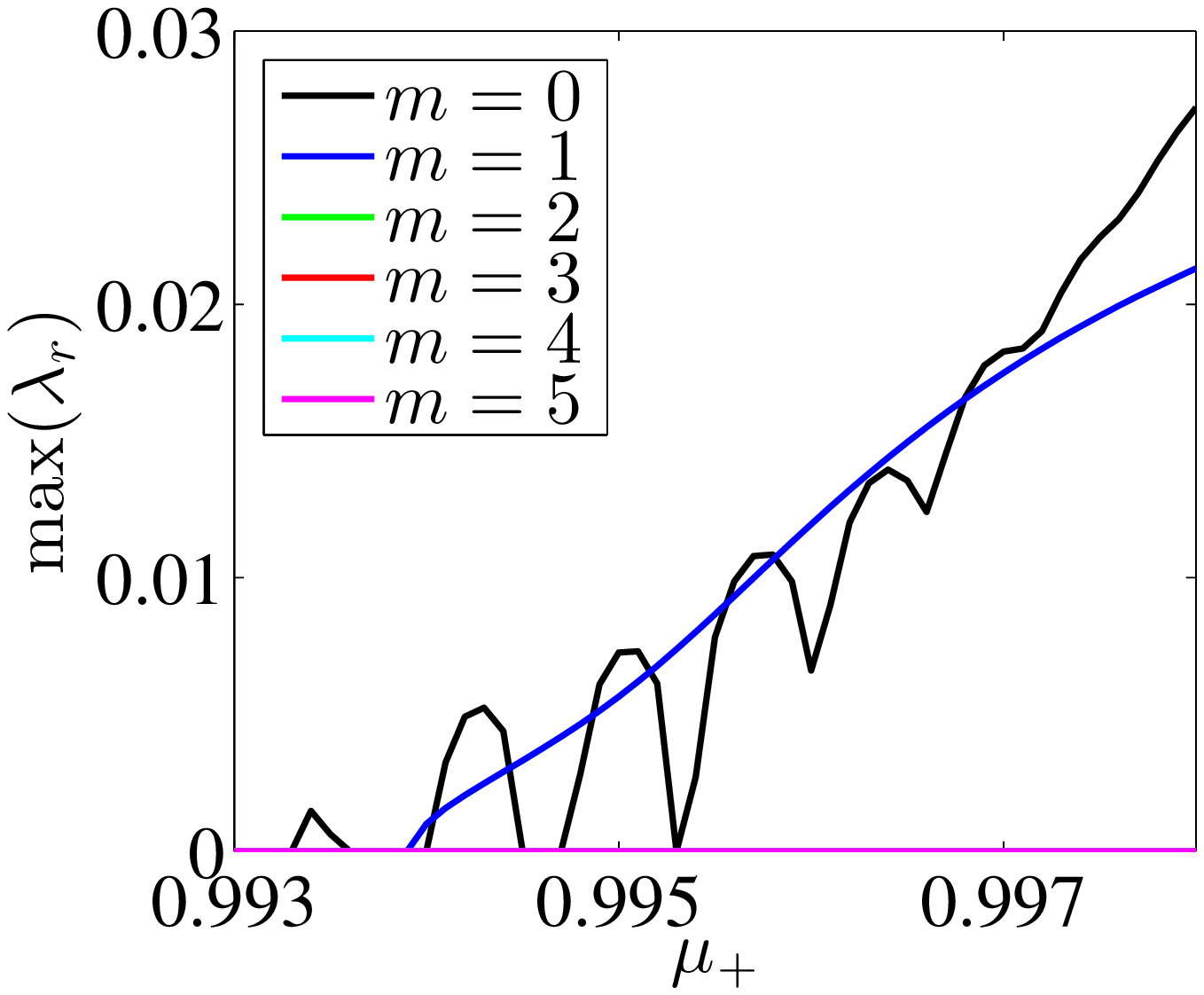}
\label{fig3c}
}
\subfigure[][]{\hspace{-0.3cm}
\includegraphics[height=.16\textheight, angle =0]{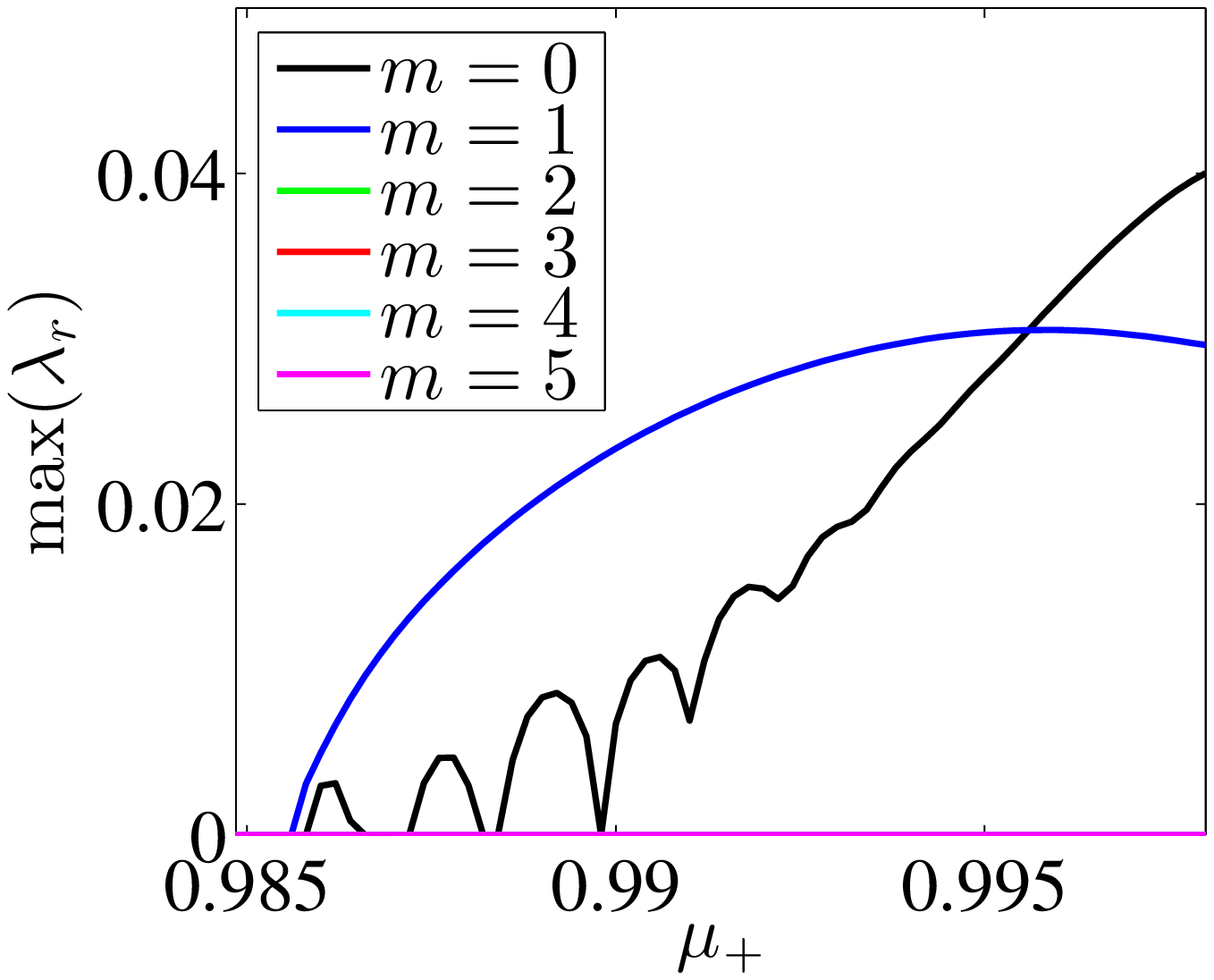}
\label{fig3d}
}
\subfigure[][]{\hspace{-0.3cm}
\includegraphics[height=.16\textheight, angle =0]{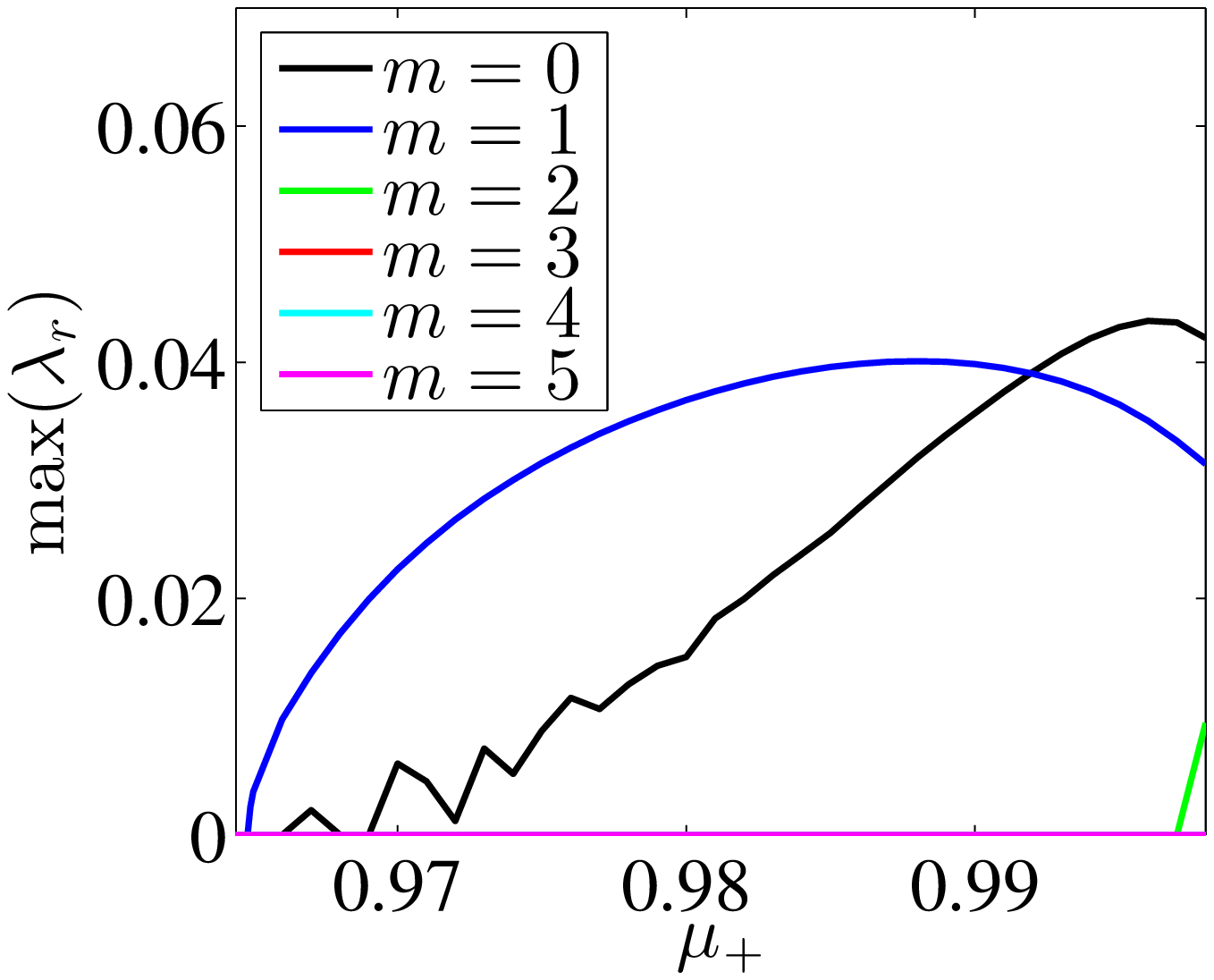}
\label{fig3e}
}
}
\mbox{\hspace{-0.1cm}
\subfigure[][]{\hspace{-0.3cm}
\includegraphics[height=.16\textheight, angle =0]{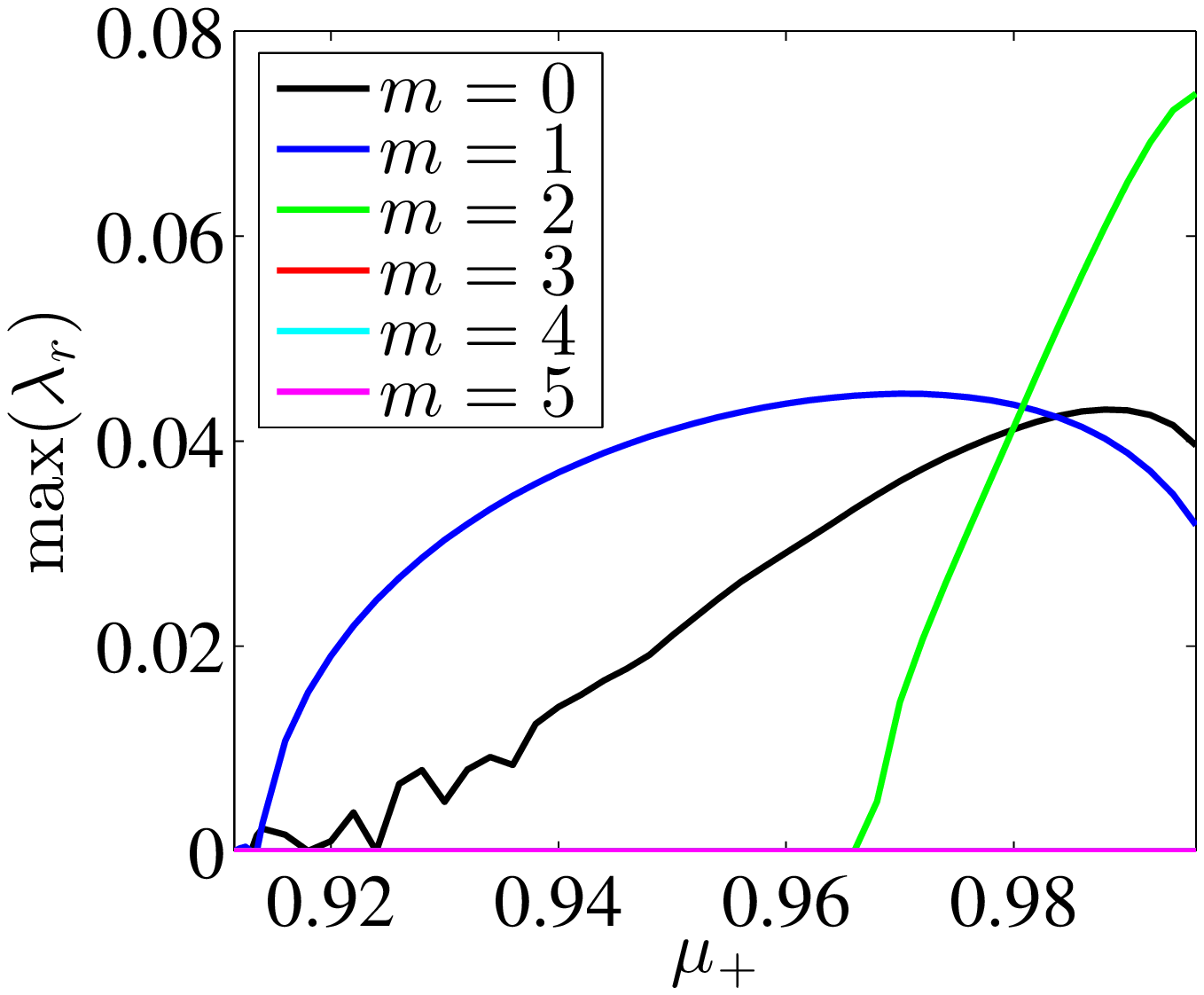}
\label{fig3f}
}
\subfigure[][]{\hspace{-0.3cm}
\includegraphics[height=.16\textheight, angle =0]{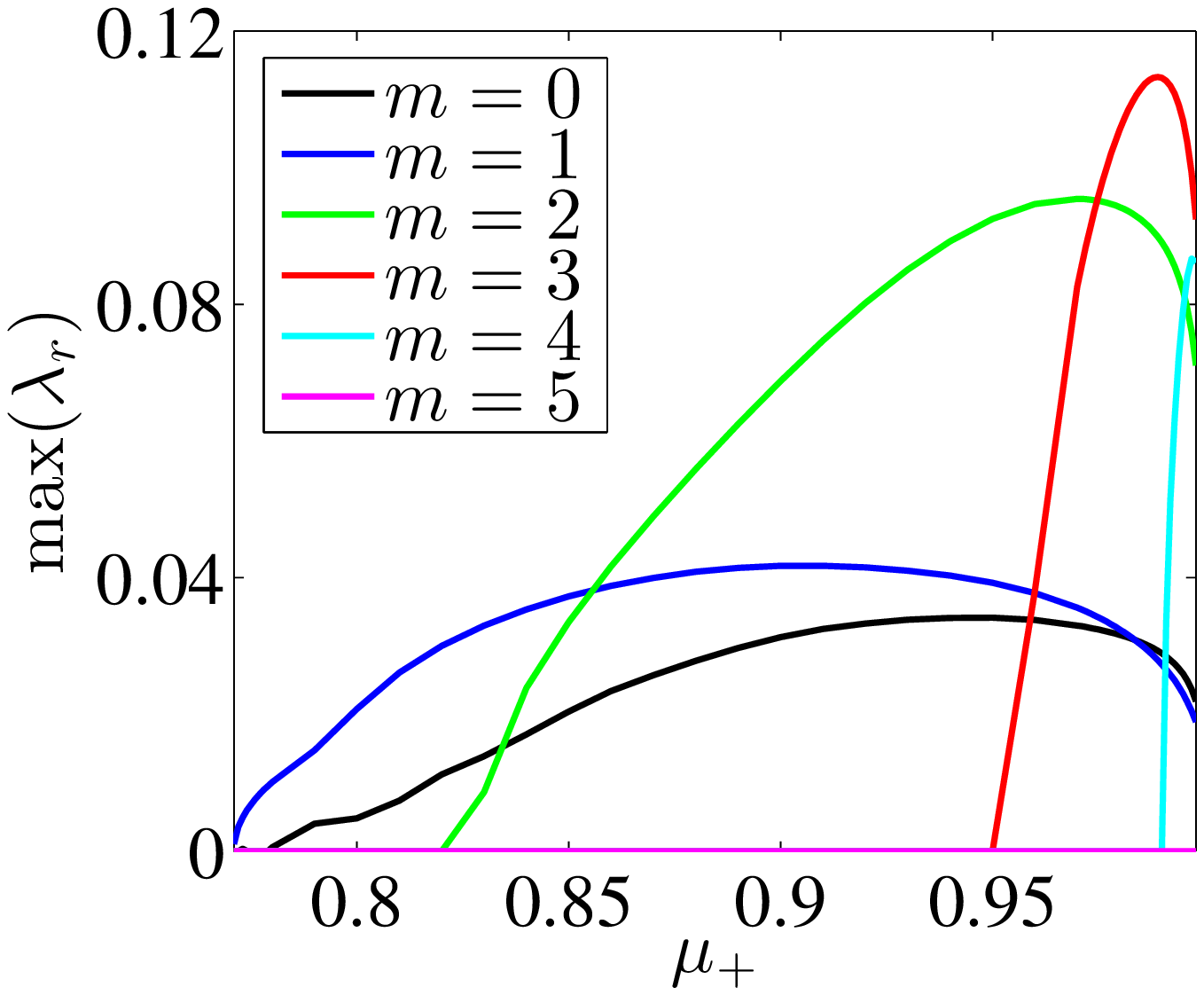}
\label{fig3g}
}
}
\end{center}
\par
\vspace{-0.55cm}
\caption{(Color online) Bound states and continuation results corresponding
to the first excited state in the bright component. \textit{Top row}: (a)
Stationary profiles of the vortex and bright components (the black and blue
lines, respectively). (b) The corresponding eigenvalue spectrum for $D=0.1$
and $\protect\mu _{+}=0.97$. \textit{Middle and bottom rows}: The largest
real part of the eigenvalues as a function of $\protect\mu _{+}$ at various
fixed values of $D$: (c) $D=0.5$, (d) $D=0.4$, (e) $D=0.3$, (f) $D=0.2$, and
(g) $D=0.1$.}
\label{fig3}
\end{figure}

The first excited state is shown in Fig.~\ref{fig3}. The profile displayed
in the top left panel demonstrates that the vortex retains its general
structure, while featuring some changes due to the presence of a more
complex spatial pattern in the bright component, with two maxima of the
local density ($|f_{+}|^{2}$), one at the center and another one at the
periphery, separated by a notch in the form of a dark ring. The typical
linearization spectrum shown in the top right panel illustrates the presence
of complex instabilities. It is relevant to stress that, both in this case
and in those considered below, the instabilities are associated with
quartets of complex eigenvalues (even when their imaginary parts are so
small that the eigenvalues may appear as real ones). The detailed spectra
shown in the middle and bottom panels of the figure (cf. Figs.~\ref{fig3c}-%
\ref{fig3g}) make it clear that the lowest perturbation eigenmodes are
always prone to instability, especially the ones with $m=0$ and $m=1$. For
smaller values of $D$, higher eigenmodes may become unstable too, and the
respective instabilities may eventually (i.e., at large $\mu _{+}$) even
dominate the respective growth rate. The enhanced instability at smaller $D$
is a natural feature to expect: indeed, as $D$ decreases, the notch shrinks, turning
into a circular quasi-1D dark soliton, whose snaking instability in two-dimensional
settings is well known \cite{siambook}. It is also relevant to stress that 
the oscillatory pattern, featured, especially, by the $m=0$ mode is associated with the
presence of gaps in the spectrum (for our finite-domain computation), which
allow the eigenmode to periodically restabilize, before it collides with
another one and destabilizes anew. Similar features for other
\textquotedblleft dark\textquotedblright\ patterns have long been known
(see, in particular, Ref.~\cite{prl8285}), and are absent in the
infinite-domain limit, where the relevant eigenvalue follows the envelope of
the respective \textquotedblleft trajectory\textquotedblright\ in the
spectral plane.
\begin{figure}[tp]
\begin{center}
\includegraphics[height=.16\textheight, angle =0]{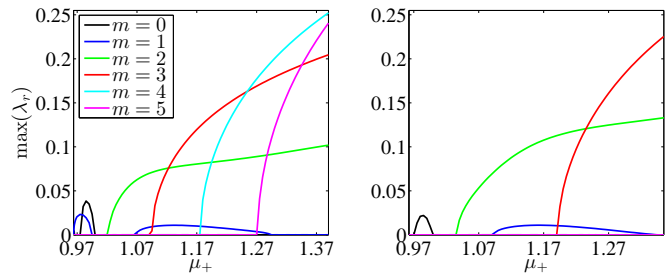}
\par
\vspace{-0.1cm}
\end{center}
\caption{(Color online) Same as Fig.~\protect\ref{fig3} but in the presence
of the harmonic-oscillator trap. The largest real part of the eigenvalues as
a function of $\protect\mu _{+}$ at $D=0.5$ is depicted in the left and
right panels for values of the trap's strength of $\Omega =0.2$ and $\Omega =0.3
$, respectively. }
\label{fig3_comp_trap}
\end{figure}
\begin{figure}[tp]
\begin{center}
\includegraphics[height=.28\textheight, angle =0]{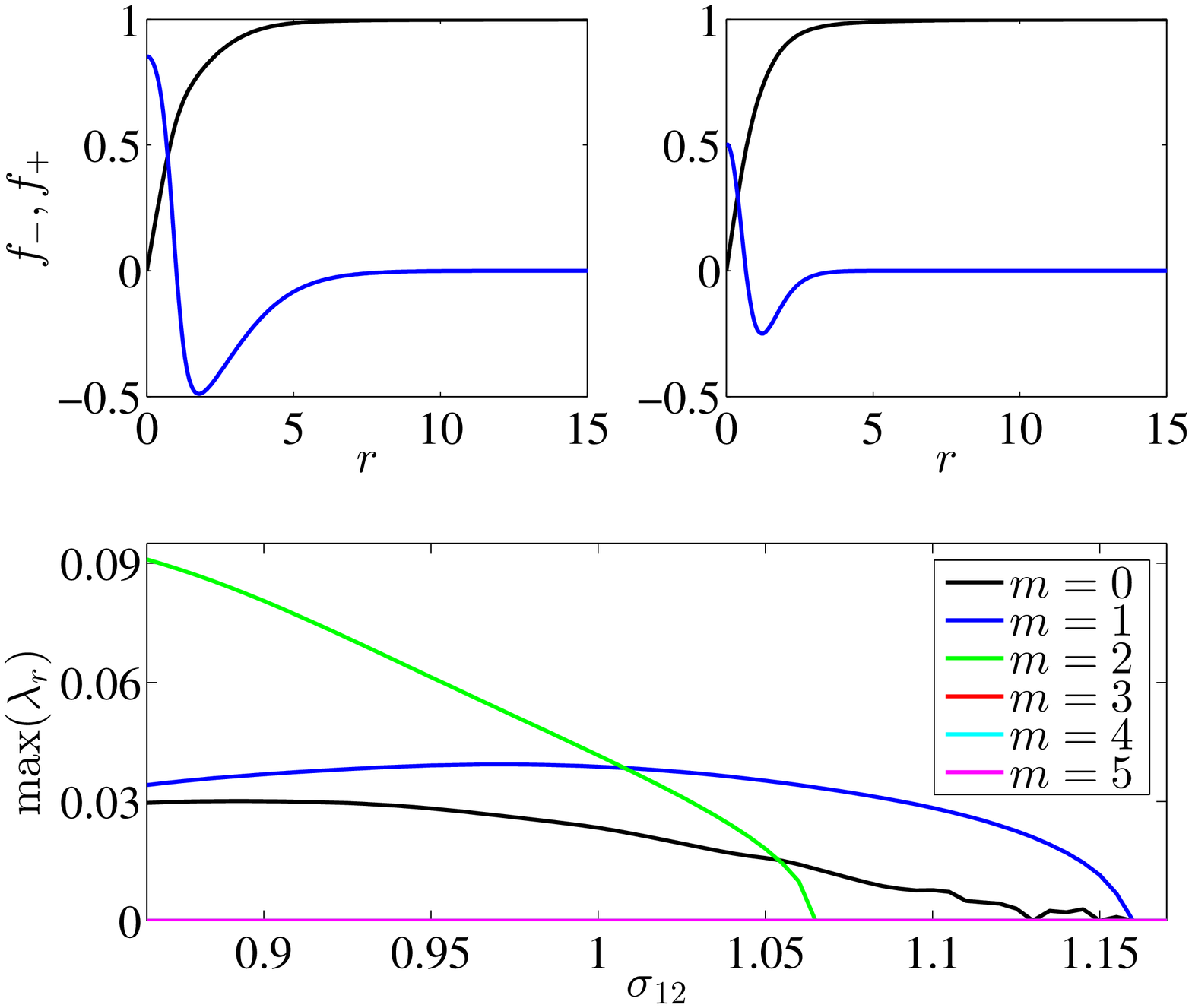}
\par
\vspace{-0.1cm}
\end{center}
\caption{(Color online) Same as Fig.~\protect\ref{fig3} but for $\protect\mu %
_{+}=0.86$ and $D=0.1$. \textit{Top row}: Stationary profiles of the vortex
and bright components for interaction coefficients $\protect\sigma _{12}=0.9$
(left panel) and $\protect\sigma _{12}=1.1$ (right panel). \textit{Bottom row%
}: The largest real part of the eigenvalues as a function of interaction
coefficient $\protect\sigma _{12}$. }
\label{fig3_comp_s12}
\end{figure}

Although our main focus in this study has been on the free space case, we
now briefly touch upon the setting involving the presence of the trapping
harmonic-oscillator potential. In particular, Fig.~\ref{fig3_comp_trap}
displays the linearization spectra of the trapped first excited state in the
bright component at $D=0.5$ for two different values of the trap's strength,
$\Omega =0.2$ and $\Omega =0.3$ in the left and right panels of the figure,
respectively. It is evident from both panels that the trap contributes to
the stability of the solution, if compared with the same branch in the free
space for the same $D$ depicted in panel (c) of Fig.~\ref{fig3} (see the
range of the values of $\mu _{+}$ as well). In addition, our findings
suggest that the stability interval (i.e., the width of the interval of $\mu
_{+}$ in which the branch is stable), is progressively wider as $\Omega $
increases (see the right panel of Fig.~\ref{fig3_comp_trap}). Thus,
gradually increasing values of the trap's strength results in wiping out the
unstable modes of the spectrum. This may be expected, as the parabolic trap
makes the linearization spectrum of the system discrete (while it is
continuous in the uniform space), gradually imposing a larger distance
between the relevant eigenvalues, thus suppressing resonant interactions
between modes that cause instabilities for such excited states~\cite%
{siambook}.

Up to now, we considered the system with all the interaction (or
nonlinearity) coefficients equal. It is also relevant to briefly touch upon
the variation of these coefficients as in realistic atomic systems they are
not precisely equal to unity; furthermore, these are parameters that can be
tuned by means of the Feshbach resonances controlling inter-atomic
scattering. In that vein, in Fig.~\ref{fig3_comp_s12} we consider the state
with the first excited state in the bright component at $\mu _{+}=0.86$ and $%
D=0.1$. In particular, the top left and right panels correspond to the
profiles for $\sigma _{12}=0.9$ and $\sigma _{12}=1.1$, respectively,
highlighting the transition from miscibility ($\sigma _{12}^{2}<g_{1}g_{2}$)
to immiscibility (for $\sigma _{12}^{2}>g_{1}g_{2}$). The bottom panel in
the figure shows the linearization spectrum in this case as a function of $%
\sigma _{12}$. In the 1D case studied in Ref. \cite{phyreva91}, it has been
observed that the stability of the individual dark-bright solitons is not 
dramatically affected by the variation of $\sigma _{12}$. Similar findings 
are observed in the present 2D case, although the instability growth rates 
start decreasing, resulting in a small stability region close to the upper 
bound of the examined window of $\sigma_{12}$ (see the bottom panel of Fig.%
~\ref{fig3_comp_s12}). It should be stressed at this point that the lower 
bound of $\sigma _{12}$ is determined by the fact that as $\sigma _{12}\,(<1)$ 
decreases, 
the expanding
bright structure eventually reaches the domain size. On the contrary, when $%
\sigma _{12}\,(>1)$ increases, the bright component becomes narrower and more
focused within the potential well induced by the dark structure. Furthermore, 
it should be pointed out that the structure decreases in amplitude with its 
amplitude being vanished, thus, determining the upper bound of the considered
interval of $\sigma_{12}$.

\begin{figure}[tp]
\begin{center}
\vspace{-0.1cm}
\mbox{\hspace{-0.1cm}
\subfigure[][]{\hspace{-0.3cm}
\includegraphics[height=.16\textheight, angle =0]{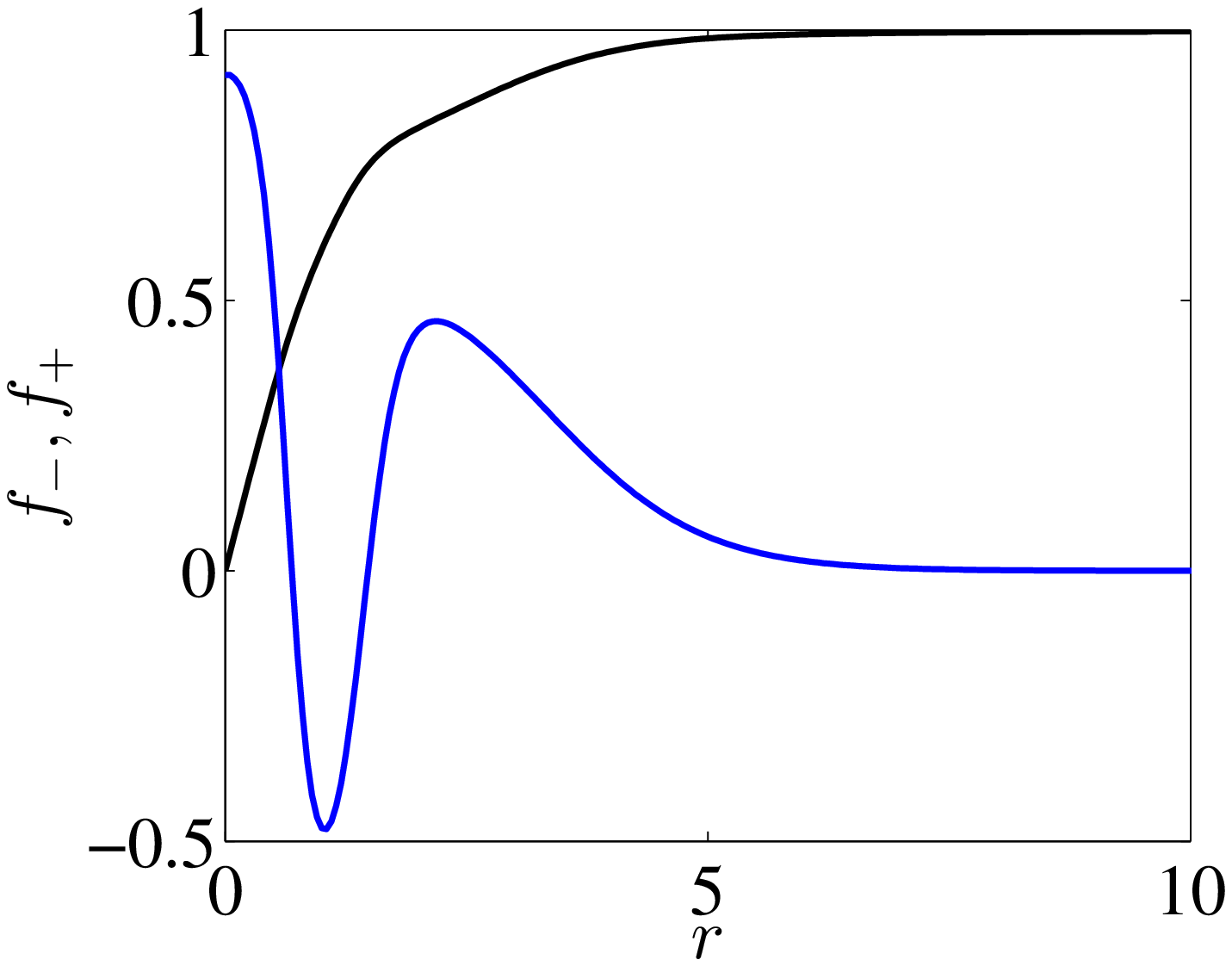}
\label{fig4a}
}
\subfigure[][]{\hspace{-0.3cm}
\includegraphics[height=.16\textheight, angle =0]{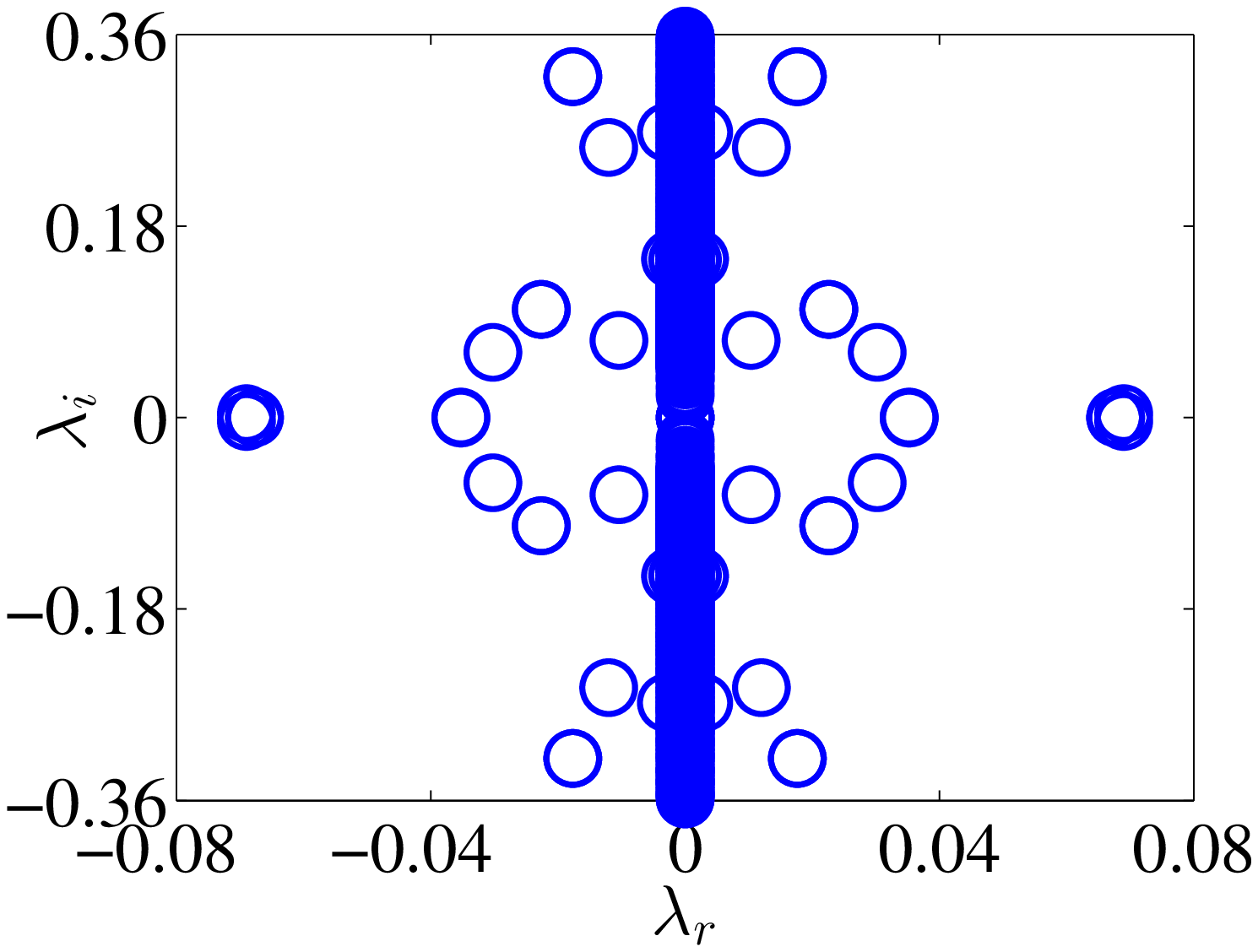}
\label{fig4b}
}
} \vspace{-0.5cm}
\mbox{\hspace{-0.1cm}
\subfigure[][]{\hspace{-0.3cm}
\includegraphics[height=.16\textheight, angle =0]{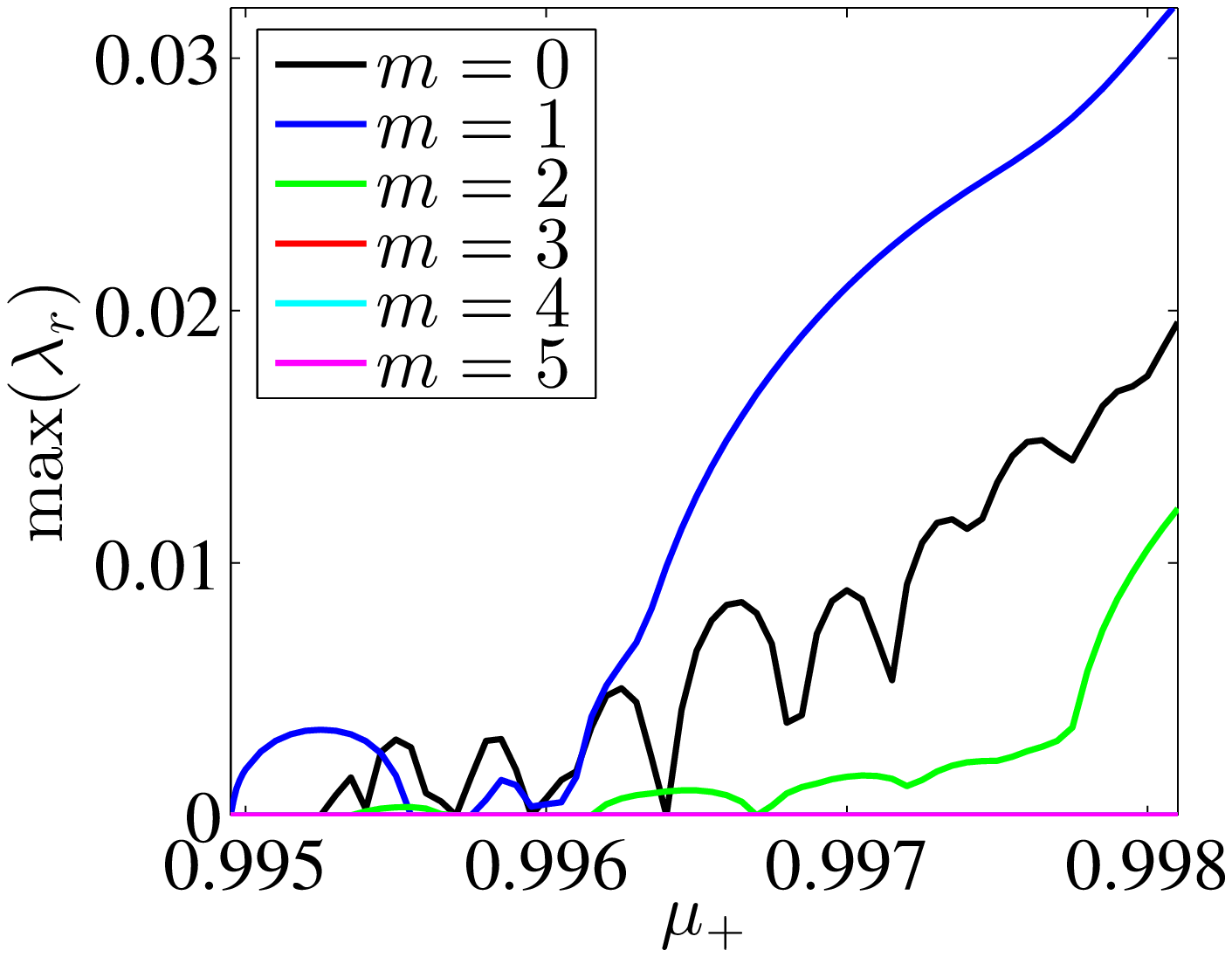}
\label{fig4c}
}
\subfigure[][]{\hspace{-0.3cm}
\includegraphics[height=.16\textheight, angle =0]{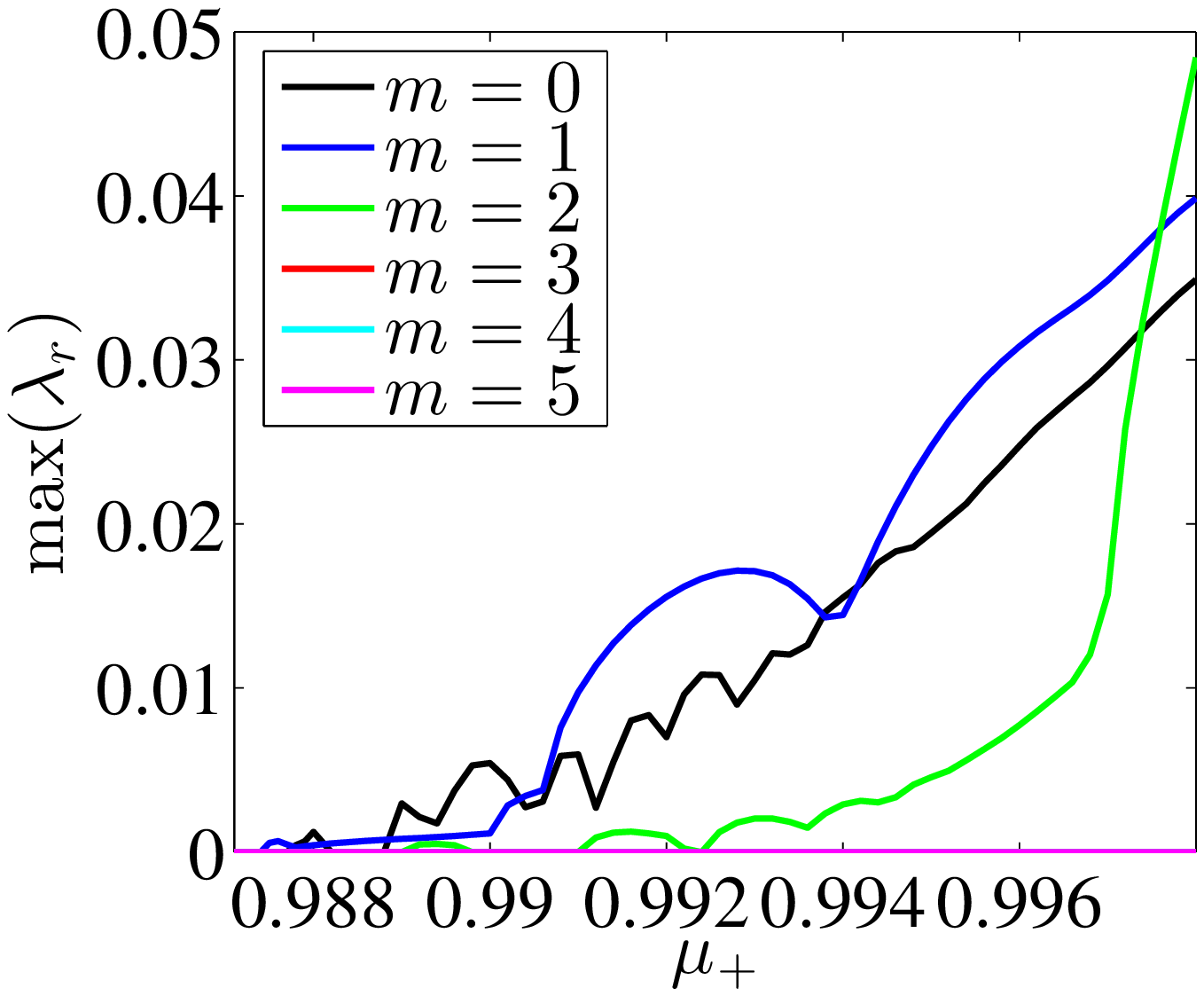}
\label{fig4d}
}
}
\mbox{\hspace{-0.1cm}
\subfigure[][]{\hspace{-0.3cm}
\includegraphics[height=.16\textheight, angle =0]{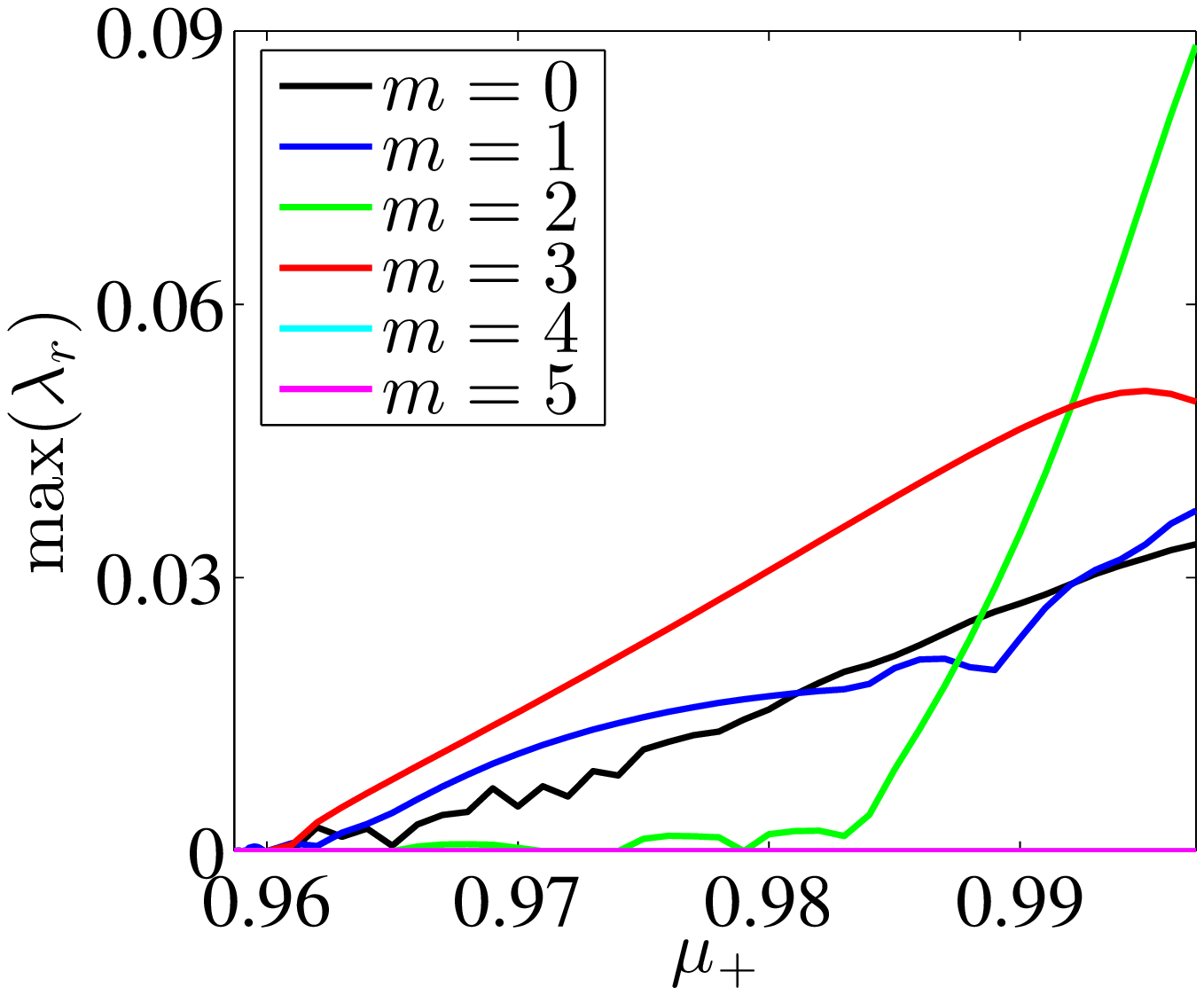}
\label{fig4f}
}
\subfigure[][]{\hspace{-0.3cm}
\includegraphics[height=.16\textheight, angle =0]{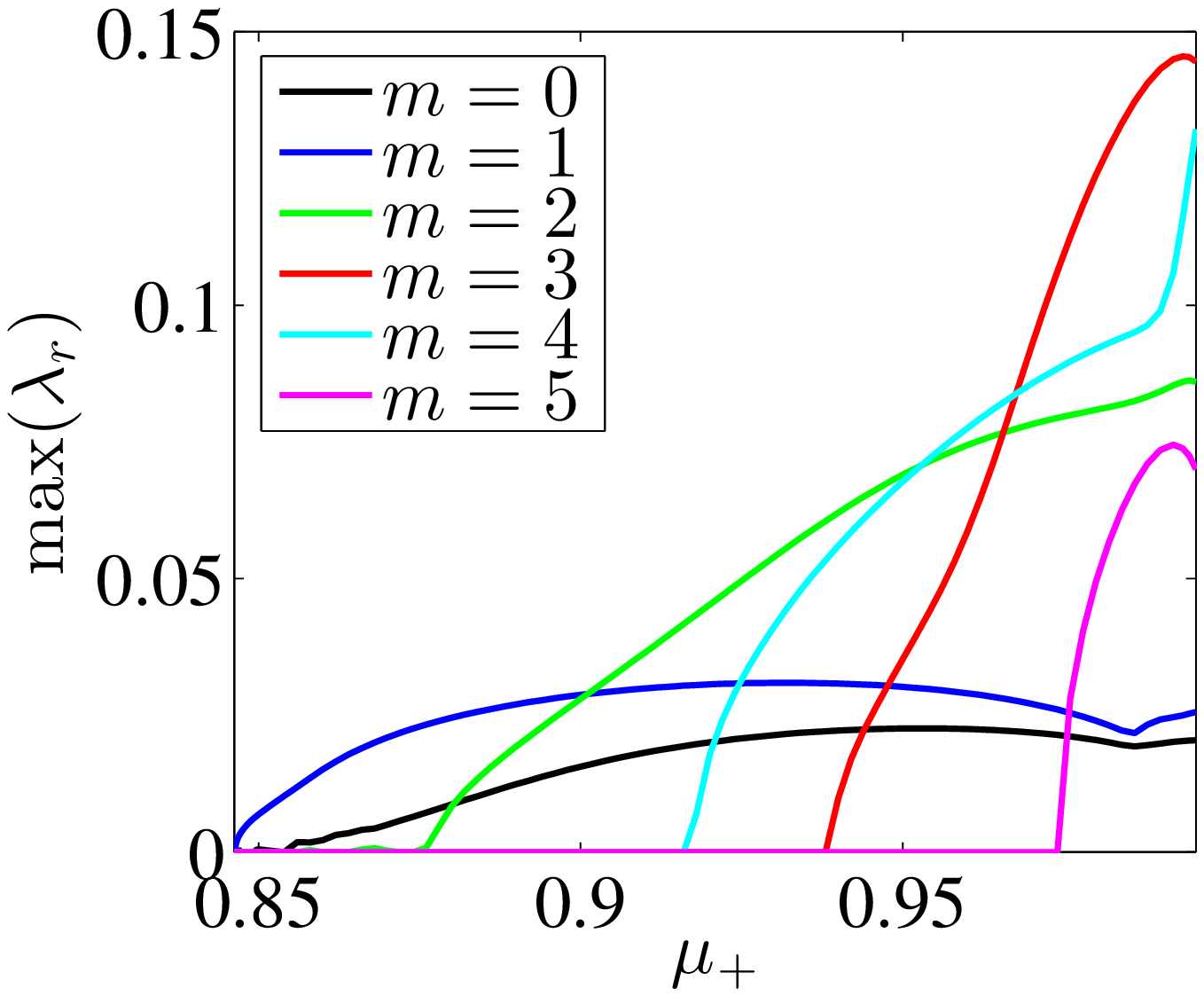}
\label{fig4g}
}
}
\end{center}
\par
\vspace{-0.55cm}
\caption{(Color online) Same as Fig.~\protect\ref{fig3} but for the second
excited states in the bright component. \textit{Top row}: (a) Steady-state
profiles of the vortex and bright components (the black and blue lines,
respectively). (b) The corresponding eigenvalue spectrum for $D=0.05$ and $%
\protect\mu _{+}=0.95$. \textit{Middle and bottom rows}: The largest real
part of eigenvalues as a function of $\protect\mu _{+}$ at fixed values of $%
D $: (c) $D=0.2$, (d) $D=0.15$, (e) $D=0.1$, and (f) $D=0.05$.}
\label{fig4}
\end{figure}

\begin{figure}[tph]
\begin{center}
\vspace{-0.1cm}
\mbox{\hspace{-0.1cm}
\subfigure[][]{\hspace{-0.3cm}
\includegraphics[height=.16\textheight, angle =0]{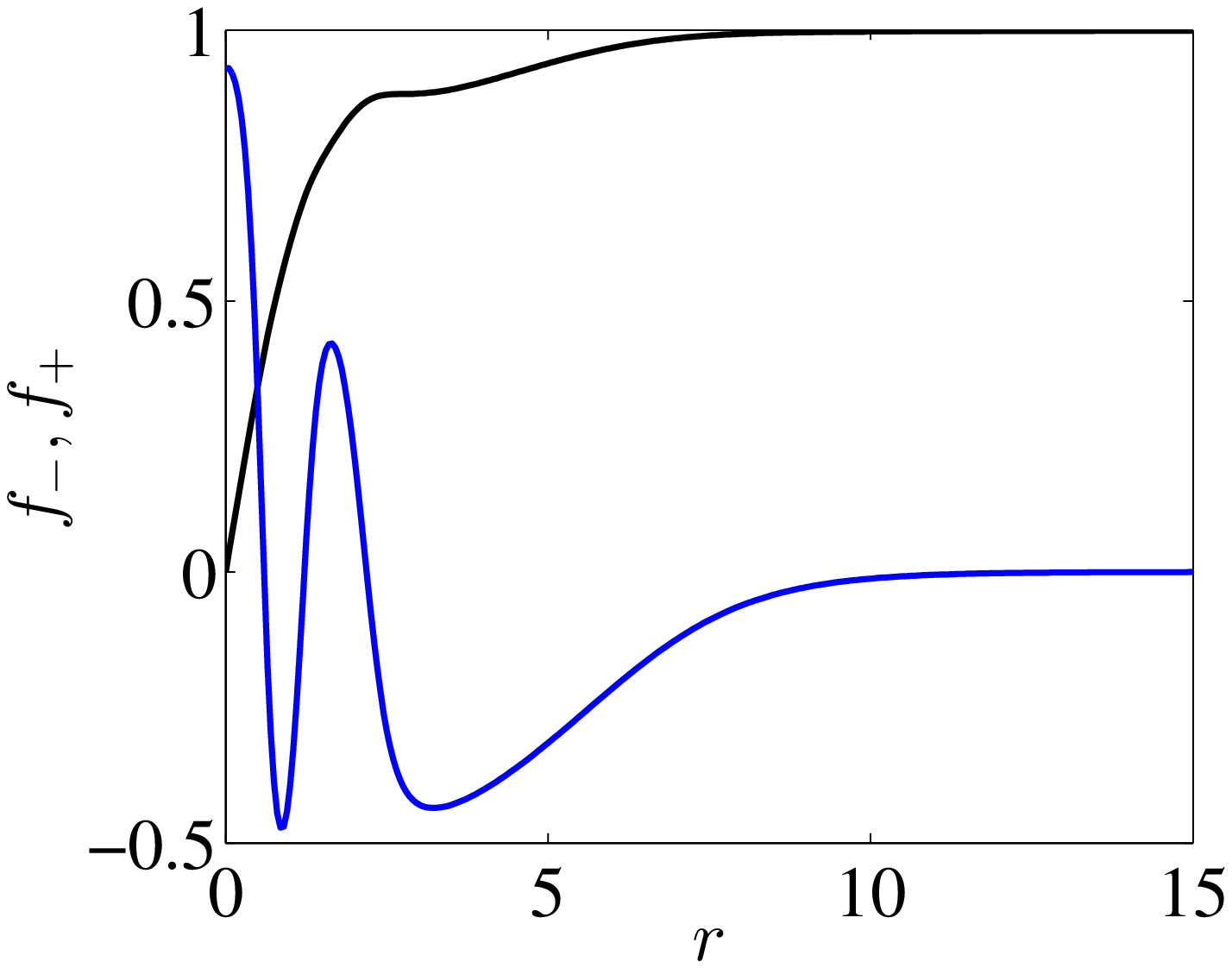}
\label{fig5a}
}
\subfigure[][]{\hspace{-0.3cm}
\includegraphics[height=.16\textheight, angle =0]{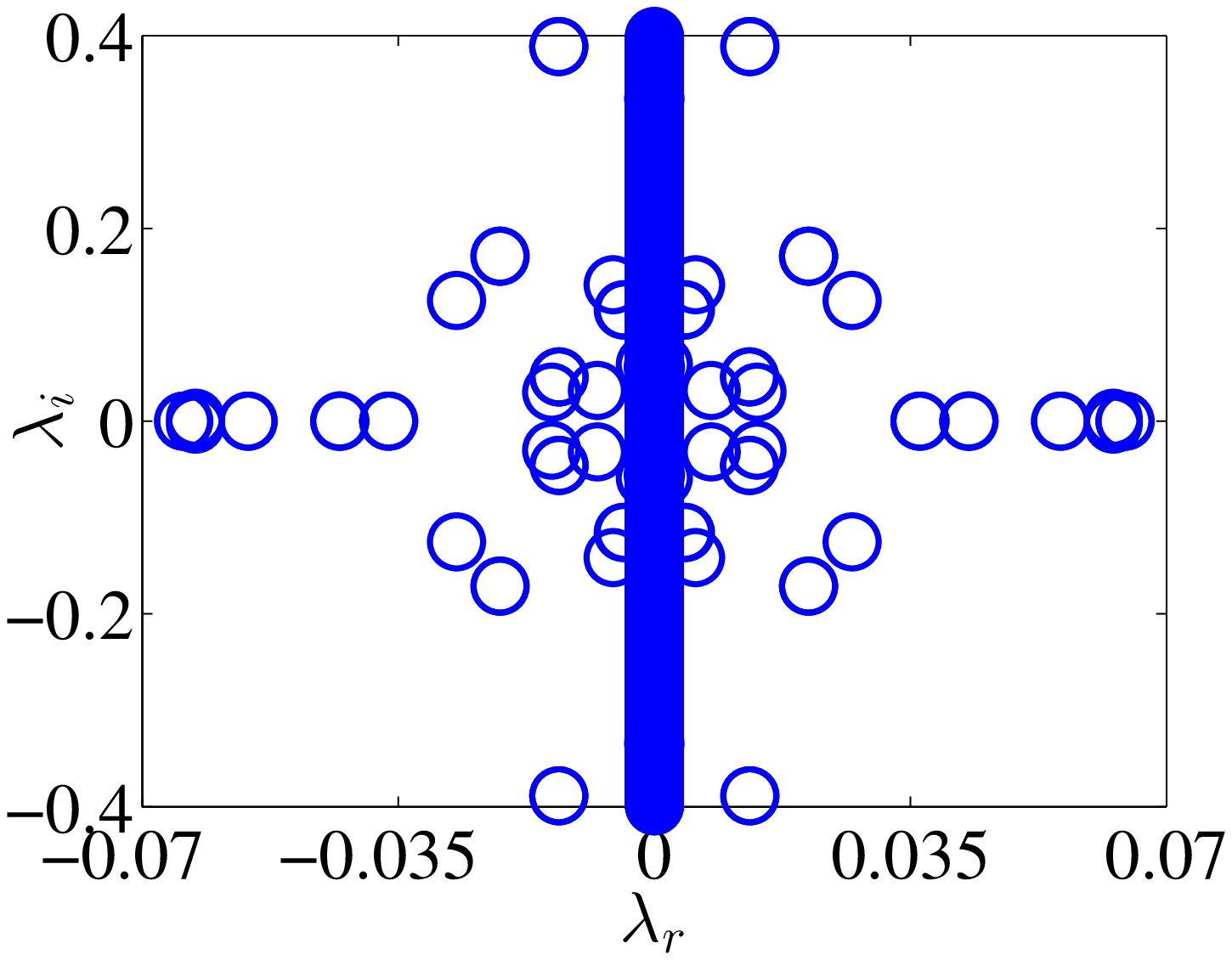}
\label{fig5b}
}
}
\mbox{\hspace{-0.1cm}
\subfigure[][]{\hspace{-0.3cm}
\includegraphics[height=.16\textheight, angle =0]{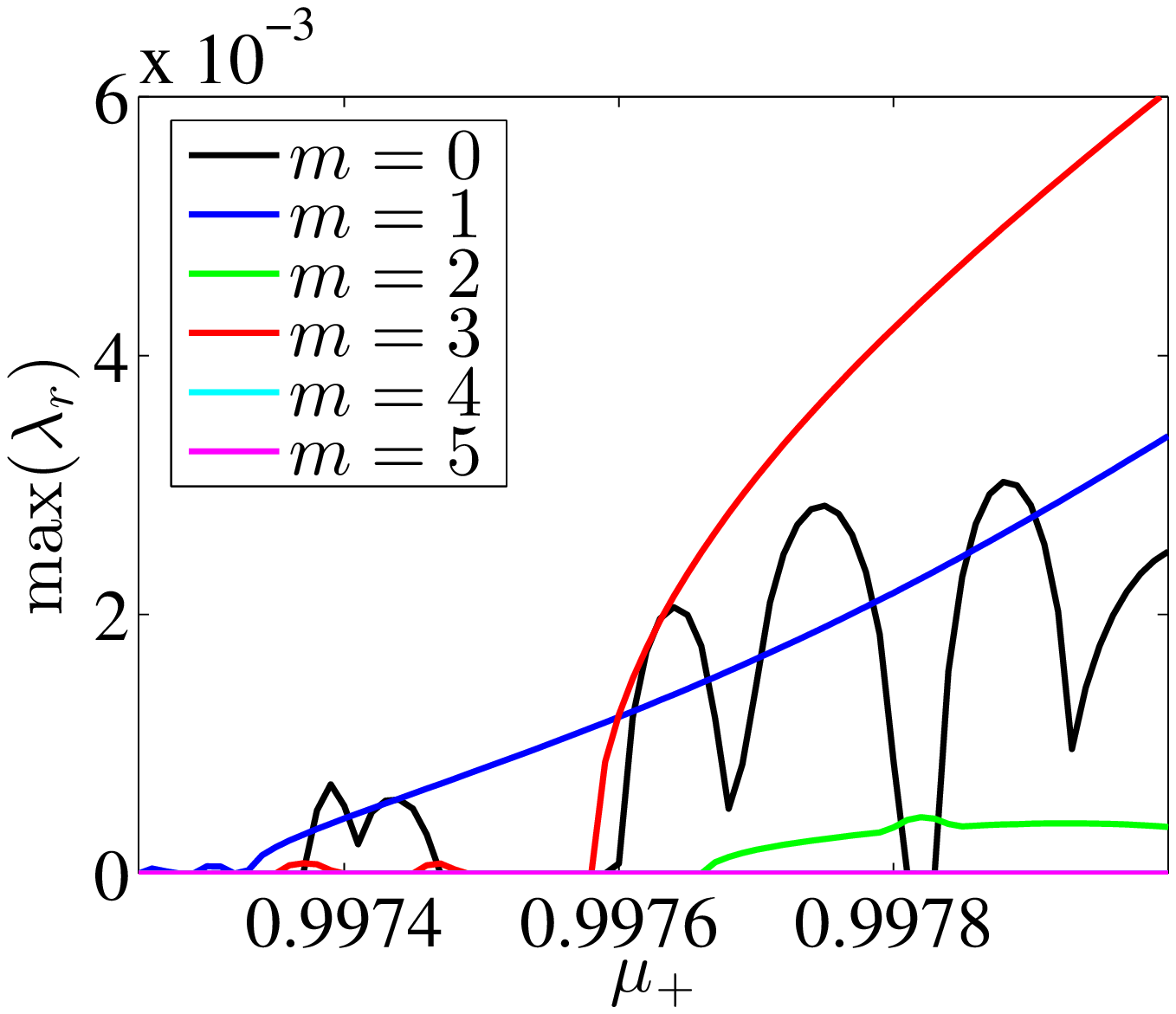}
\label{fig5c}
}
\subfigure[][]{\hspace{-0.3cm}
\includegraphics[height=.16\textheight, angle =0]{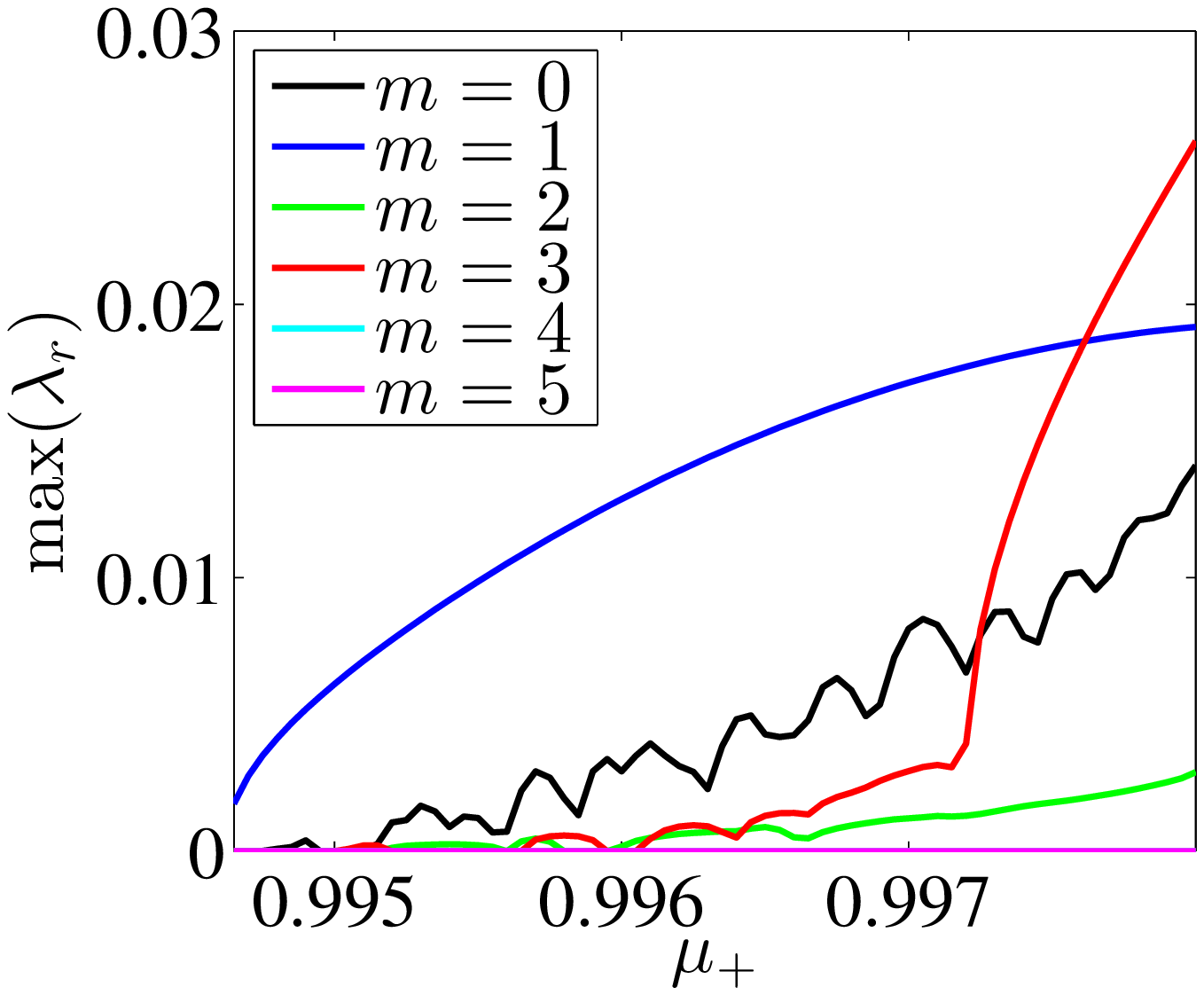}
\label{fig5d}
}
\subfigure[][]{\hspace{-0.3cm}
\includegraphics[height=.16\textheight, angle =0]{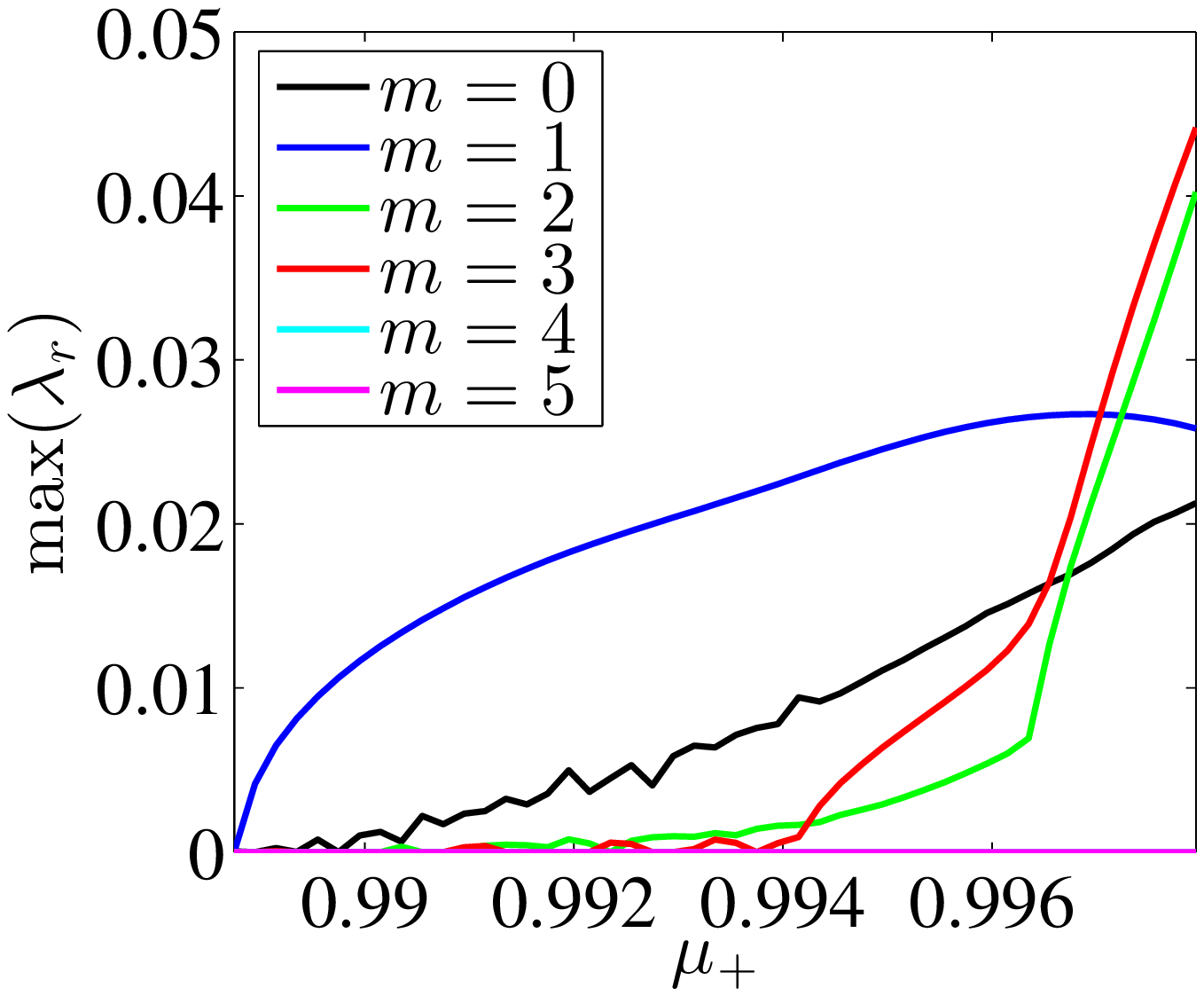}
\label{fig5e}
}
}
\mbox{\hspace{-0.1cm}
\subfigure[][]{\hspace{-0.3cm}
\includegraphics[height=.16\textheight, angle =0]{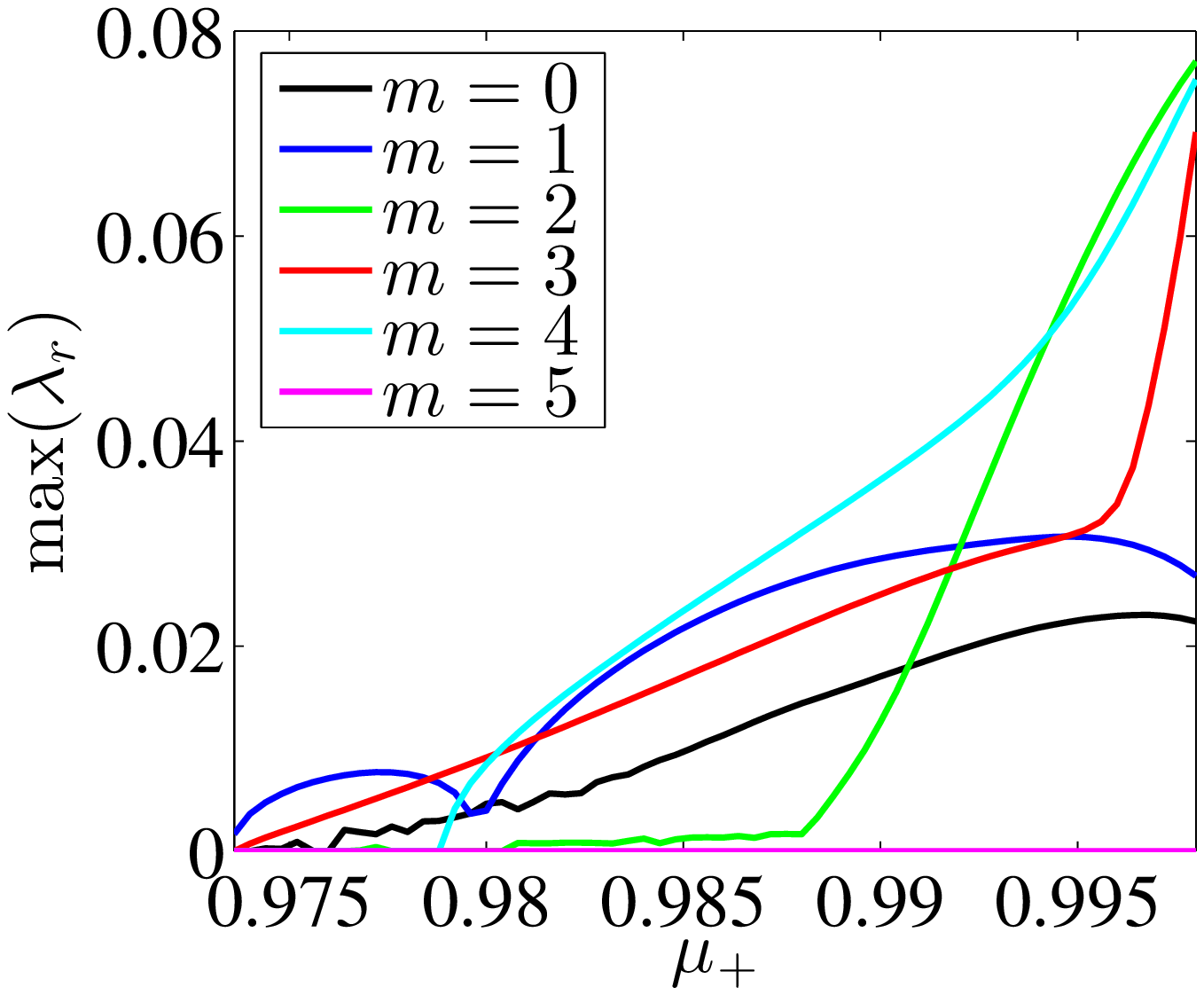}
\label{fig5f}
}
\subfigure[][]{\hspace{-0.3cm}
\includegraphics[height=.16\textheight, angle =0]{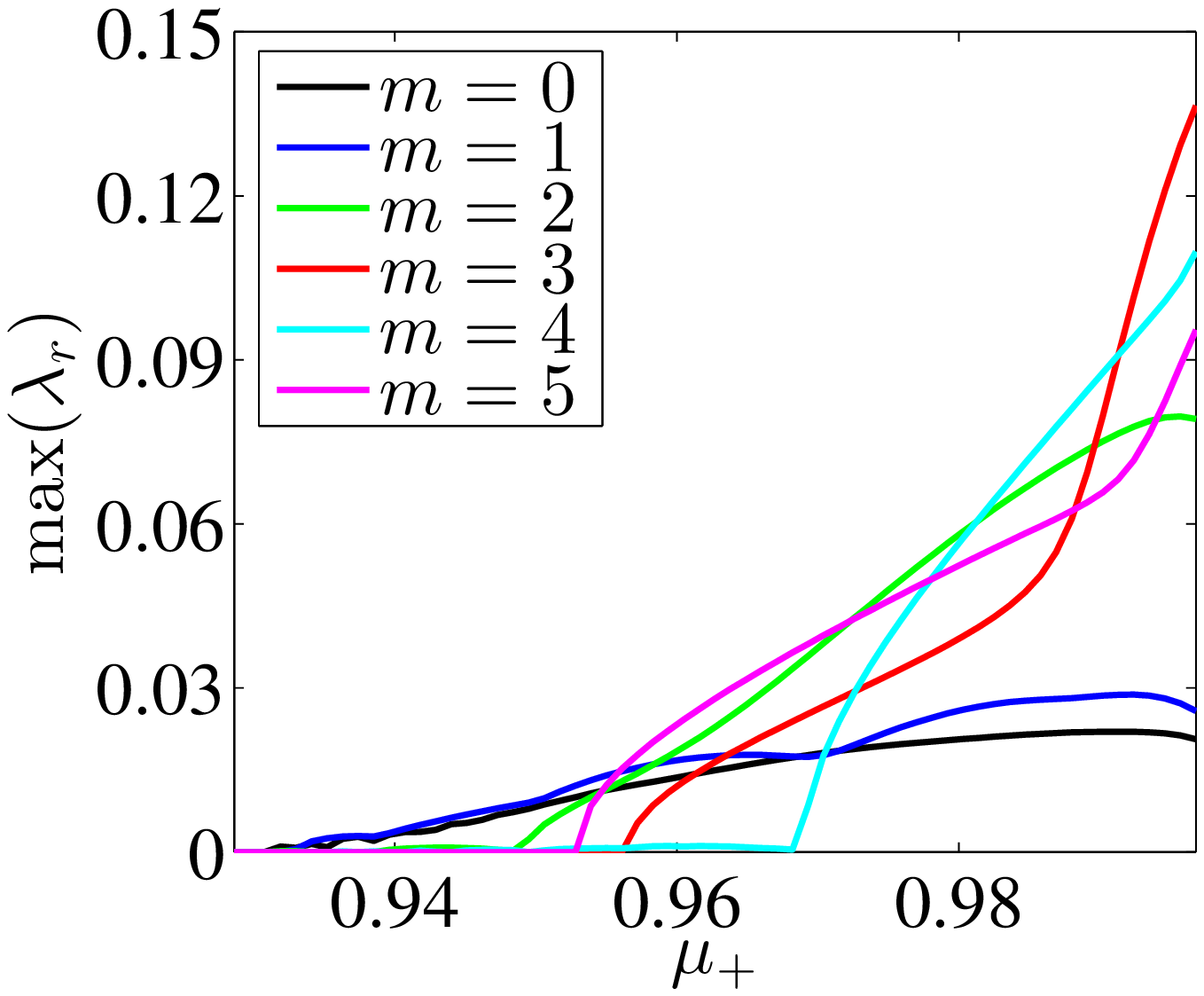}
\label{fig5g}
}
}
\end{center}
\par
\vspace{-0.55cm}
\caption{(Color online) Same as Fig.~\protect\ref{fig3}, but for the third
excited states in the bright component. \textit{Top row}: (a) Steady-state
profiles of the vortex and bright components (black and blue lines,
respectively). (b) The corresponding eigenvalue spectrum for $D=0.04$ and $%
\protect\mu _{+}=0.9825$. \textit{Middle and bottom rows}: The largest real
eigenvalue as a function of $\protect\mu _{+}$ at fixed values of $D$: (c) $%
D=0.12$, (d) $D=0.1$, (e) $D=0.08$, (f) $D=0.06$, and (g) $D=0.04$.}
\label{fig5}
\end{figure}

The case of the second and third excited states in the bright component is
considered in Figs.~\ref{fig4} and \ref{fig5}, respectively. The former
state features a triple local density maximum in the bright component, with
these maxima separated by two dark rings; the latter state has four local
maxima, separated by three dark rings, as shown in the respective top left
panels of the figures. One can also observe in the corresponding top right
panels, which showcase typical examples of the spectral plane, $(\lambda
_{r},\lambda _{i})$, of eigenvalues $\lambda =\lambda _{r}+i\lambda _{i}$,
that the number of unstable modes is getting progressively larger with the
increase of the order of the state. Some additional relevant observations
regarding these figures are as follows. In Fig.~\ref{fig4}, we observe that,
for sufficiently small dispersion coefficient $D$, eigenvalues of
higher-order perturbation modes, including ones for $m=3$ (and even $m=4$
and $5$) grow fast with $\mu _{+}$, so that they play a dominant role in the
resulting dynamics, making it different from that in more typical cases of $%
m=0$ and $m=1$. As for the waveform in Fig.~\ref{fig5}, on the other hand,
it is relevant to point out that the internal structure of the ring state
has a conspicuous feedback effect on the spatial profile of the vortex. In
this case, the vortex features a nearly non-monotonic profile. Here, too,
higher perturbation eigenmodes, including most notably $m=3$, but also, in
some cases, $m=2$, $m=4$, etc., may result in the largest growth rate of the
instability.

\begin{figure}[tbp]
\begin{center}
\vspace{-0.1cm}
\mbox{\hspace{-0.1cm}
\subfigure[][]{\hspace{-0.3cm}
\includegraphics[height=.16\textheight, angle =0]{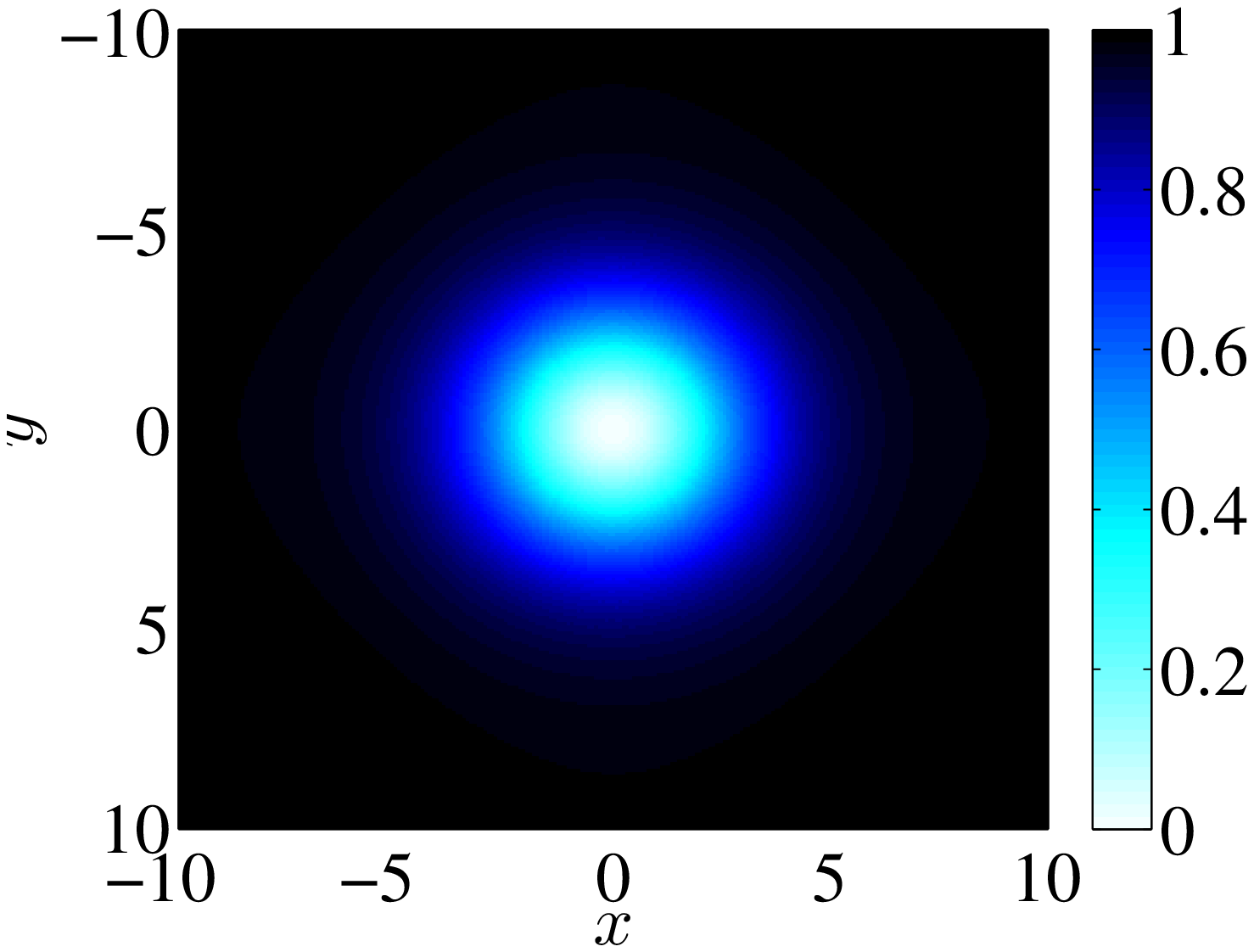}
\label{fig6a}
}
\subfigure[][]{\hspace{-0.3cm}
\includegraphics[height=.16\textheight, angle =0]{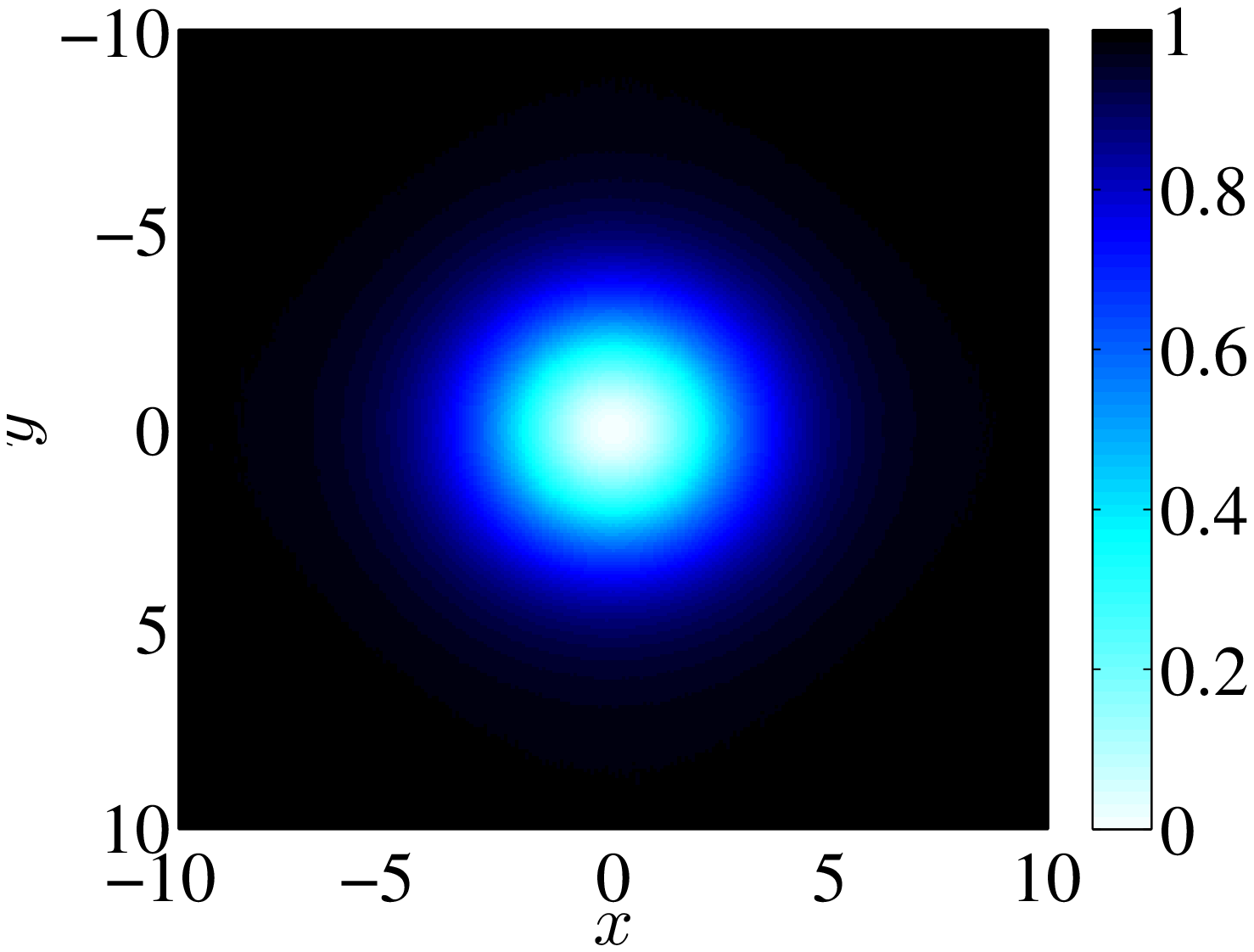}
\label{fig6b}
}
\subfigure[][]{\hspace{-0.3cm}
\includegraphics[height=.16\textheight, angle =0]{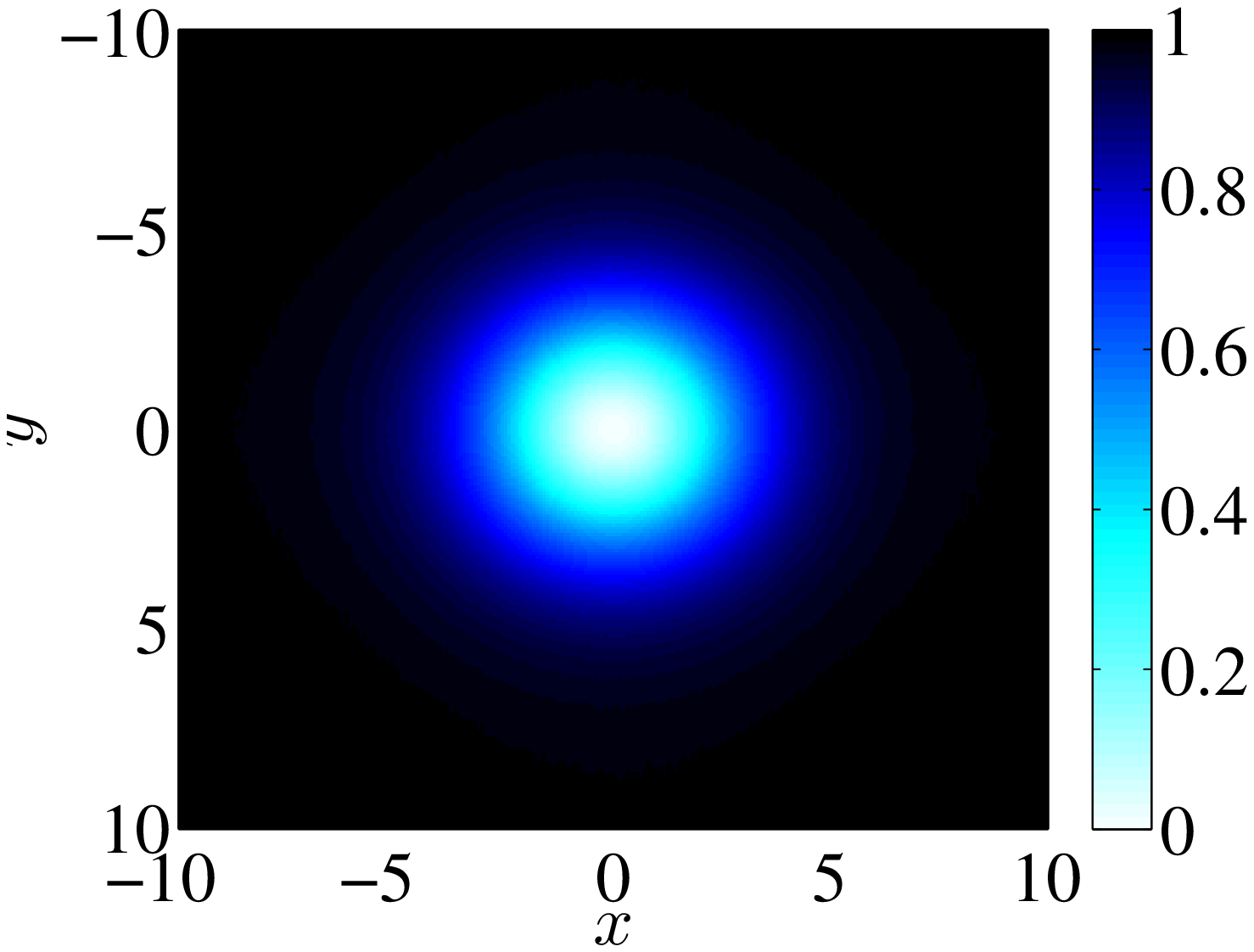}
\label{fig6c}
}
}
\mbox{\hspace{-0.1cm}
\subfigure[][]{\hspace{-0.3cm}
\includegraphics[height=.16\textheight, angle =0]{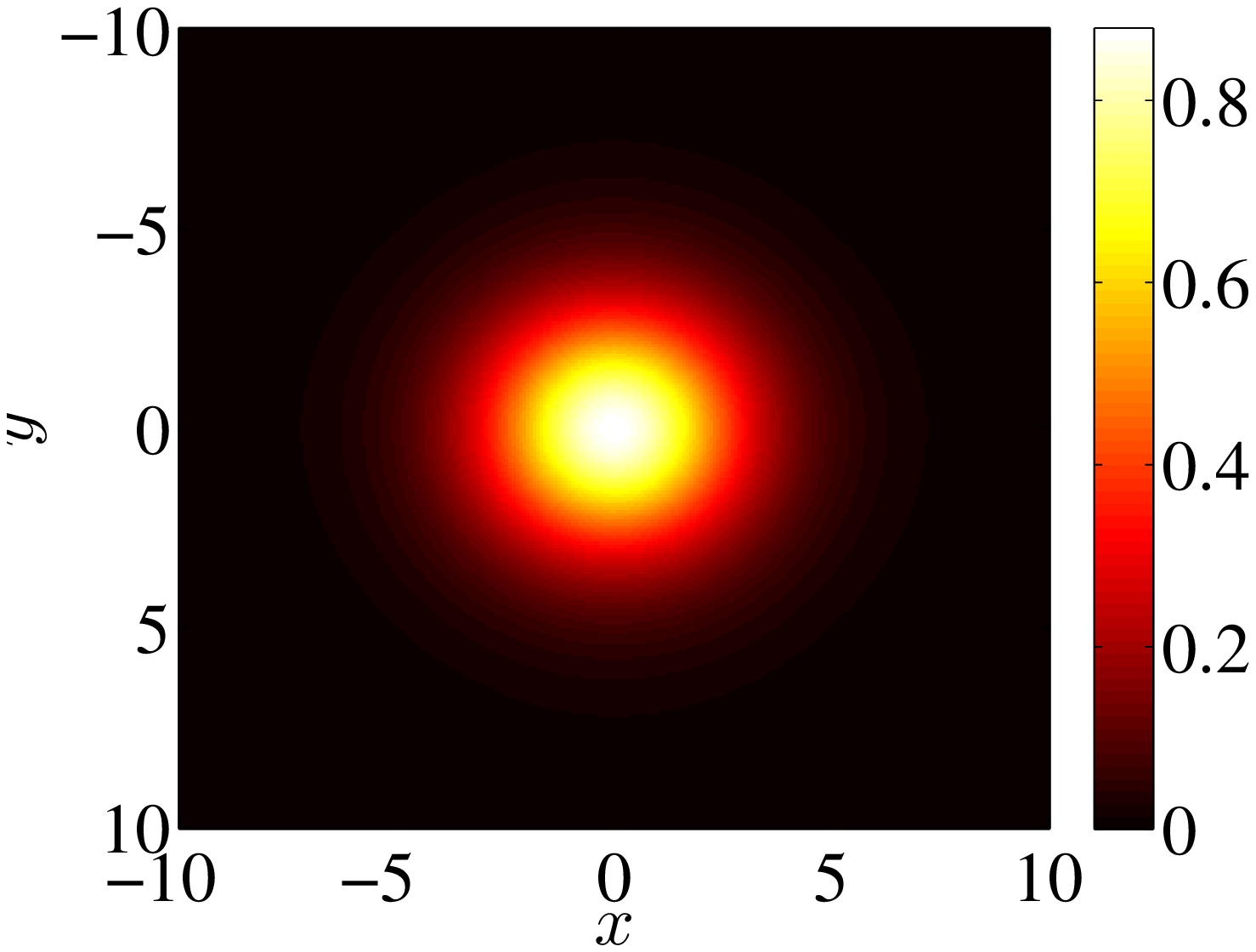}
\label{fig6d}
}
\subfigure[][]{\hspace{-0.3cm}
\includegraphics[height=.16\textheight, angle =0]{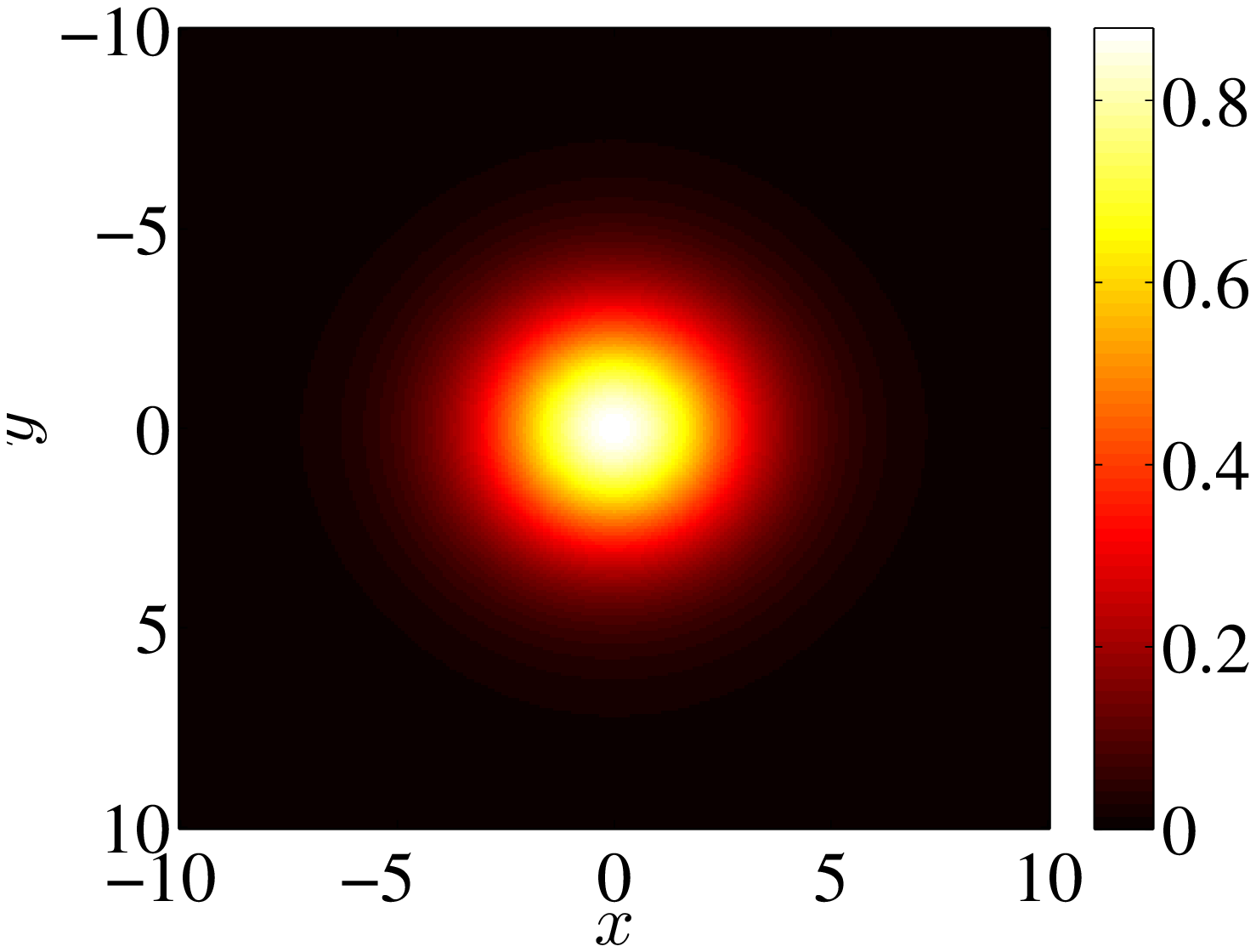}
\label{fig6e}
}
\subfigure[][]{\hspace{-0.3cm}
\includegraphics[height=.16\textheight, angle =0]{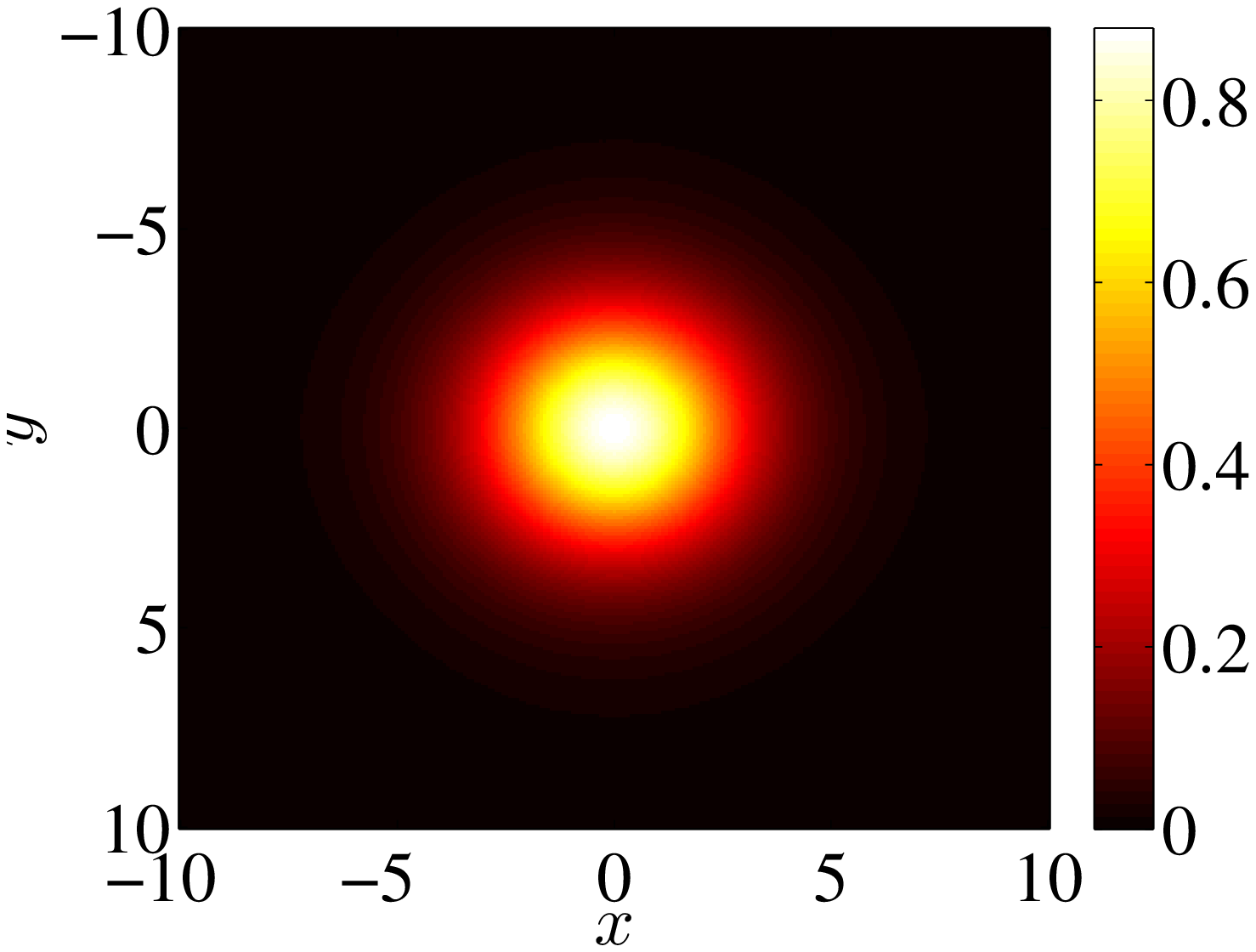}
\label{fig6f}
}
}
\end{center}
\par
\vspace{-0.55cm}
\caption{(Color online) The evolution of densities $|\Phi _{-}(x,t)|^{2}$
and $|\Phi _{+}(x,t)|^{2}$ (the top and bottom rows), displayed at different
instants of time: $t=0$ (left panels), $t=1000$ (middle panels), and $t=2000$
(right panels), for perturbed complexes with the bright component in the
form of the ground state, at $D=0.6$ and $\protect\mu _{+}=0.95$.}
\label{fig6}
\end{figure}

Having examined the spectral stability of the different states, we now turn
to direct simulations to study the evolution of these states. First, in Fig.~%
\ref{fig6}, we confirm that the evolution of the fundamental VB soliton
branch (where the bright component is the GS of the vortex-induced
potential) does not exhibit any instability in long simulations (up to $%
t=2000$), even though the solution is initially perturbed.

\begin{figure}[tbp]
\begin{center}
\vspace{-0.1cm}
\mbox{\hspace{-0.1cm}
\subfigure[][]{\hspace{-0.3cm}
\includegraphics[height=.16\textheight, angle =0]{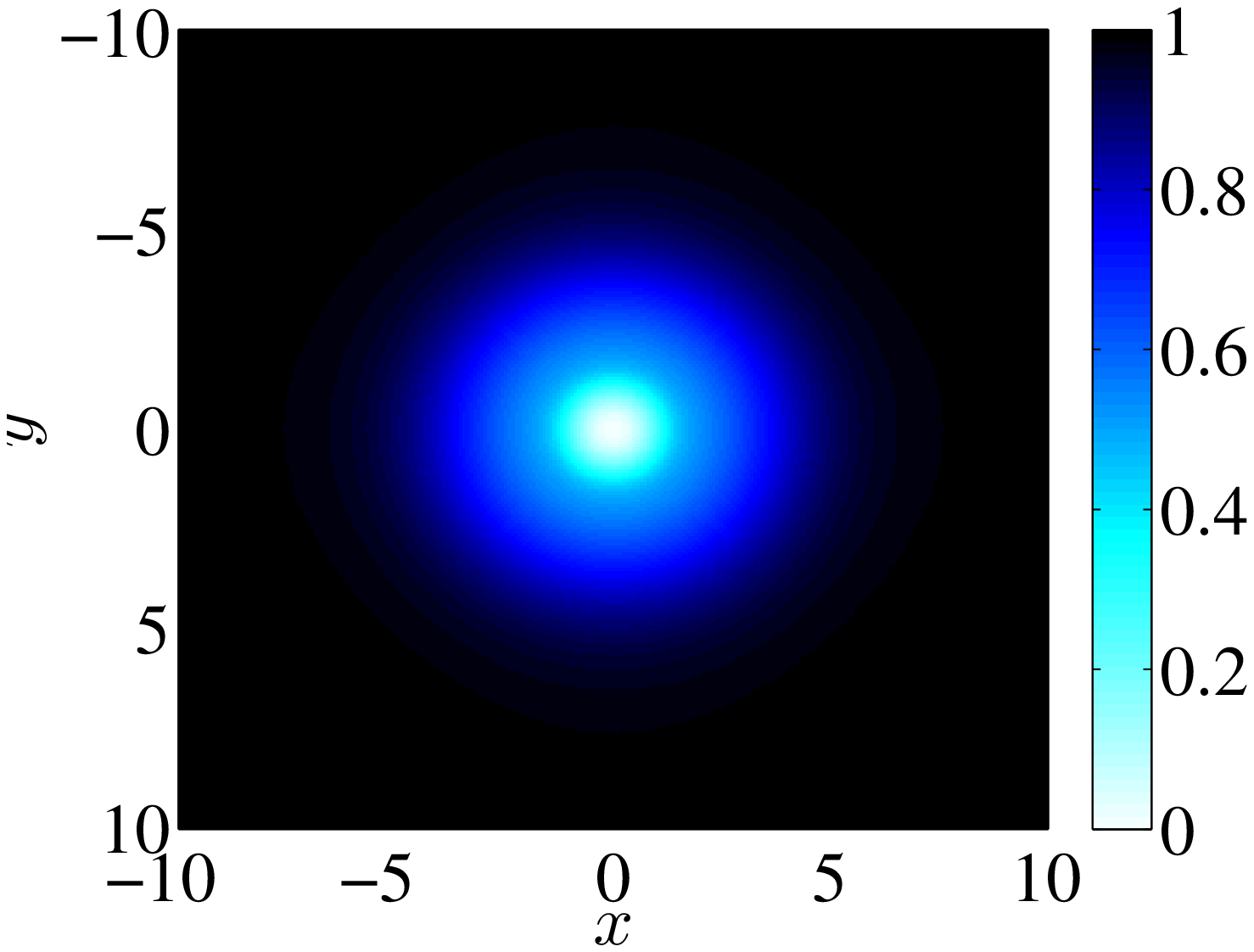}
\label{fig7a}
}
\subfigure[][]{\hspace{-0.3cm}
\includegraphics[height=.16\textheight, angle =0]{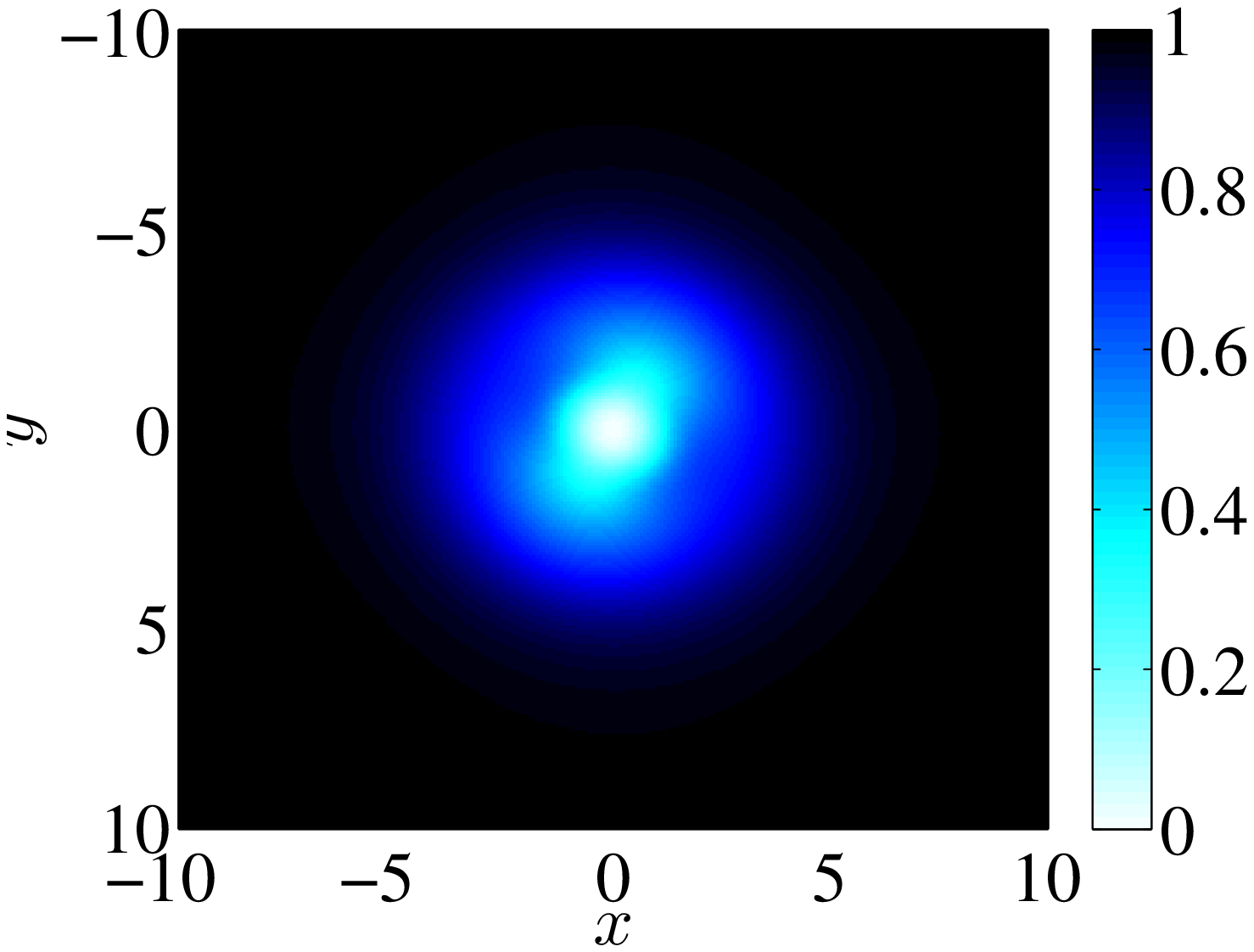}
\label{fig7b}
}
\subfigure[][]{\hspace{-0.3cm}
\includegraphics[height=.16\textheight, angle =0]{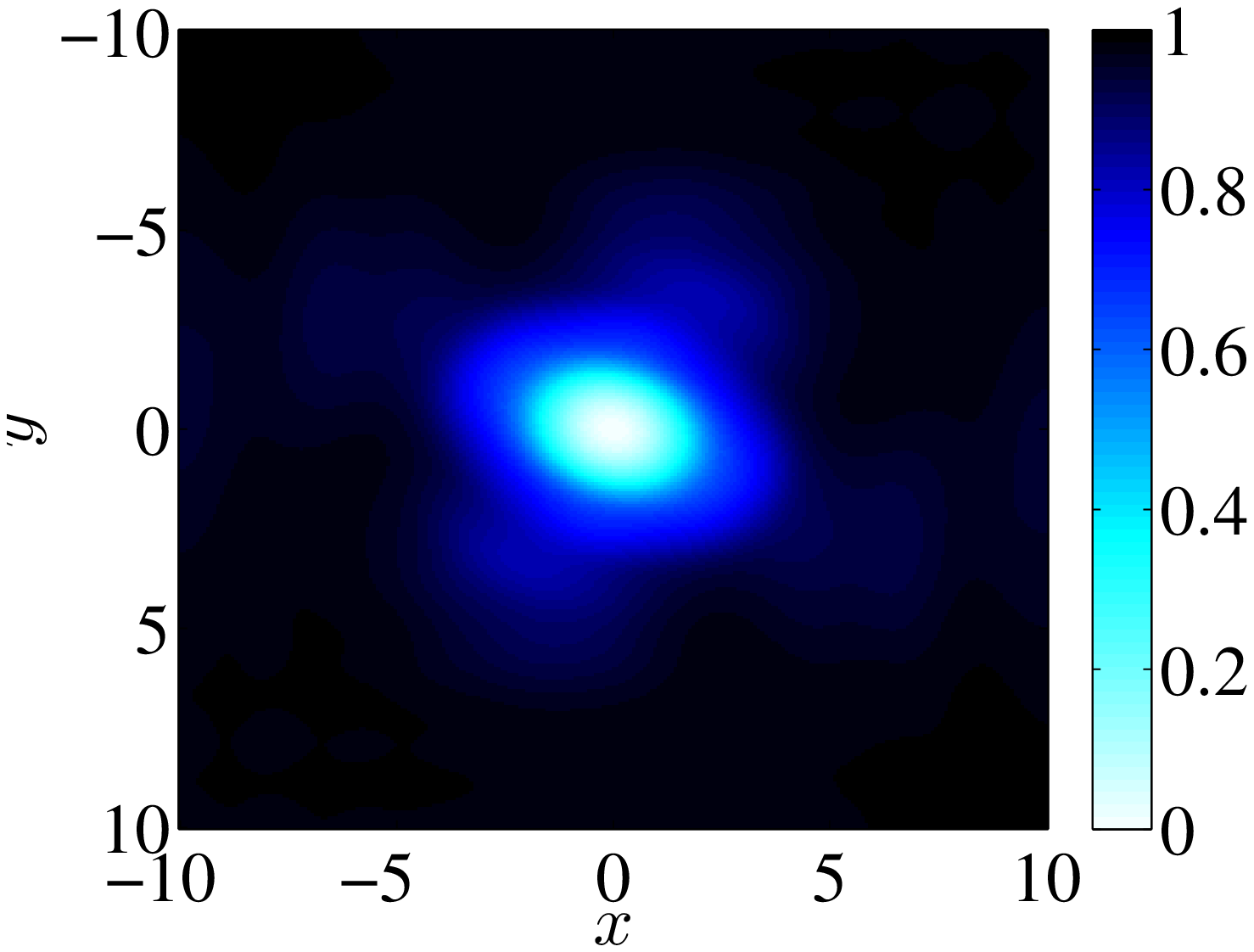}
\label{fig7c}
}
}
\mbox{\hspace{-0.1cm}
\subfigure[][]{\hspace{-0.3cm}
\includegraphics[height=.16\textheight, angle =0]{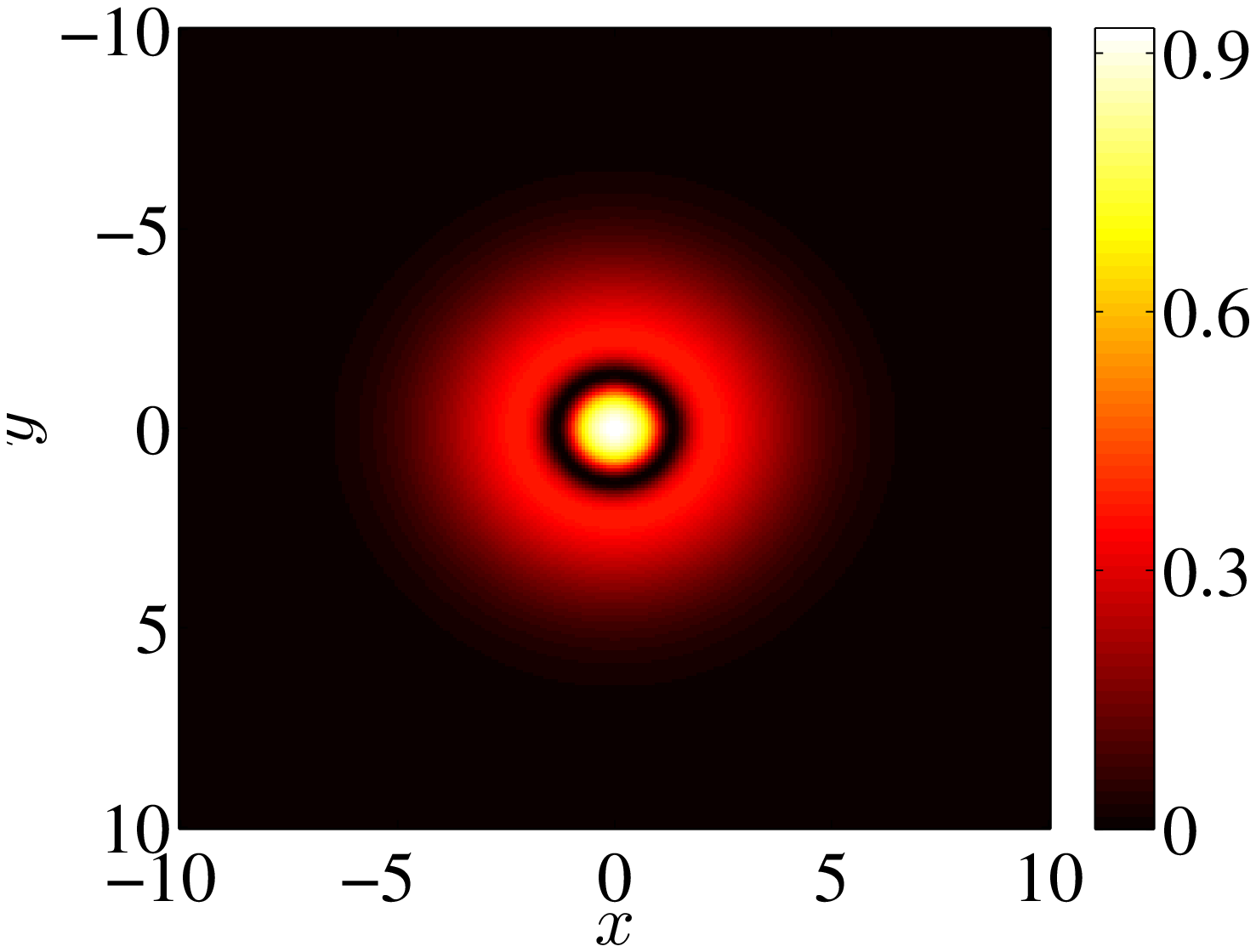}
\label{fig7d}
}
\subfigure[][]{\hspace{-0.3cm}
\includegraphics[height=.16\textheight, angle =0]{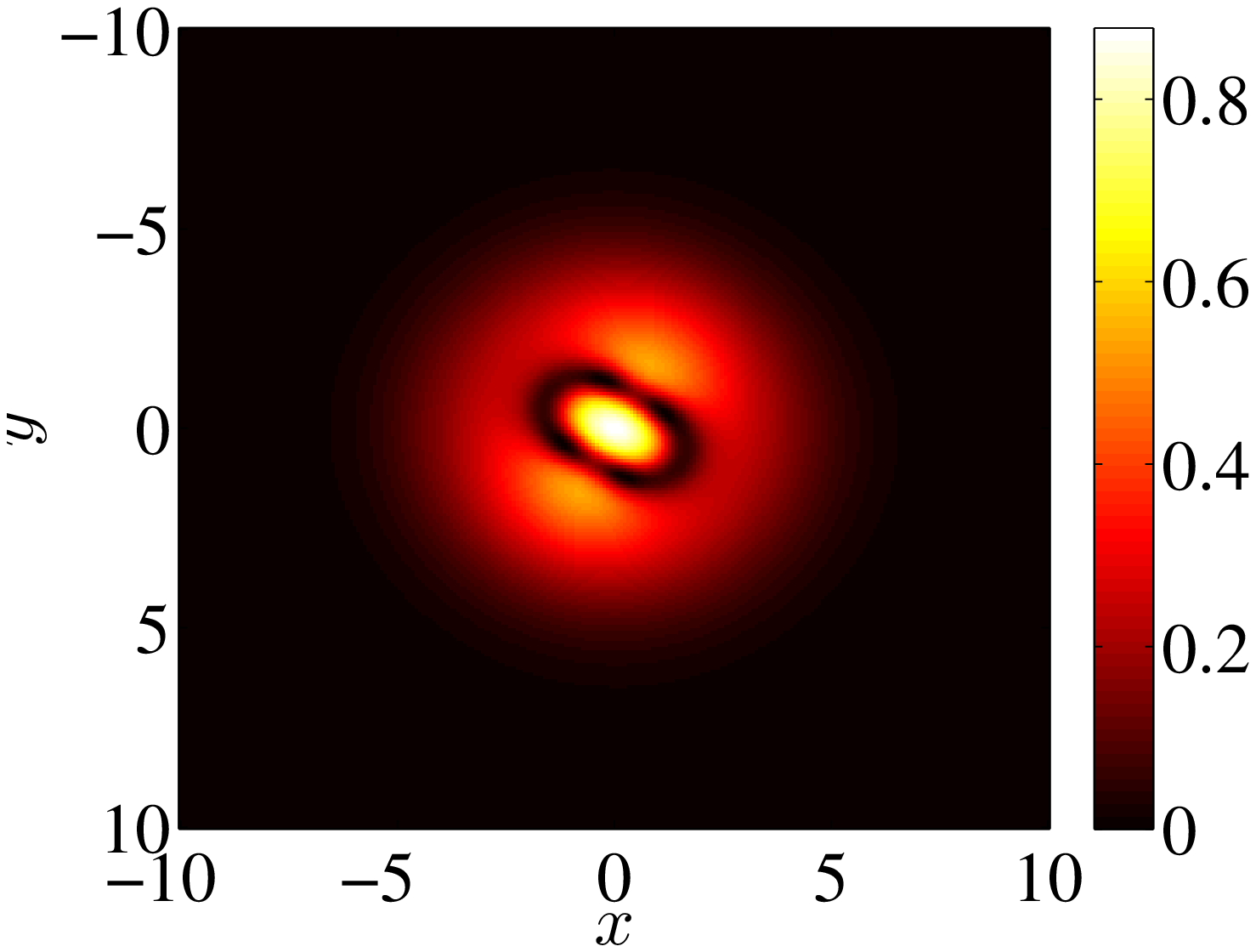}
\label{fig7e}
}
\subfigure[][]{\hspace{-0.3cm}
\includegraphics[height=.16\textheight, angle =0]{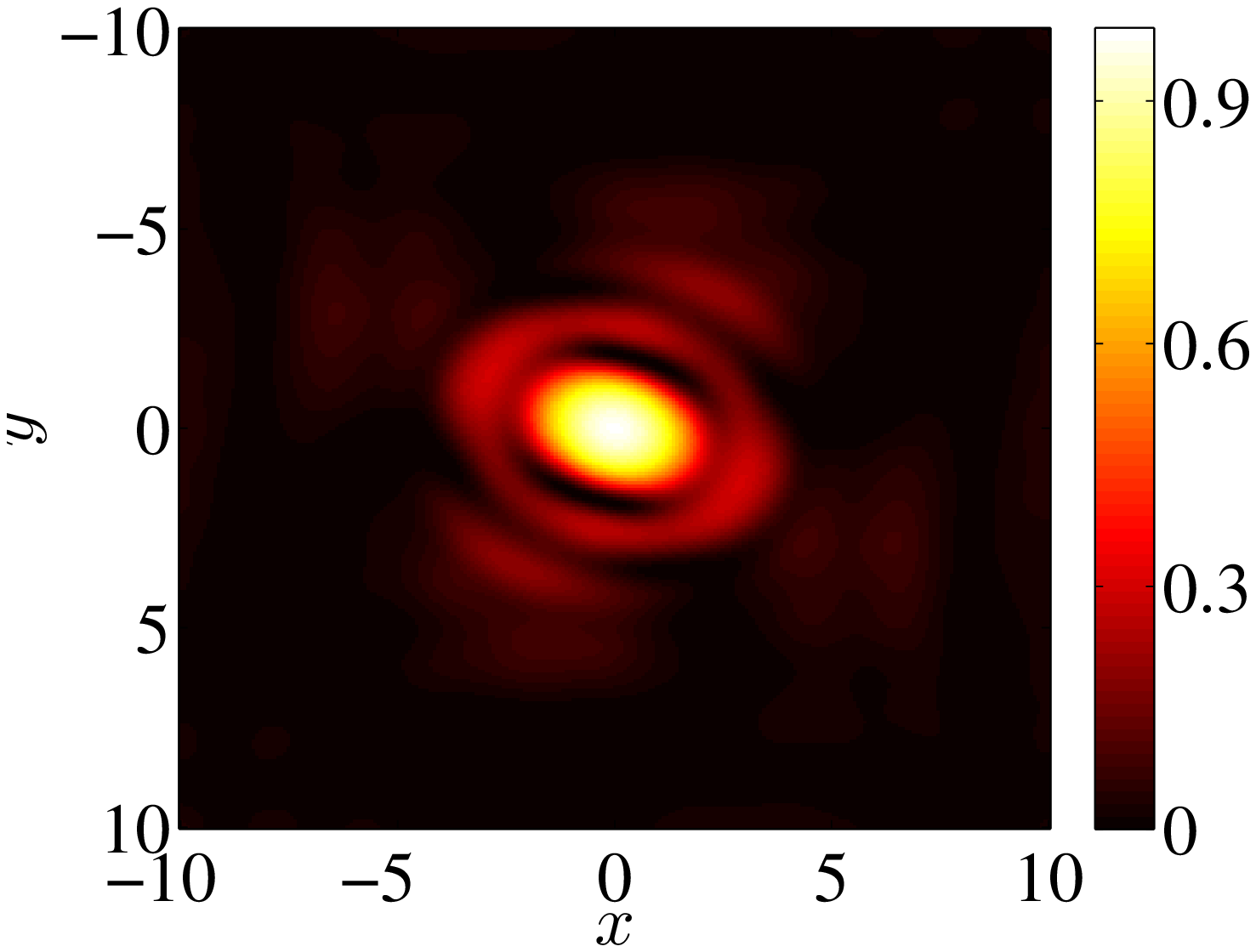}
\label{fig7f}
}
}
\end{center}
\par
\vspace{-0.55cm}
\caption{(Color online) The same as Fig.~\protect\ref{fig6}, but for the
complexes with the bright component in the form of the first excited state.
Top and bottom rows display densities $|\Phi _{-}(x,t)|^{2}$ and $|\Phi
_{+}(x,t)|^{2}$, respectively, at $t=0$ (left panels), $t=80$ (middle
panels), and $t=160$ (right panels) for $D=0.1$, $\protect\mu _{+}=0.97$ and
$m=2$.}
\label{fig7}
\end{figure}

\begin{figure}[tbp]
\begin{center}
\vspace{-0.1cm}
\mbox{\hspace{-0.1cm}
\subfigure[][]{\hspace{-0.3cm}
\includegraphics[height=.16\textheight, angle =0]{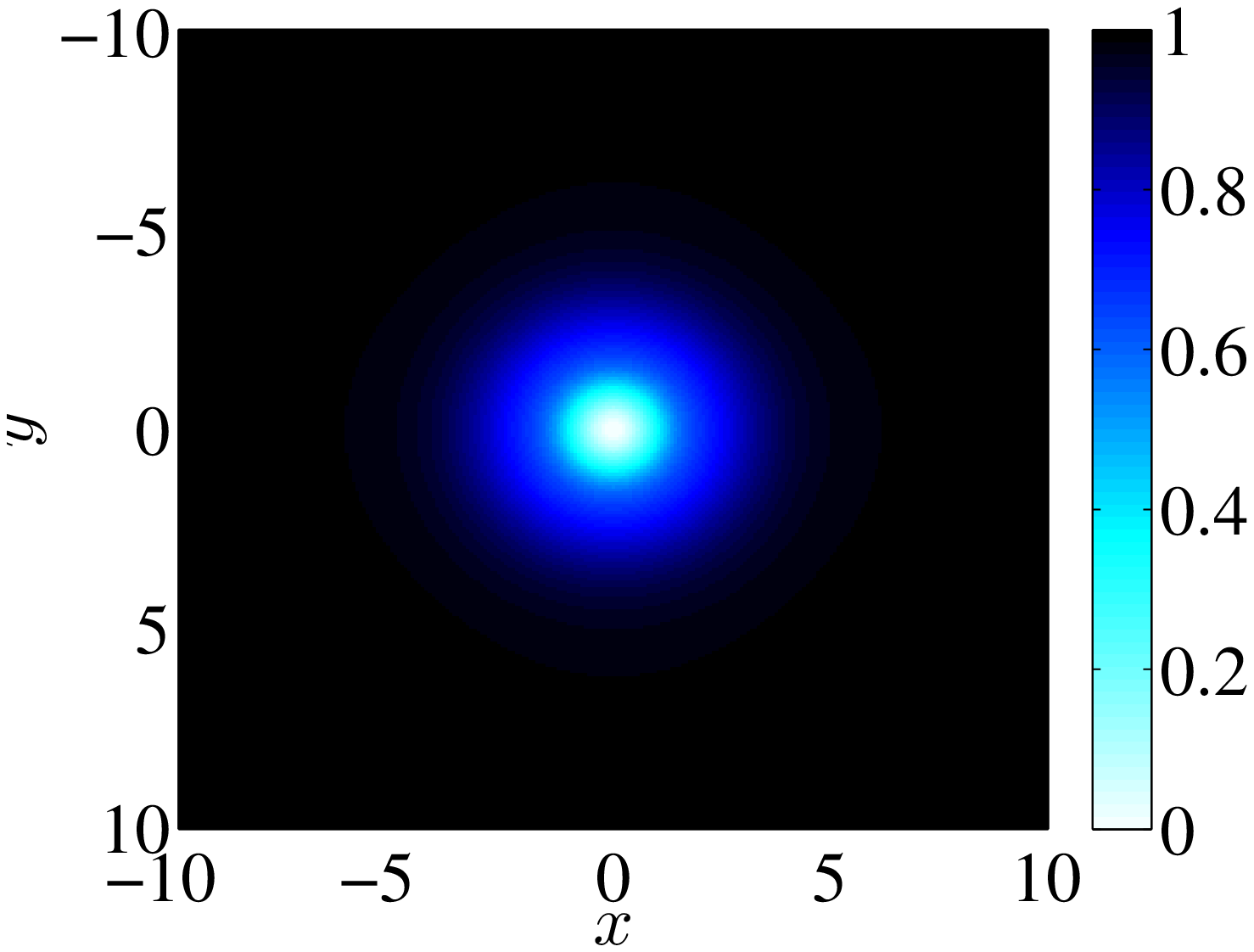}
\label{fig8a}
}
\subfigure[][]{\hspace{-0.3cm}
\includegraphics[height=.16\textheight, angle =0]{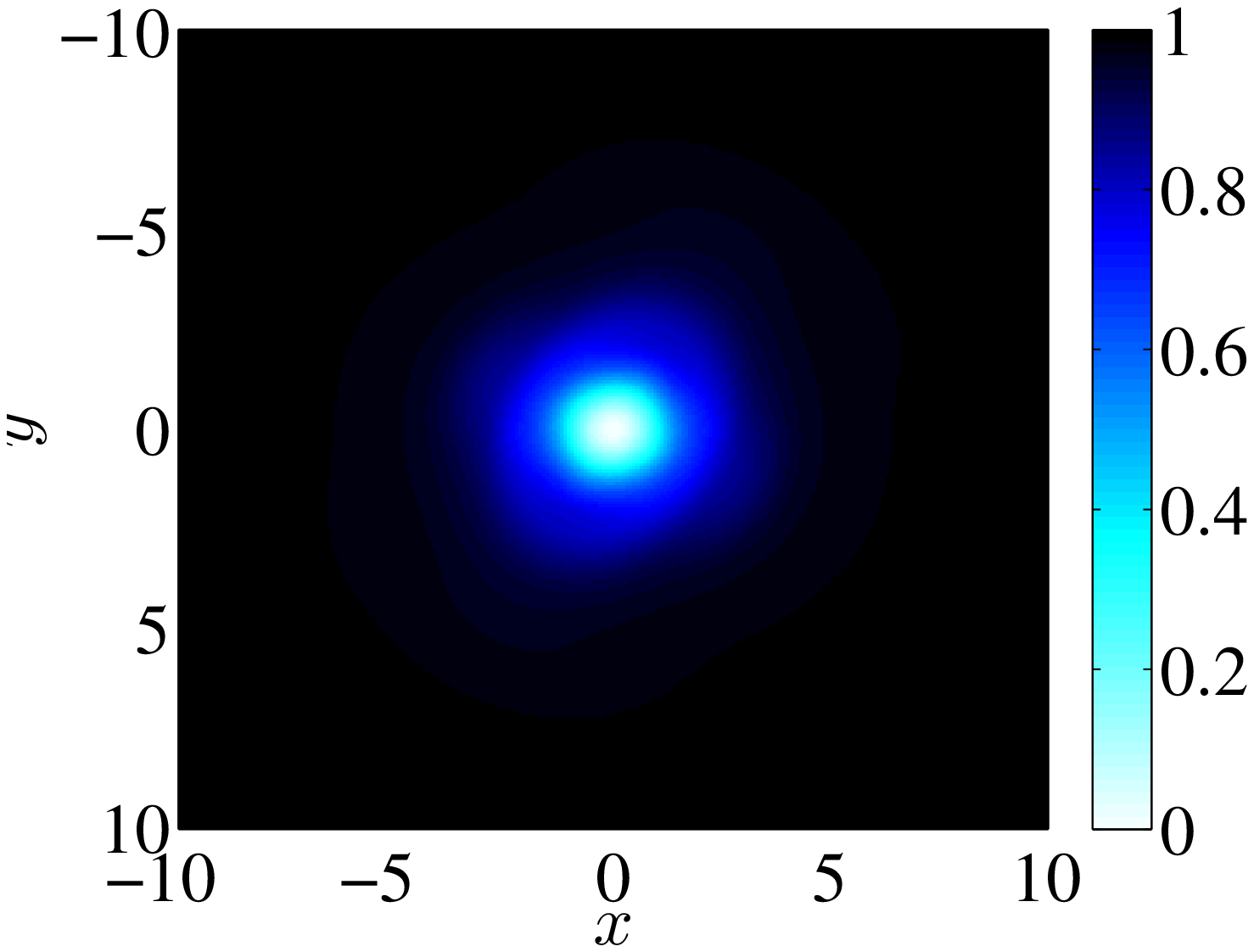}
\label{fig8b}
}
\subfigure[][]{\hspace{-0.3cm}
\includegraphics[height=.16\textheight, angle =0]{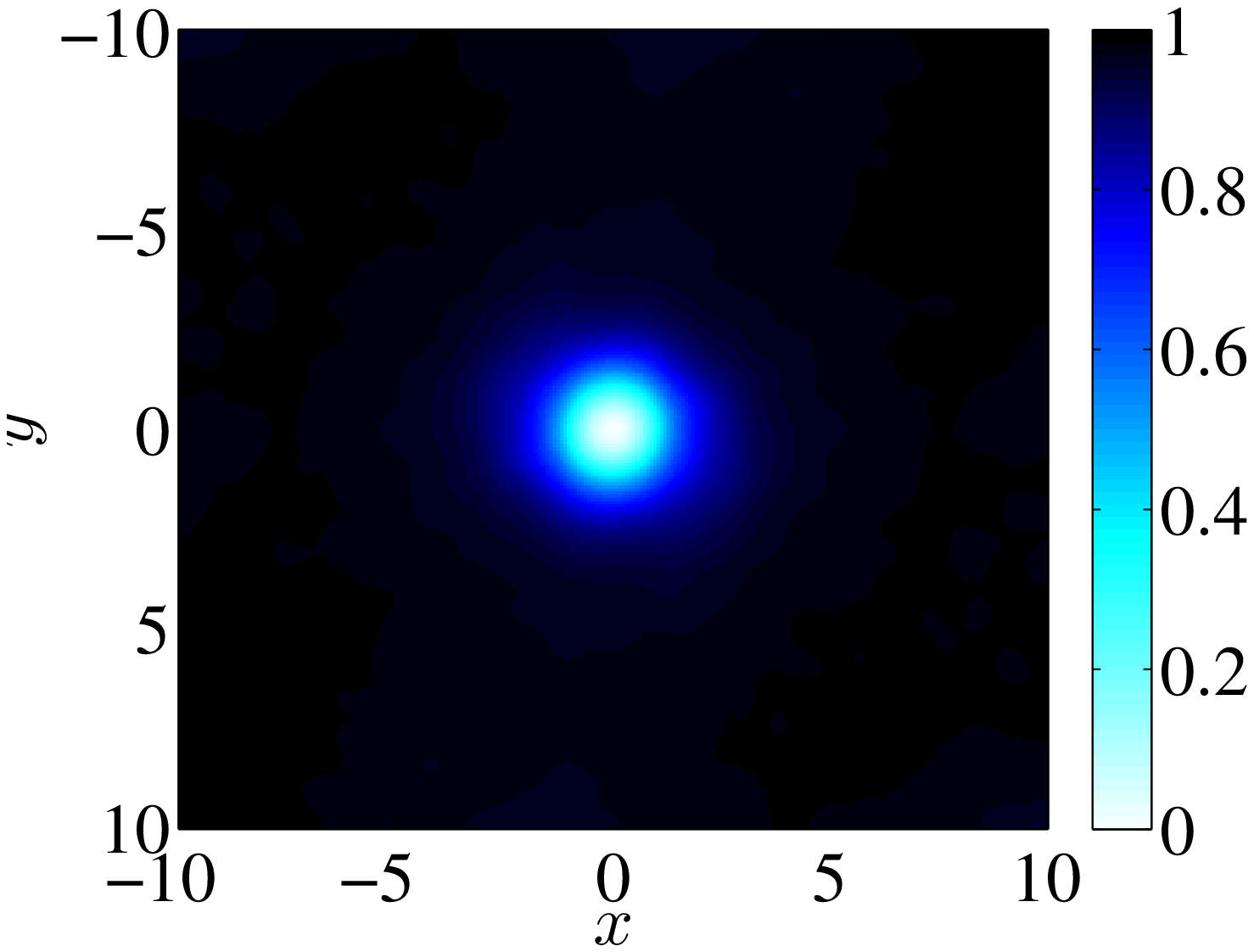}
\label{fig8c}
}
}
\mbox{\hspace{-0.1cm}
\subfigure[][]{\hspace{-0.3cm}
\includegraphics[height=.16\textheight, angle =0]{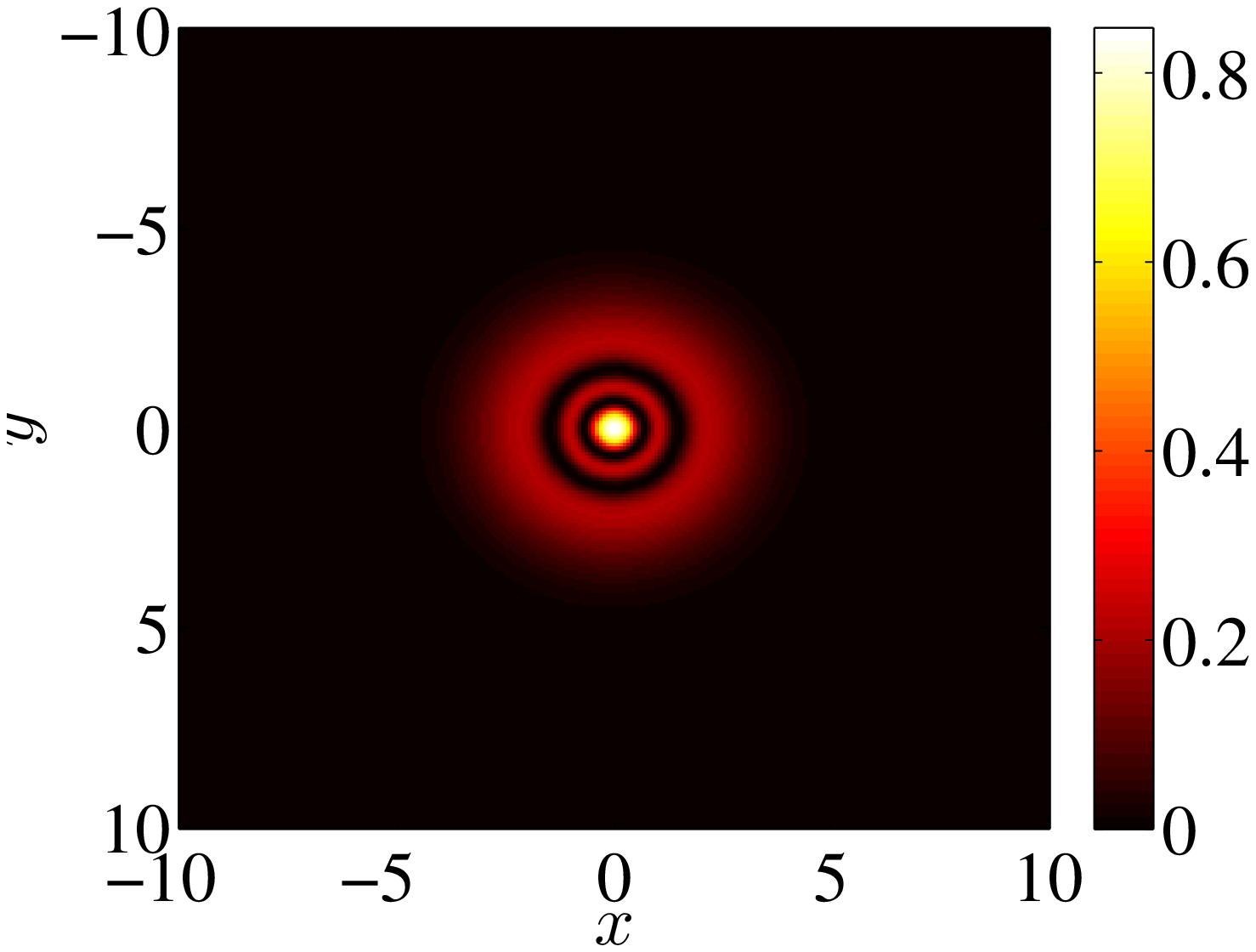}
\label{fig8d}
}
\subfigure[][]{\hspace{-0.3cm}
\includegraphics[height=.16\textheight, angle =0]{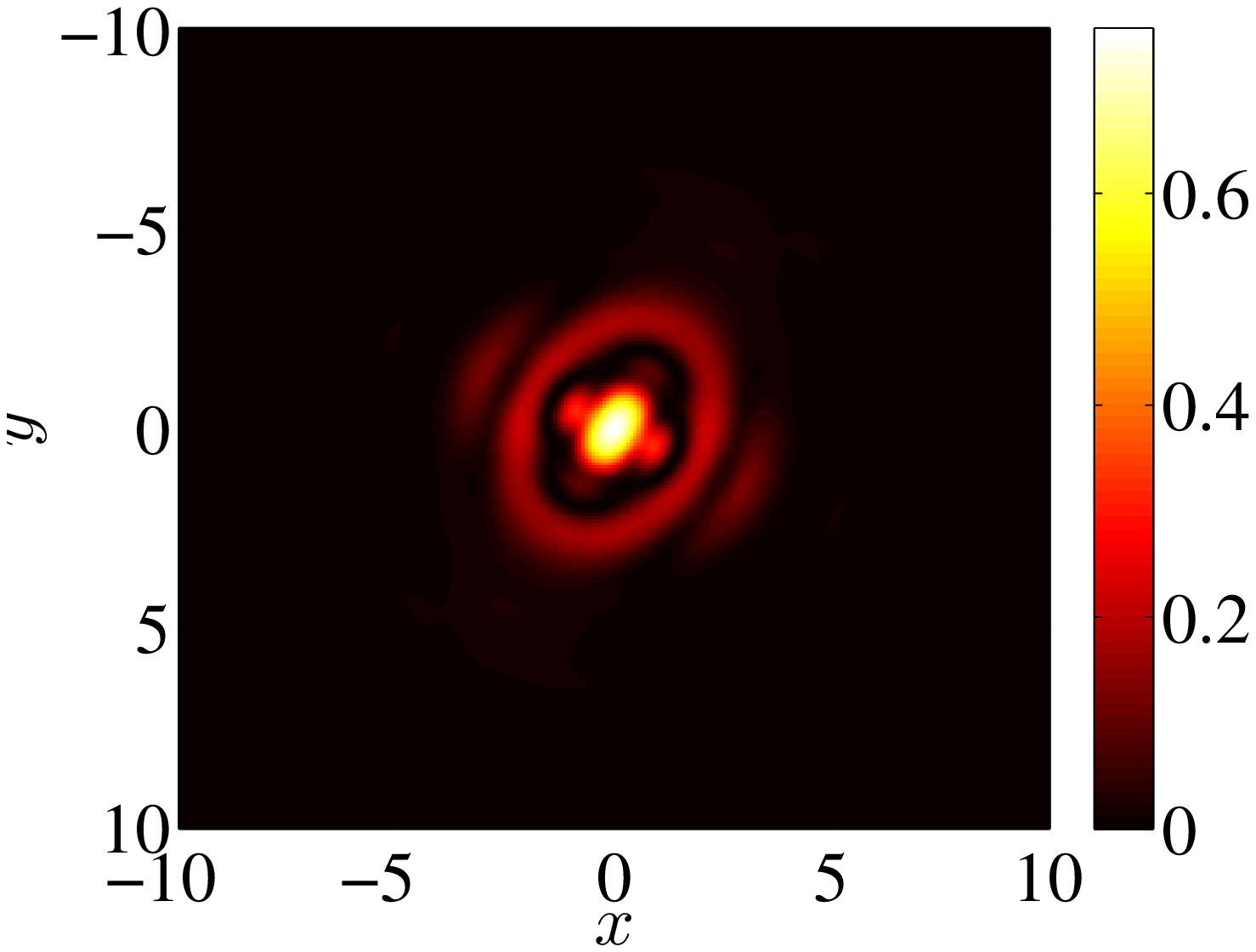}
\label{fig8e}
}
\subfigure[][]{\hspace{-0.3cm}
\includegraphics[height=.16\textheight, angle =0]{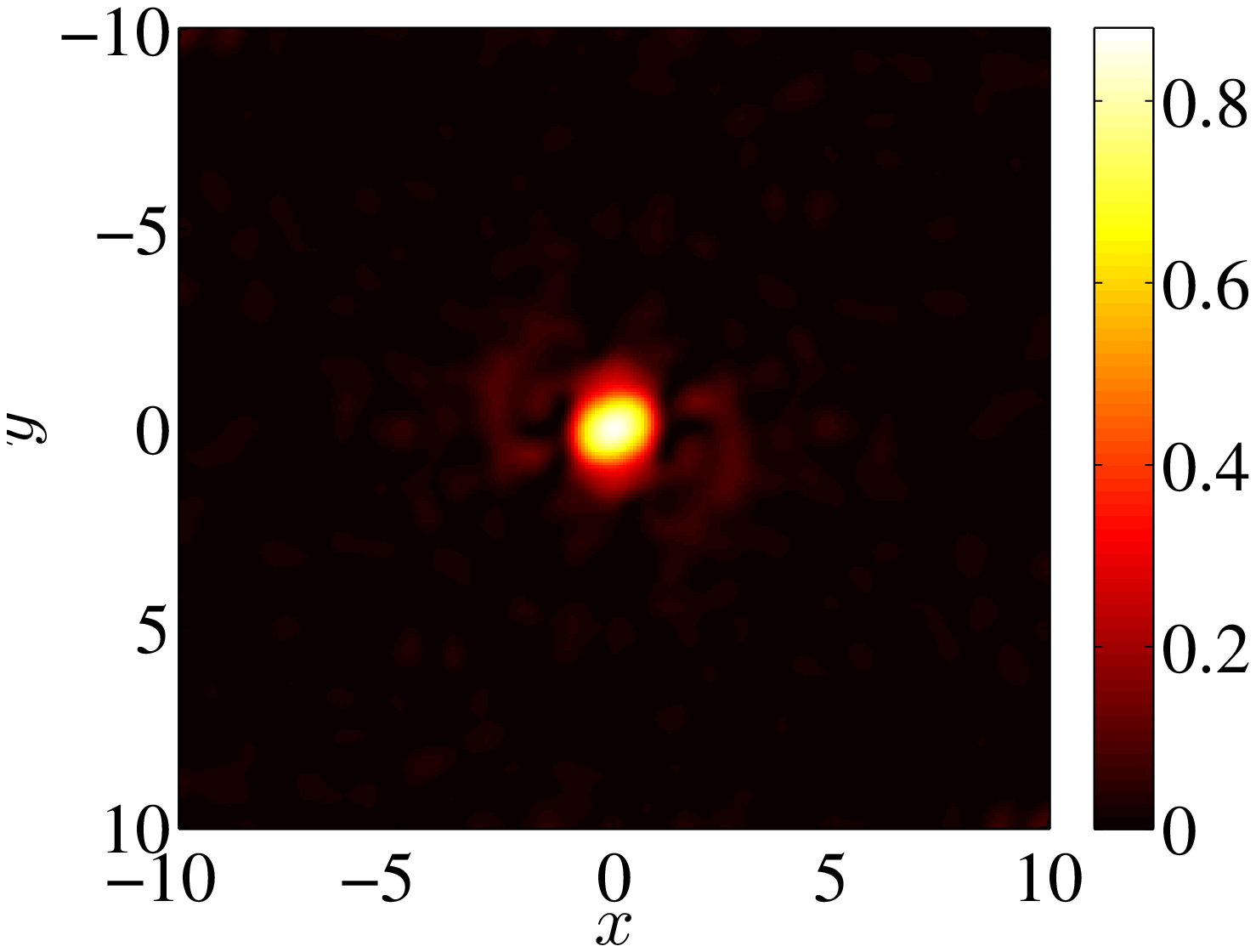}
\label{fig8f}
}
}
\end{center}
\par
\vspace{-0.55cm}
\caption{(Color online) Same as Fig.~\protect\ref{fig6} but for the
complexes with the bright component in the form of the second excited state.
Top and bottom rows display densities $|\Phi _{-}(x,t)|^{2}$ and $|\Phi
_{+}(x,t)|^{2}$, respectively, at $t=0$ (left panels), $t=210$ (middle
panels), and $t=430$ (right panels), for $D=0.05$, $\protect\mu _{+}=0.95$
and $m=2$.}
\label{fig8}
\end{figure}

\begin{figure}[tbp]
\begin{center}
\vspace{-0.1cm}
\mbox{\hspace{-0.3cm}
\subfigure[][]{\hspace{-0.3cm}
\includegraphics[height=.142\textheight, angle =0]{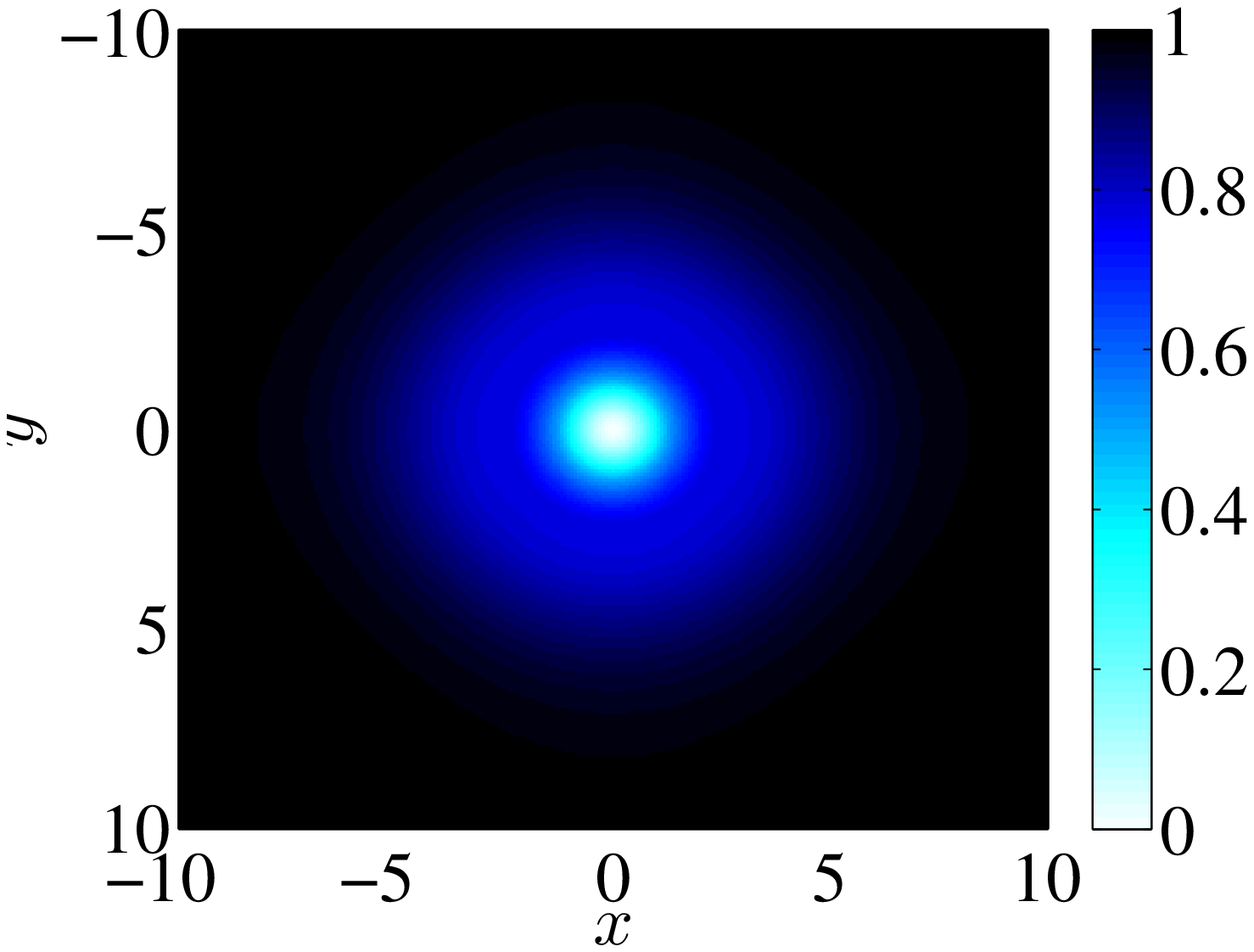}
\label{fig9a}
}
\subfigure[][]{\hspace{-0.3cm}
\includegraphics[height=.142\textheight, angle =0]{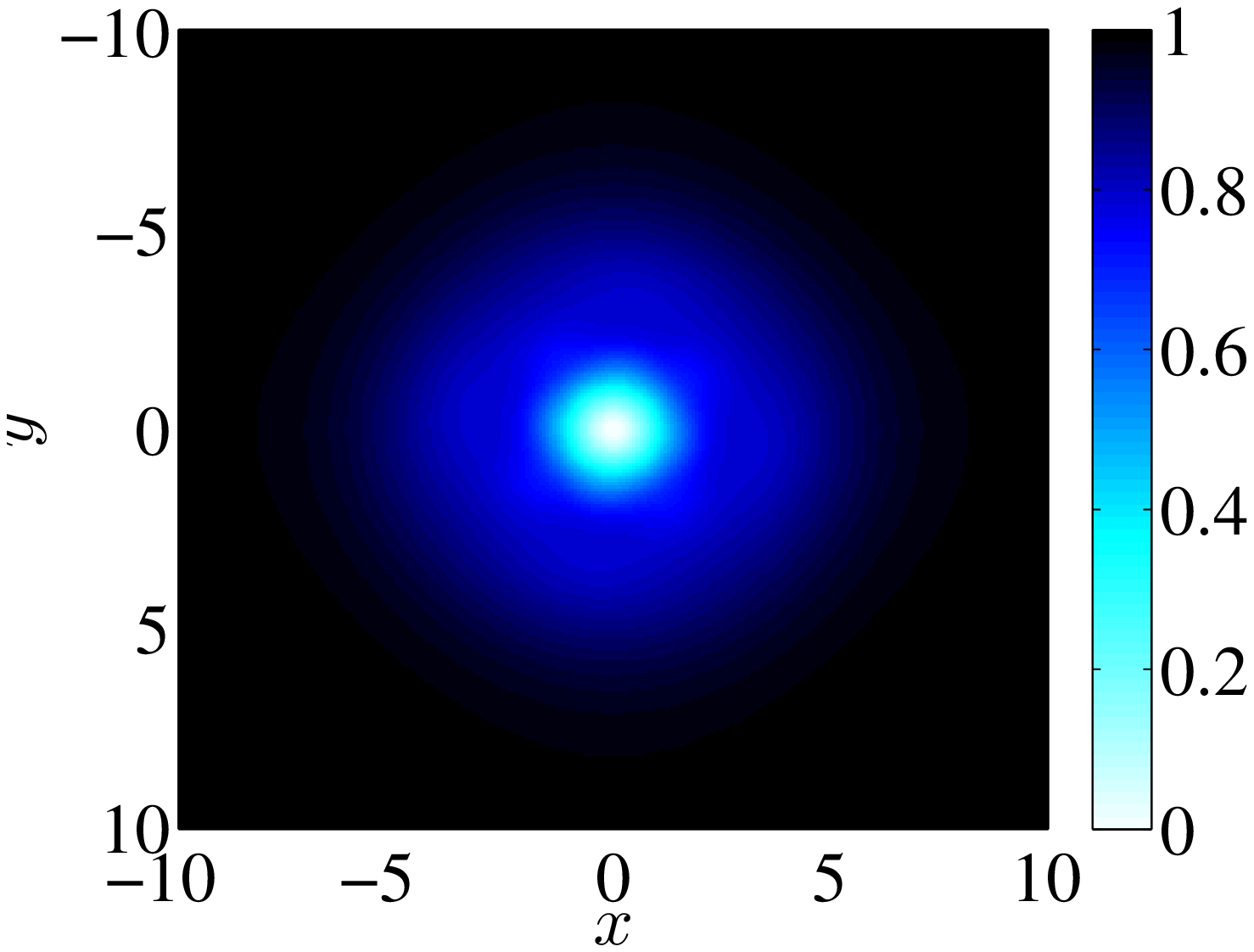}
\label{fig9b}
}
\subfigure[][]{\hspace{-0.3cm}
\includegraphics[height=.142\textheight, angle =0]{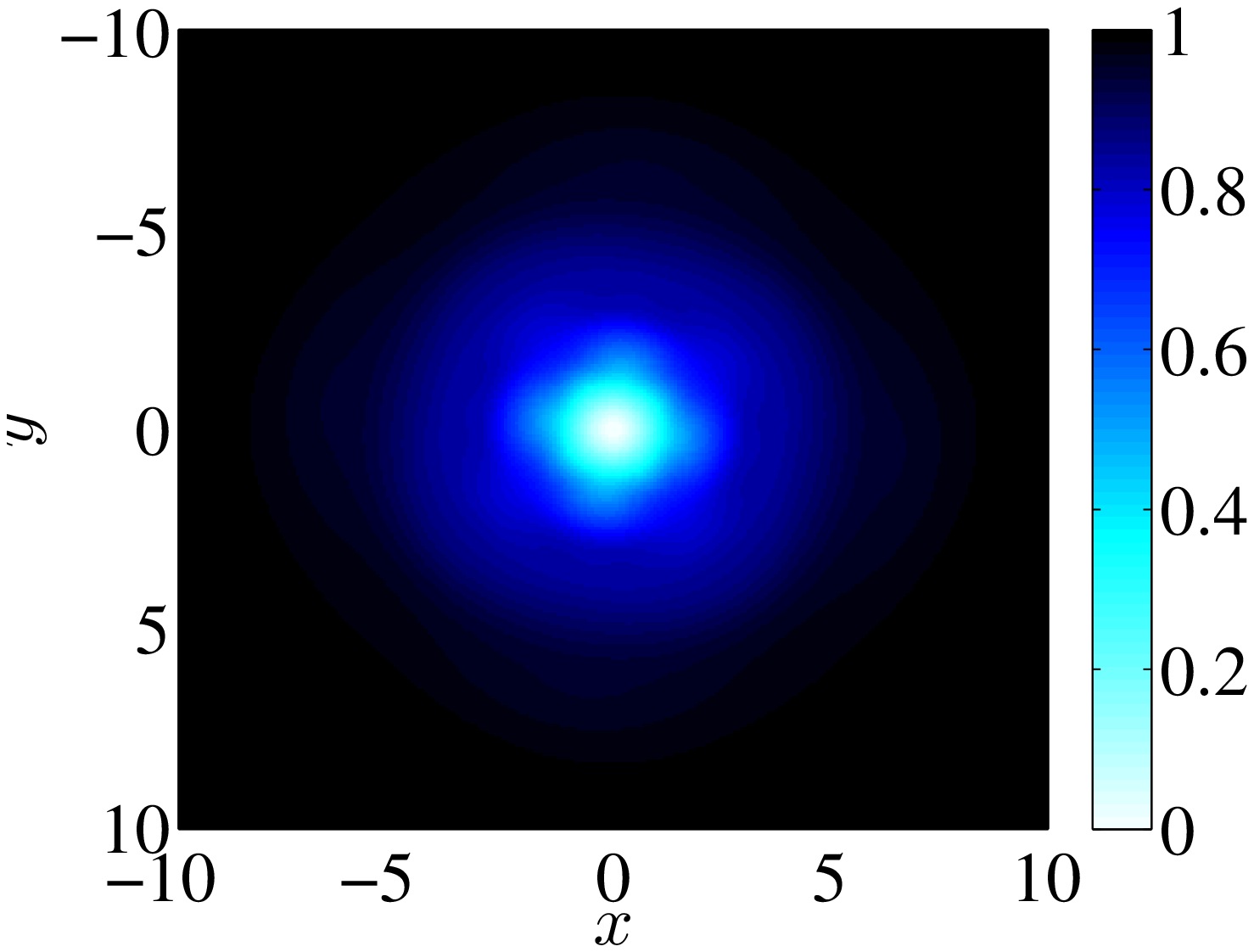}
\label{fig9c}
}
\subfigure[][]{\hspace{-0.3cm}
\includegraphics[height=.142\textheight, angle =0]{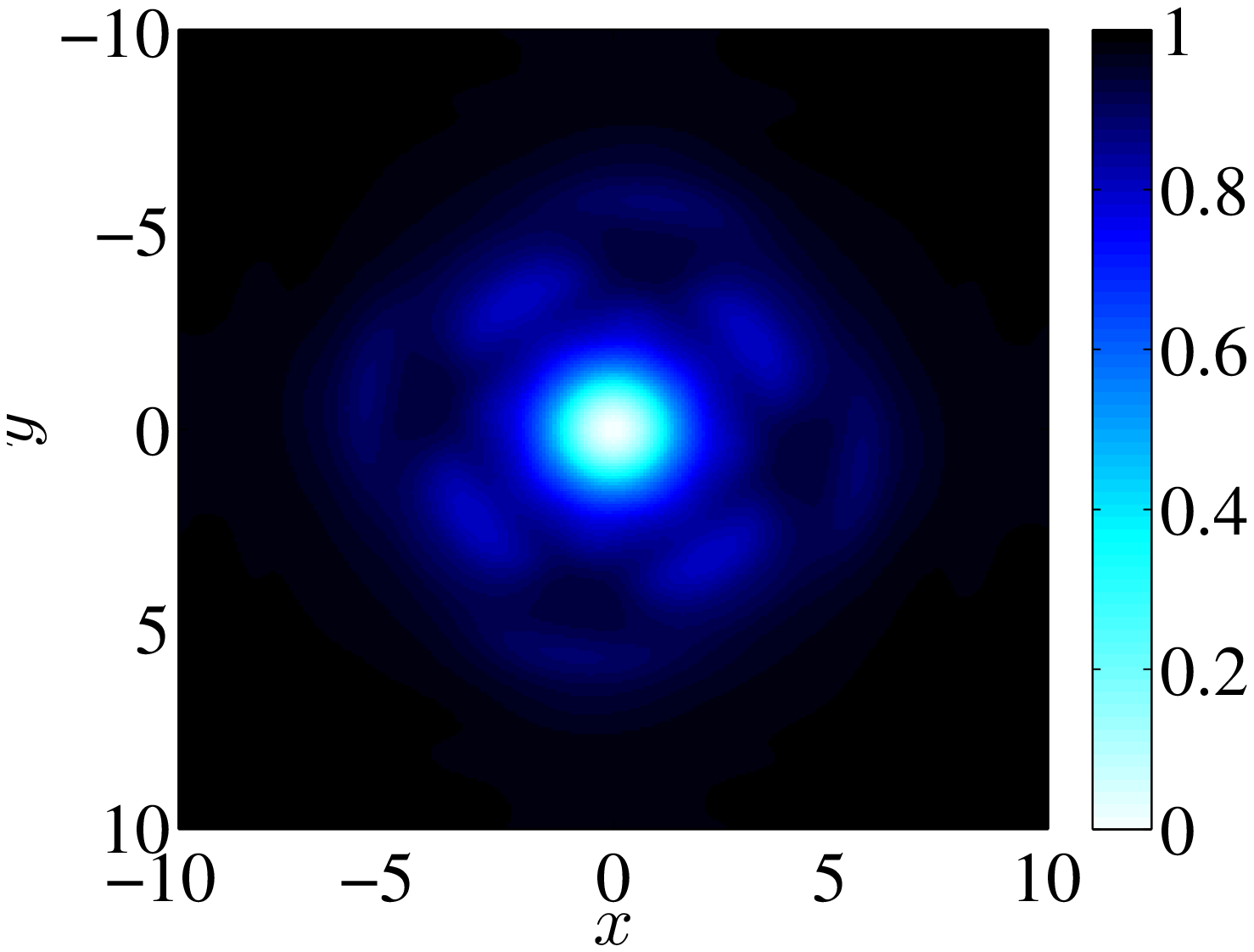}
\label{fig9d}
}
}
\mbox{\hspace{-0.3cm}
\subfigure[][]{\hspace{-0.3cm}
\includegraphics[height=.142\textheight, angle =0]{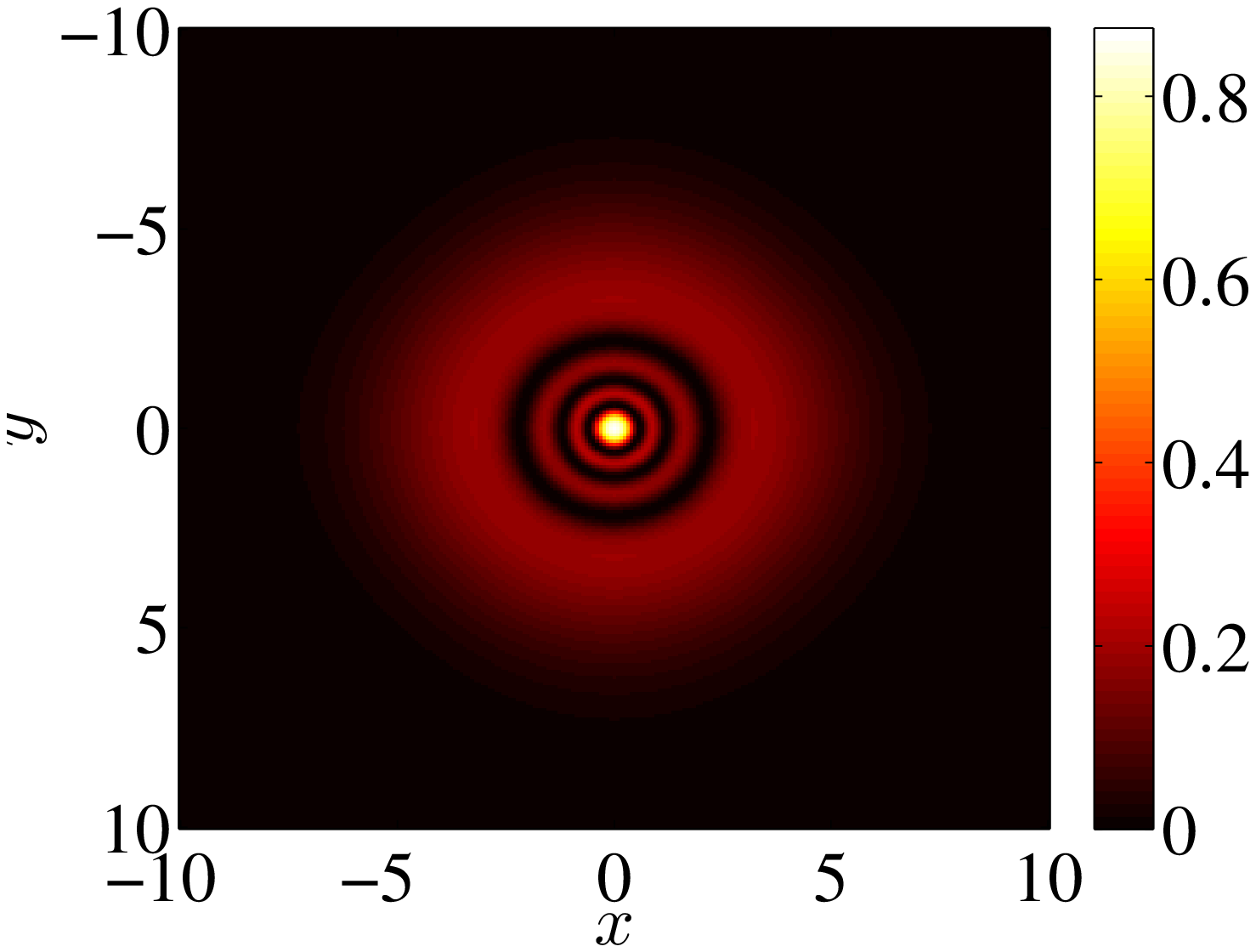}
\label{fig9e}
}
\subfigure[][]{\hspace{-0.3cm}
\includegraphics[height=.142\textheight, angle =0]{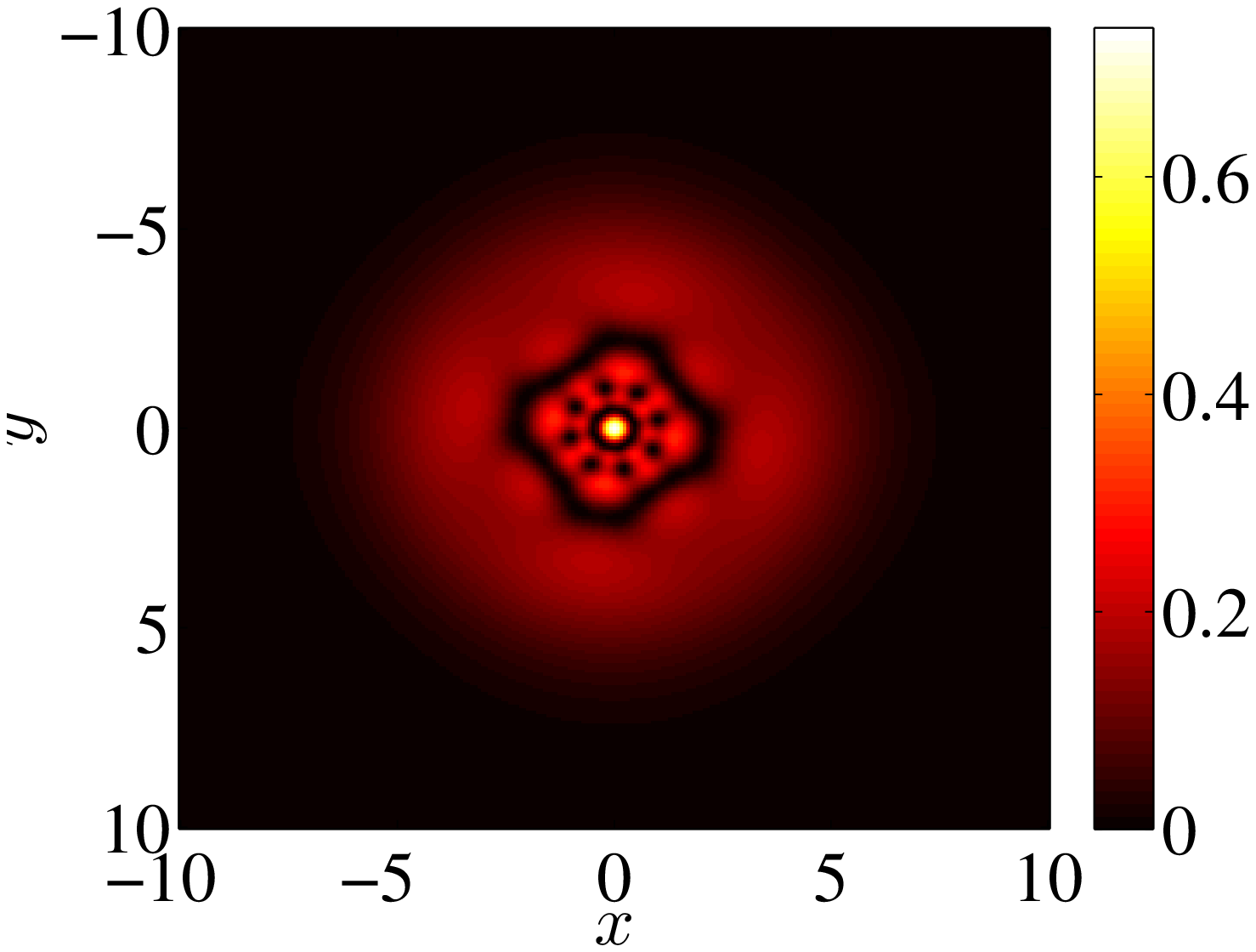}
\label{fig9f}
}
\subfigure[][]{\hspace{-0.3cm}
\includegraphics[height=.142\textheight, angle =0]{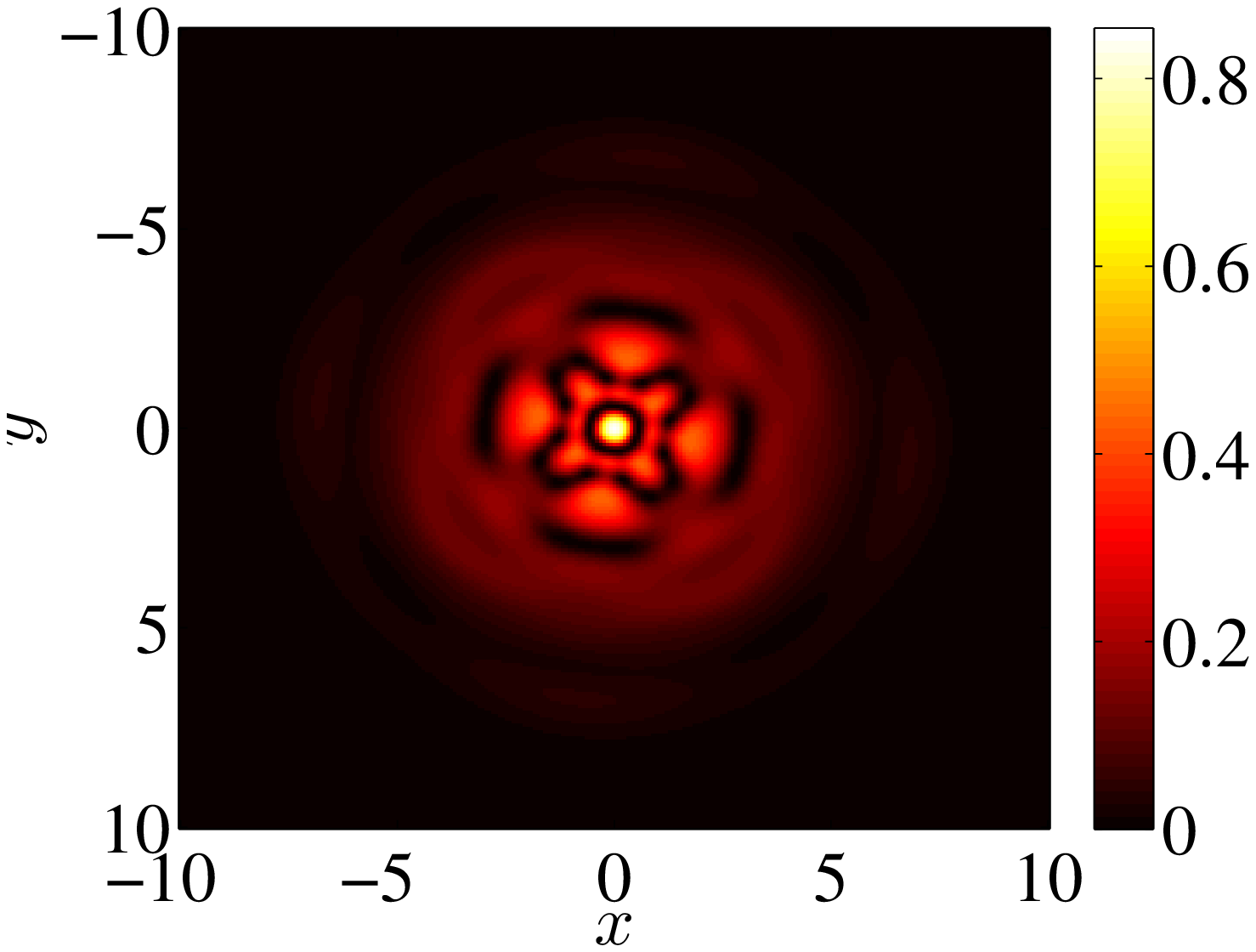}
\label{fig9g}
}
\subfigure[][]{\hspace{-0.3cm}
\includegraphics[height=.142\textheight, angle =0]{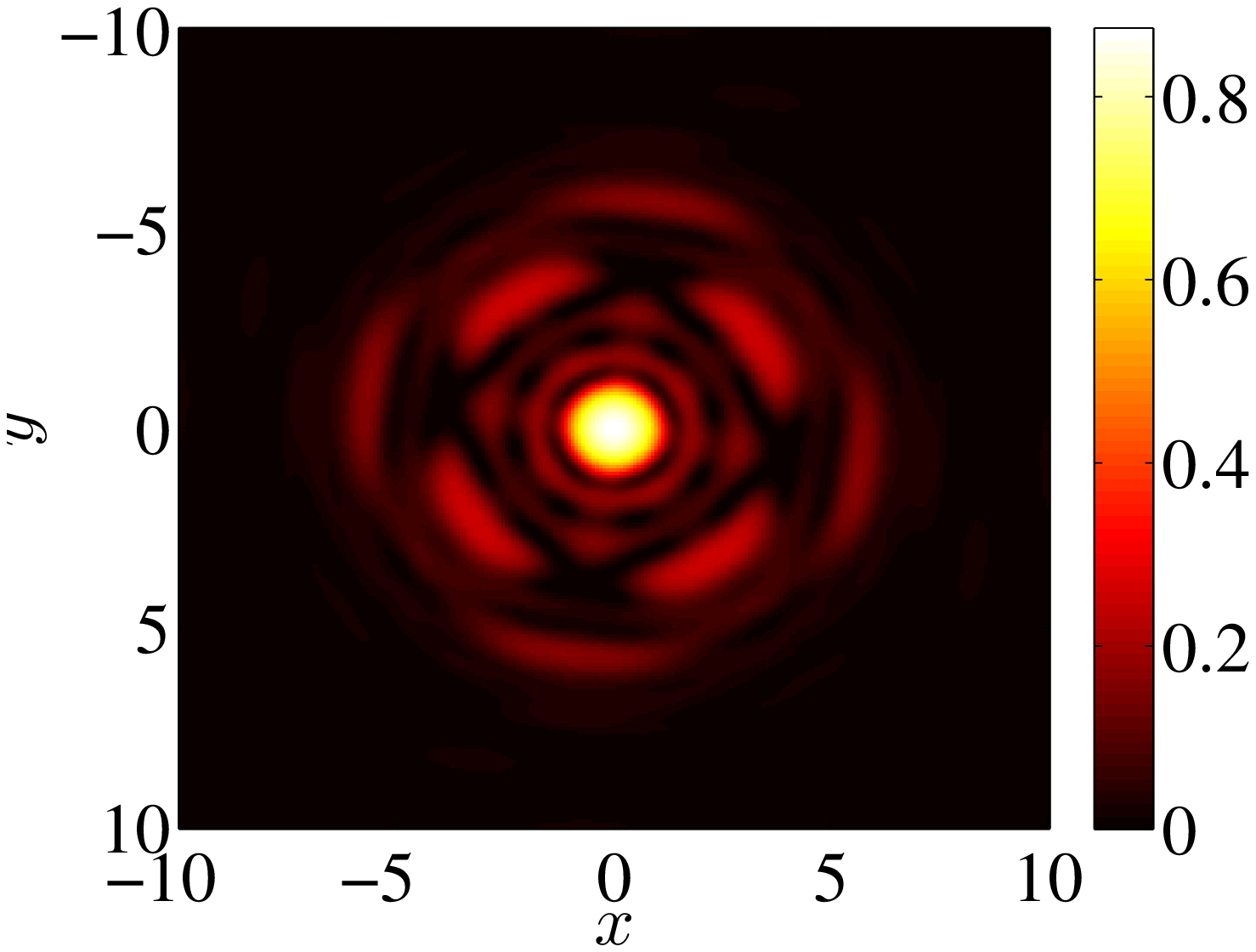}
\label{fig9h}
}
}
\end{center}
\par
\vspace{-0.55cm}
\caption{(Color online) Same as Fig.~\protect\ref{fig6}, but for the
complexes with the bright component represented by the third excited state.
Top and bottom rows display densities $|\Phi _{-}(x,t)|^{2}$ and $|\Phi
_{+}(x,t)|^{2}$, respectively, at $t=0$ (panels (a) and (e)), $t=130$
(panels (b) and (f)), $t=180$ (panels (c) and (g)), and $t=230$ (panels (d)
and (h)), for $D=0.04$, $\protect\mu _{+}=0.9825$ and $m=4$.}
\label{fig9}
\end{figure}

The situation is different for the excited states. This is observed, in
particular, in the evolution of the structure with the bright component
represented by the first excited state displayed in Fig.~\ref{fig7}; for the
second and third excited states in the bright component the same is shown in
Figs.~\ref{fig8} and \ref{fig9}, respectively. In the case of the first
excited state, we see in Fig.~\ref{fig7} that the shape becomes elongated,
resulting in the breakup of the dark density ring embedded into the bright
component. As a result, the bright component gradually transforms into the
GS (see, e.g., the right panel in the figure).

In the case of the second excited state shown in Fig.~\ref{fig8}, the
instability breaks the two dark rings embedded into the bright component. As
a result, more norm (or power, in terms of the optical model) migrates from
the outside rings towards the mode's core, pulled into the potential well
induced by the vortex in the mate component. In this case, the vortex
structure is only weakly affected by the instability of the bright
component. Eventually (see the panel on the right side of the figure), the
bright waveform builds a conspicuous maximum at the center, having shed off
considerable amount of radiation. Thus, this solution approaches the GS in
the bright component too, as a result of the instability development.

Finally, the waveform featuring the third excited state in the bright
component, characterized by a triple dark ring, exhibits complex evolution,
as seen in Fig.~\ref{fig9}. The rings get distorted, as is shown in the
second column of the figure --the outer nodal line is no longer a ring,
while the middle one has already been broken up. In the third column, the
outer and middle dark-ring patterns are severely distorted, resulting,
eventually (in the right column), in the transfer of the norm of the bright
component towards the center, although surrounded by a complex pattern
involving multiple nodal structures. Thus, one again sees a trend for the
aggregation of the norm of the bright component at the center, implying
spontaneous rearrangement of the mode into the GS. Here (as well as in the
case of the bright component shaped as the first excited state), the vortex
component suffers a more significant feedback from the instability
development in the bright one, resulting in complex patterns observed in the
dark component too. Nevertheless, the central core of the vortex remains
intact, thus maintaining the effective potential trapping the bright
waveform.

\section{Conclusion}

We have considered the two-component GP/NLS 2D system, chiefly with equal
strengths of the self- and cross-defocusing cubic nonlinearities, in which a
vortex in one component induces an effective trapping potential for the
other (bright) component. The system models heteronuclear BEC mixtures and
the copropagation of optical beams carried by different wavelengths.

Depending on the relative dispersion parameter of the second component, the
effective potential can trap not only the GS (ground state) in the bright
component, but also the first, second and even third excited radial states.
This results in complexes with multi-ring solitons in the bright component,
which produce a feedback on the vortex in the first component. Among these
complexes, the VB (vortex-bright) one, with the GS in the bright component,
has been identified as a spectrally stable state, which is, accordingly,
robust in the direct evolution.

On the other 
hand, the complexes with the bright component represented by the excited
states are unstable, although the instability growth rate is broadly
tunable, and may be made very small, by means of the variation of the
relative dispersion parameter of the bright component. If the
harmonic-oscillator trapping potential, which is experimentally relevant in
atomic BECs, is included, the complete stabilization of the structures with
the excited state of the bright component can be achieved in suitable
parametric regions. The stabilization may also be provided by unequal
strengths of self- and cross-repulsive interactions, as shown in Fig. \ref%
{fig3_comp_s12}. The spectral stability and instability, predicted by the
analysis of small perturbations, were corroborated by direct simulations. In
particular, the unstable complexes with the excited bright component have
been shown to spontaneously rearrange into VB modes with the GS in the
bright constituent.

This work paves the way for exploration of related systems. First, we
actually considered only the bright component with zero vorticity, $n=0$ in
Eq.~(\ref{+}) (i.e., without the angular momentum). The existence and
stability of complexes with a vortical bright component, $n\neq 0$, is a
very relevant generalization. In particular, the stability may be quite
different for the same shape of the vortical bright mode with opposite signs
of the vorticity, $n=\pm 1$, while $S=+1$ is fixed in the first component,
cf. the stability of two-component trapped modes with the \textit{hidden
vorticity} studied in Ref. \cite{Nal}. Further, in this work we restrict the
considerations to 2D settings, while recent work \cite{adhi} has shown~that
3D vortical structures are capable of trapping bright states. Another
possibility, that we only briefly broached here, is to systematically
consider the states formed in the presence of the harmonic-oscillator
trapping potential, which is necessarily present in experiments with atomic
BECs. Finally, the present analysis is restricted to axially symmetric
states trapped by the vortex-induced effective potential. It is also
interesting to check if azimuthally modulated states (\textit{azimuthons}~%
\cite{desya}) may be produced in the present setup. Some of these extensions
are presently under consideration, and will be reported elsewhere.

\begin{acknowledgments}
P.G.K. and D.J.F. gratefully acknowledge the support of NPRP8-764-1-160. E.G.C. is 
indebted to the Department of Physical Electronics, School of Electrical Engineering
at the Tel Aviv University for hospitality. This author thanks Hans Johnston (UMass)
for providing help in connection with the parallel computing performed in this work. 
P.G.K. acknowledges support from the National Science Foundation under Grant
DMS-1312856 and from FP7-People under Grant No. IRSES-605096. The work of
D.J.F. was partially supported by the Special Account for Research Grants of
the University of Athens. The work of P.G.K., E.G.C., and B.A.M. was
supported in part by the U.S.-Israel Binational Science Foundation through
Grant No. 2010239.
\end{acknowledgments}


\end{document}